\documentclass[12pt]{article}
\usepackage{amsmath,amssymb,color,epsfig,subfigure,a4wide}

\makeatletter
\renewcommand{\theenumi}{\Roman{enumi}}

\renewcommand{\p@enumii}{\theenumi.}
\makeatother

\begin{document}
\begin{titlepage}
\hspace*{\fill}\parbox[t]{5cm}
{CP3-07-29\\
\today} \vskip2cm
\begin{center}
{\Large \bf  Top pair invariant mass distribution:\\[10pt] 
a window on new physics} \\
\medskip
\bigskip\bigskip\bigskip\bigskip
{\large  {\bf Rikkert Frederix} and  {\bf Fabio Maltoni}}\\
\bigskip
Centre for Particle Physics and Phenomenology (CP3) \\
Universit\'{e} Catholique de Louvain\\[1mm]
Chemin du Cyclotron 2 \\
B-1348 Louvain-la-Neuve, Belgium \\
\end{center}

\bigskip\bigskip\bigskip

\begin{abstract}
We explore in detail the physics potential of a measurement of the $t
\bar t$ invariant mass distribution. First, we assess the accuracy of
the best available predictions for this observable and find that in
the low invariant mass region, the shape is very well predicted and
could be used to perform a top mass measurement. Second, we study the
effects of a heavy $s$-channel resonance on the $t \bar t$ invariant
mass distribution in a model independent way.  We provide the
necessary Monte Carlo tools to perform the search and outline a simple
three-step analysis.
\end{abstract}

\end{titlepage}

\section{Introduction}
\label{sec:intro}

The top quark is unique among the so-far-discovered matter
constituents: it is the only fermion whose mass is very close to the
scale of electroweak symmetry breaking (EWSB), $m_t \simeq
v/\sqrt{2}$, and therefore it has a naturally strong coupling,
$\lambda_t \simeq 1$, to the Higgs boson in the standard
model.\footnote{Another numerical coincidence is also often mentioned,
namely $m_t\simeq m_W+m_Z$, with somewhat less inspirational effects.}
It is because of this large coupling, for instance, that the Higgs
mass can be predicted via precision measurements, and the Higgs can be
copiously produced at hadron colliders operating at the TeV scale, via
its top-loop mediated interactions to gluons.  Thanks to its large
mass, the top quark has also been exploited in many scenarios that go
beyond the standard model (BSM). The simplest one is SUSY, where top
has the important role of triggering EWSB and the historical merit of
having escorted the MSSM to the LHC era by
allowing the Higgs mass to survive to the LEP bounds.

Top has also been the source of inspiration for many alternative
mechanisms of EWSB, where it has a direct or indirect role in the
boson as well as fermion mass generation:
Topcolor~\cite{Hill:1991at,Hill:1994hp,Hill:1993hs,Harris:1999ya}, top
see-saw~\cite{Dobrescu:1997nm,Chivukula:1998wd,He:2001fz} and other
refinements~\cite{Hill:2002ap}.  More recently, in other theoretical
constructions that aim at the stabilization of the Higgs mass, the
presence of the top quark requires a mechanism that compensates for
its large (negative) radiative corrections to the Higgs mass.  In
general such mechanisms entail the existence of new particles, {\it
i.e.}, top partners, that can be scalars, like in SUSY, as well as
fermions, like in the Little
Higgs~\cite{ArkaniHamed:2001nc,ArkaniHamed:2002pa,ArkaniHamed:2002qx,ArkaniHamed:2002qy,Low:2002ws,Han:2003wu,Azuelos:2004dm,Schmaltz:2005ky}
or models with extra
dimensions~\cite{ArkaniHamed:1998rs,Randall:1999ee}.  In addition, in
many such models gauge interactions exist whose coupling with the
third generation quarks and in particular to the top quark are
enhanced. These include Kaluza-Klein (KK) excitations of the
graviton~\cite{Fitzpatrick:2007qr,Arai:2004yd,Arai:2007ts} as well as
the weak~\cite{McMullen:2001zj} and the strong gauge
bosons~\cite{Agashe:2003zs,Agashe:2006hk,Lillie:2007yh,Djouadi:2007eg,Fitzpatrick:2007qr,Ghavri:2006kc,Burdman:2006gy,Lillie:2007ve,Agashe:2007ki}
which couple to top quarks.  Such particles could show up as
resonances in the $pp \to X \to t\bar t$ production channel and not in
other channels, like di-jets or di-leptons, due to their small couplings to
light particles.

If on the one hand, the fact that many rich and theoretically
motivated models predict new physics in connection with the top quark
provides a strong motivation for detailed experimental investigations,
on the other it makes difficult to perform the analyses corresponding
to all the suggested scenarios.  A way to avoid such an ``explosion''
of models is a bottom-up approach, where first a physical observable,
which carries some potential for BSM studies, is identified and then
the effects due to the existence of new physics are systematically
explored. The aim of this paper is to present such a model independent
approach to the discovery and identification of new physics in
$m_{t\bar t}$, the top-antitop invariant mass distribution.  We do not
focus on specific models, instead we assume the existence of heavy
particles, whose masses and quantum numbers are unknown but are such
that interactions with the top quark are privileged, a feature common
to many scenarios where the top quark plays a role in the EWSB, as
discussed above. To simulate the physics associated with the new
states, we have developed a dedicated ``model'' in the multipurpose
Monte Carlo generator
MadGraph/MadEvent~\cite{Stelzer:1994ta,Maltoni:2002qb,Alwall:2007st}.\footnote{The Monte Carlo and a wide collection of parton-level data
samples (Les Houches format) suitable for further experimental
analysis on the scenarios presented in this work, are available on the
MadGraph servers, {\it e.g.}, {\tt http://madgraph.phys.ucl.ac.be}.}

The strategies and the difficulties associated to the accuracy which
will be needed for the reconstruction of $m_{t\bar t}$ have been the
subject of several investigations~\cite{Barger:2006hm,Baur:2007ck} and
are left to dedicated experimental studies. However, we briefly comment
on them in Appendix A.

The paper is organized as follows. In Section~\ref{sec:SM}, we perform
an analysis of the theoretical uncertainties on the QCD predictions
for $m_{t\bar t}$, in the low- as well as high-mass regions.  We find
that the shape of the $m_{t\bar t}$ distribution has very little
theoretical uncertainties (in contrast to its normalization, {\it
i.e.}, the total cross section) and we argue it could provide a new
handle on the top mass determination.

In Sections~\ref{sec:BSM}, \ref{sec:spin_info} and~\ref{sec:spin_corr} we
explore a three-step analysis. In Section~\ref{sec:BSM} the
effects on $m_{t\bar t}$ induced by new heavy
resonances in the s-channel, $pp\to X \to t\bar t$, from the presence
of simple peaks, to non-trivial patterns arising from interference
between signal and SM background.  In the following section we assume that
a resonance is found and study
how the spin structure of a resonance affects the angular
distributions of the top and anti-top quarks.  In
Section~\ref{sec:spin_corr} we discuss how spin-correlations are affected by
nature of the coupling of the resonances to SM particles,
by simulating the full matrix elements including the decay,
$pp \to X \to t\bar t \to 6f $.
We leave to Section~\ref{sec:con} the discussion of the
results and our conclusions.

\section{SM theoretical predictions at NLO}\label{sec:SM}

\begin{figure*}[t]
  \begin{center}
    \mbox{
      \subfigure[]{
        \epsfig{file=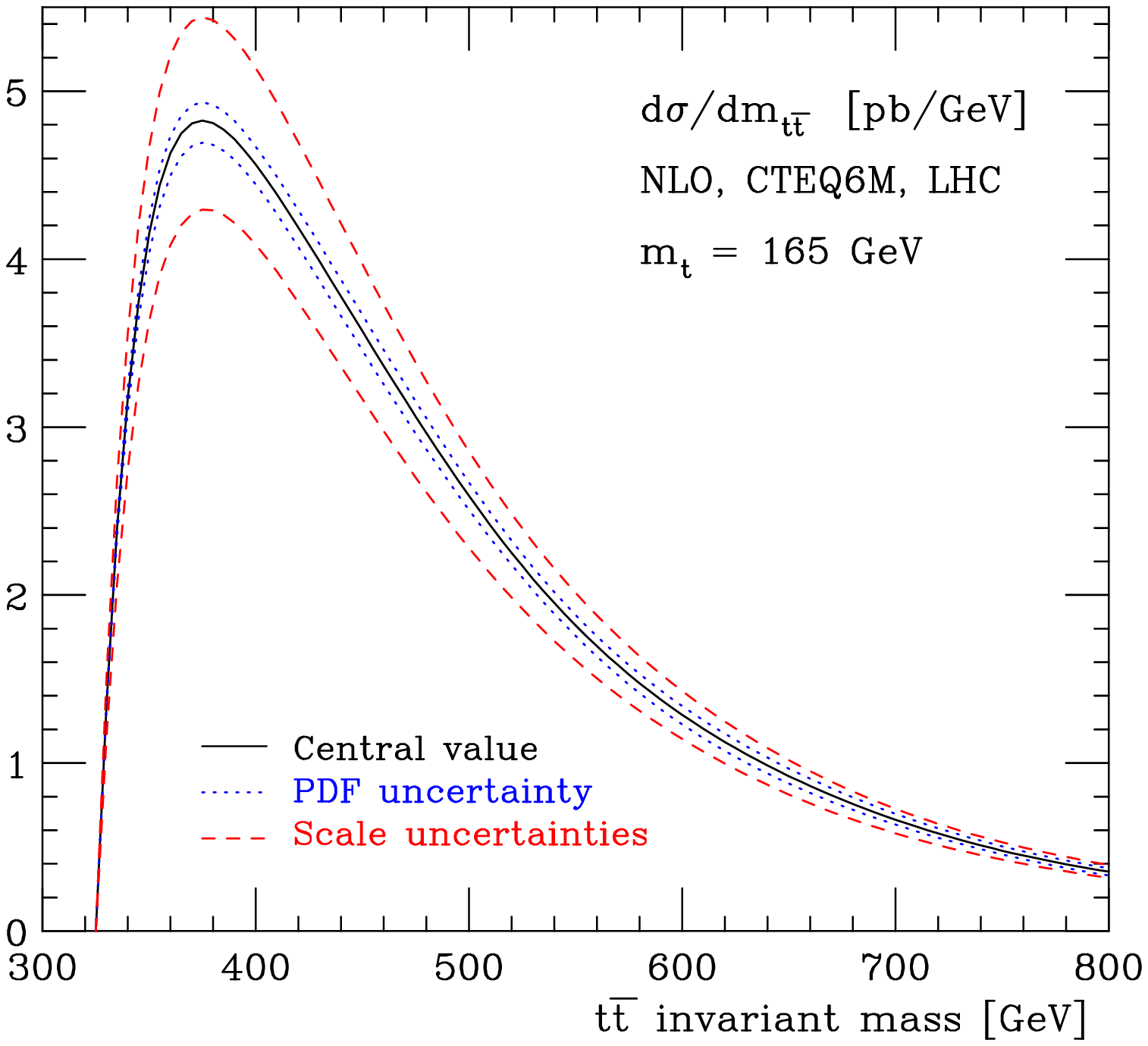, angle=90, width=0.30\textwidth}
      }
      \subfigure[]{
        \epsfig{file=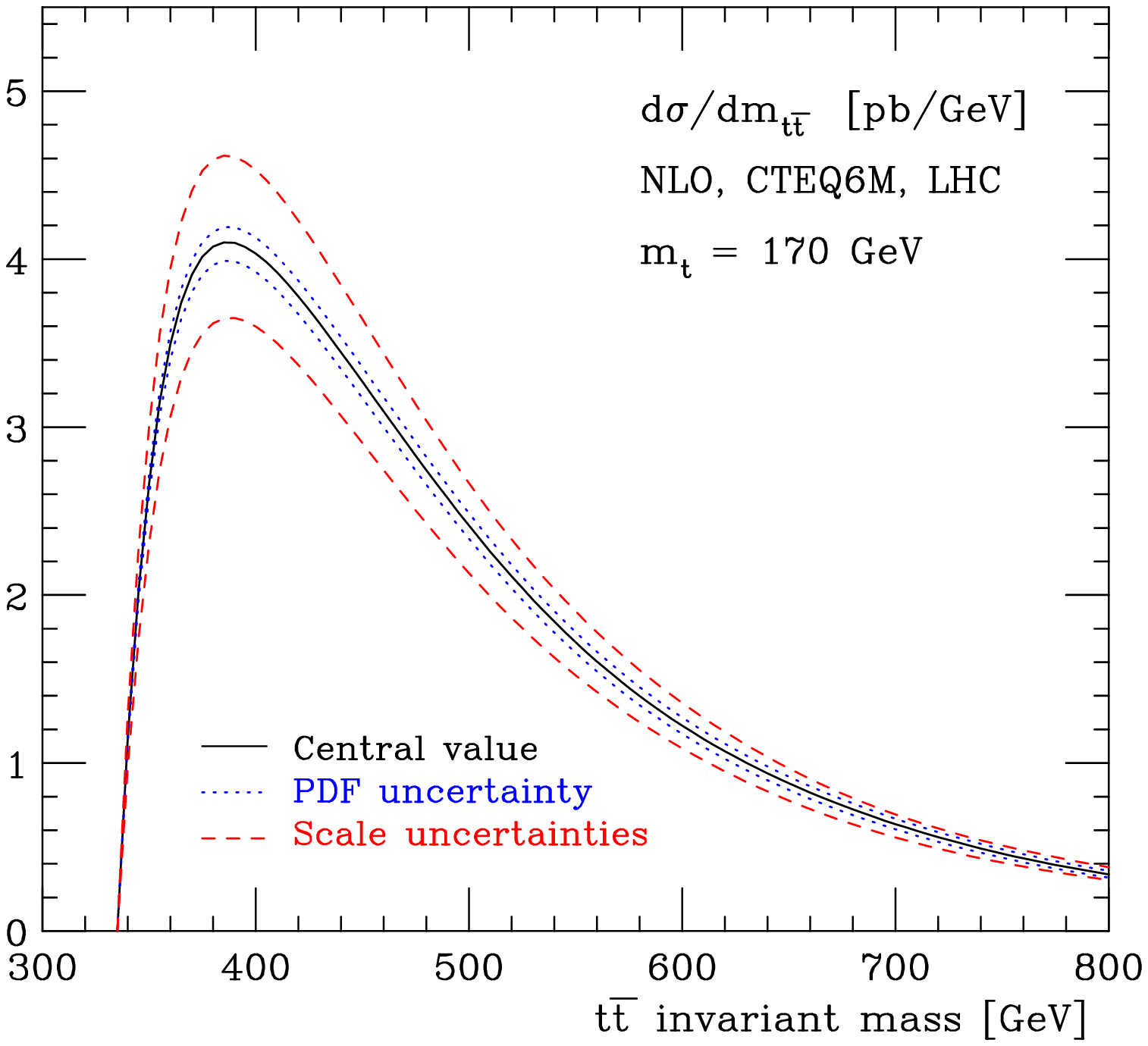, angle=90, width=0.30\textwidth}
      }
      \subfigure[]{
        \epsfig{file=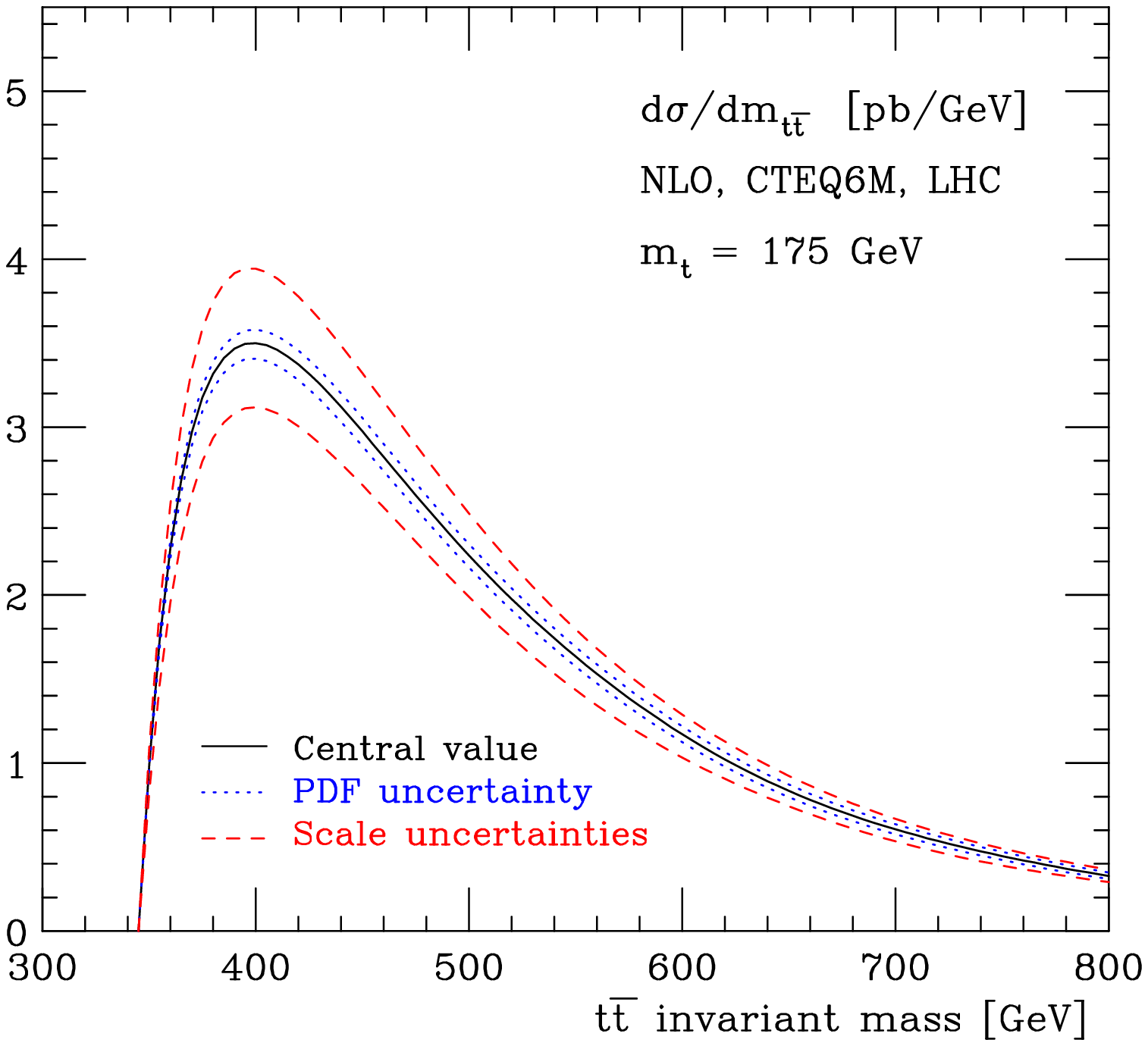, angle=90, width=0.30\textwidth}
      }
    }
  \end{center}
  \vspace{-20pt}
      \caption{
Scale (\emph{dashed}) and PDF (\emph{dotted}) uncertainties in the
$t\bar{t}$ invariant mass spectrum for top masses
\mbox{(a) $m_t=$ 165 GeV},
(b) $m_t=$ 170 GeV and
(c) $m_t=$ 175 GeV at NLO for the LHC using the CTEQ6M pdf set.}
      \label{scale_pdf}
\end{figure*}

In this section we study the theoretical uncertainties on the
available predictions for the invariant mass spectrum and their
dependence on the top mass. We mainly focus on the LHC and refer to
the results of Ref.~\cite{Cacciari:2003fi} for the Tevatron.  We start
by considering the invariant mass spectrum of the $t\bar{t}$ pair
calculated up to next-to-leading order (NLO), as implemented in MCFM
\cite{Campbell:1999ah}.  We use the CTEQ6M PDF-set
\cite{Pumplin:2002vw} and do not apply cuts on the final state
particles.  Here and in the following we always assume that the
invariant mass can be fully reconstructed, see also the Appendix.
In Fig.~\ref{scale_pdf}
results at the LHC are plotted for three different top quark masses
with uncertainty bands associated to the PDF errors and
renormalization and factorization scale variations.

The PDF uncertainty is estimated by running the 41 members of CTEQ6
PDF set, with the scales set equal to $\mu_R=\mu_F=m_t$, and found to
be about $\pm3.2$\%.  The scale uncertainty is obtained by varying
independently the renormalization and factorization scales in the
region between $\mu_R=\mu_F=m_t/2$ and $\mu_R=\mu_F=2m_t$.  The
associated total scale uncertainty at NLO is about $\pm13$\%.  Thus,
the theoretical errors at the LHC are completely dominated by the
scale uncertainty.  This is contrast to the Tevatron where scale and
PDF errors are comparable, of the order $6$\%~\cite{Cacciari:2003fi}.

NLL resummed calculations suggest that the dependence on the scales
could go down to $\pm6$\% \cite{Bonciani:1998vc,Beneke:2000hk},
however in that analysis the scales were not varied independently. A
more recent study suggests that changing the scales independently
might increase the error to a size similar to that estimated through
the NLO fixed order calculation~\cite{matteo}.

\begin{figure*}[t]
  \begin{center}
    \mbox{
      \subfigure[]{
    \epsfig{file=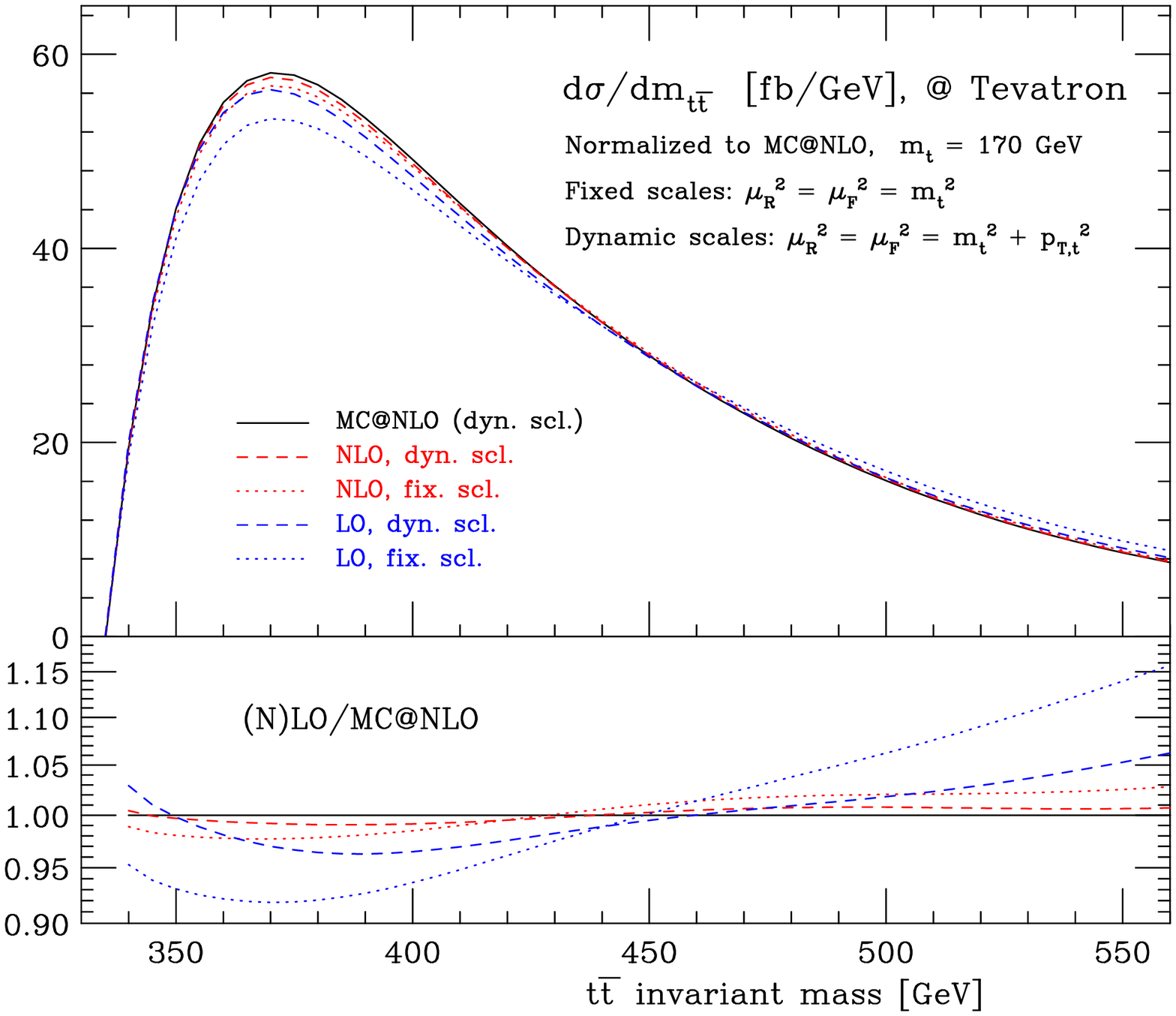, angle=90, width=0.45\textwidth}
      }\quad
      \subfigure[]{
    \epsfig{file=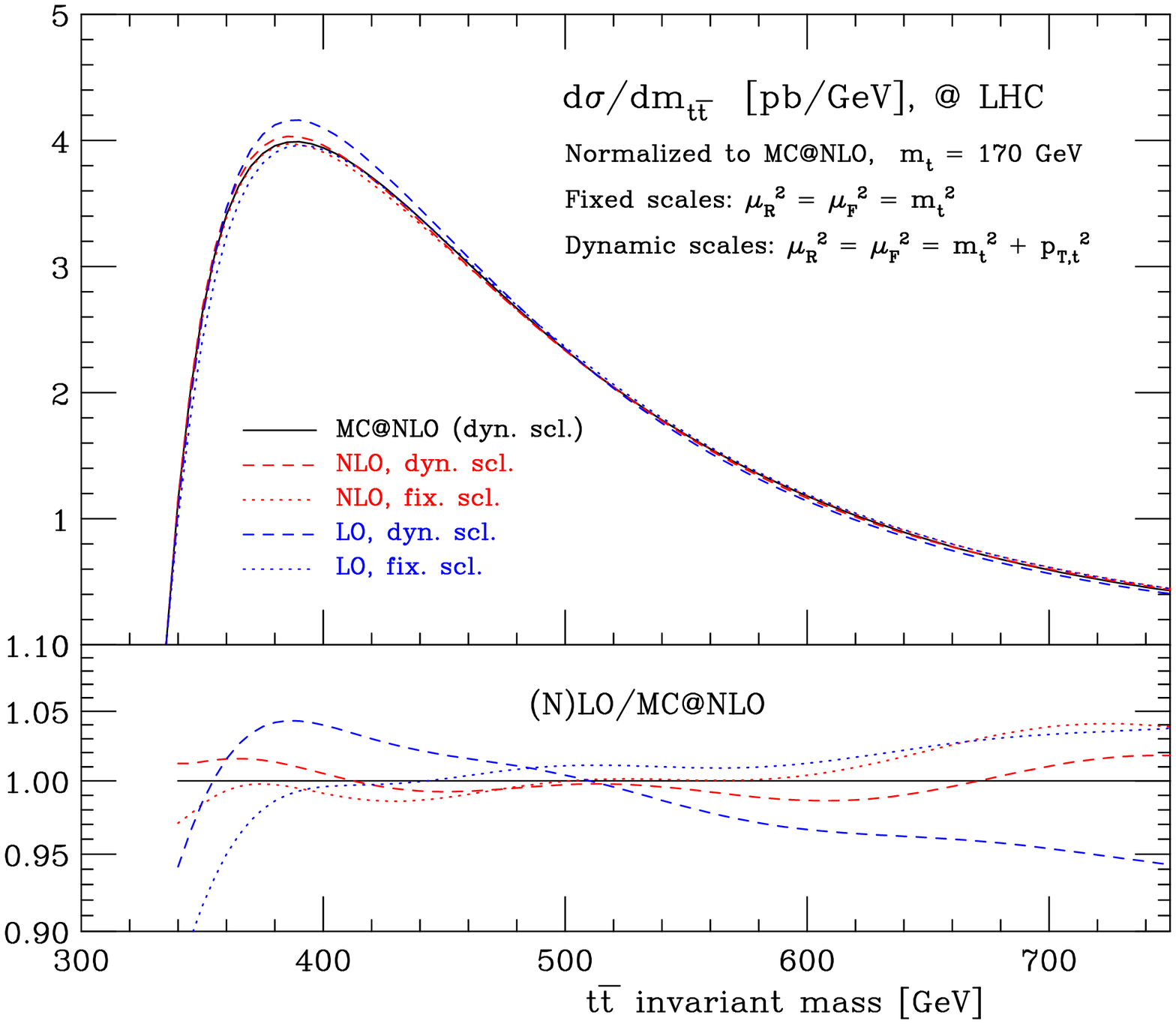, angle=90, width=0.45\textwidth}
      }
    }
  \end{center}
  \vspace{-20pt}
      \caption{
The MC@NLO $m_{t\bar{t}}$ distribution compared with the LO ({\it
blue}) and NLO ({\it red}) fixed order predictions. The distributions
are normalized to the MC@NLO cross section. We set $m_t=$ 170 GeV,
include no cuts, and use CTEQ6M for the NLO and MC@NLO, and CTEQ6L1
for the LO calculations.}
      \label{scale_pdf_MCatNLO}
\end{figure*}

Next we compare the NLO shapes for the invariant mass distribution
with those obtained at LO and MC@NLO \cite{Frixione:2003ei},
Fig.~\ref{scale_pdf_MCatNLO}, both at the Tevatron and the LHC.  We
find that the differences are minimal at the LHC, and well within the
uncertainty bands of the theoretical errors on the NLO cross
section. On the other hand, at the Tevatron the differences between LO
and NLO, fixed and dynamic renormalization and factorization scales
are larger.

\begin{figure*}[t]
  \begin{center}
    \epsfig{file=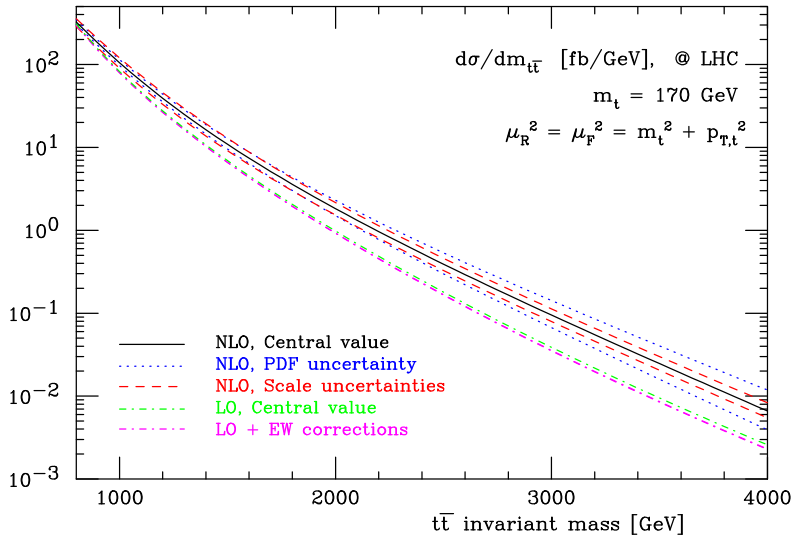, width=0.65\textwidth}
  \end{center}
  \vspace{-20pt}
      \caption{
Scale (\emph{dashed}) and PDF (\emph{dotted}) uncertainties in the
$t\bar{t}$ invariant mass spectrum for $m_t=$ 170 GeV at NLO in QCD,
with the CTEQ6M PDF-sets. Also plotted are the LO distribution
(\emph{light dash-dotted}), the LO including NLO Electro-Weak
corrections (\emph{dark dash-dotted}) with CTEQ6L1 PDF-set. The LO
distribution is normalized to the NLO total cross section. }
      \label{scale_pdf_high}
\end{figure*}

In the high invariant mass region for the LHC,
Fig.~\ref{scale_pdf_high}, the LO approximation starts to deviate from
the NLO order and clearly underestimates the NLO distribution (note
that curves here are normalized to the total cross section at
NLO). Also, as expected, the PDF uncertainties start to increase and
dominate the theoretical errors as the most important contributions
come from the large $x$ region.  Next-to-leading order electroweak
corrections to the LO distribution are also included in this
figure~\cite{Kuhn:2006vh,Bernreuther:2006vg}. Their effect is to decrease the cross
section by a few percent for invariant masses below 1000 GeV and up to
15\% for invariant masses around 4 TeV (the Higgs mass dependence is
mild).  This means that EW effects on this distribution are negligible
compared to the current PDF uncertainties and give only a minor deviation 
from the LO curve.

We conclude this section by mentioning the other sources of
potentially large uncertainties in the determination of the $t\bar{t}$
invariant mass. The first is related to its reconstruction from the decay
products. In general the uncertainty on the $m_{t \bar t}$
distribution will depend on the final state signature (fully-hadronic,
single-lepton and double-lepton final states), which determine the
reconstruction technique and, more importantly, on the detector
efficiencies and resolutions. For completeness we briefly discuss the
current proposals for reconstruction in the various decay channels in
the Appendix. 
The second is due to both QCD backgrounds, {\it i.e.} multi-jet,
$W,Z$+jets and $WW$+jets, and top backgrounds, {\it i.e.} single-top
and $t\bar t$ itself as coming from a final state different signature
than the one considered. While the QCD backgrounds at the Tevatron are
severe but very well studied, it has been shown that at the LHC their
impact at low $t\bar{t}$ invariant mass is
negligible when at least one lepton is present 
in the final state~\cite{Ferrari:2007qf}.  
In the high invariant mass tail, some QCD
backgrounds, and in particular $W$+ one or two jets, become important
due to the fact that the tops are highly boosted and can give rise to
single jet-like topologies when they decay hadronically. The interested reader
can find a detailed study for the single lepton final state signature in
Refs.~\cite{Baur:2007ck,Baur:2008uv}.

\subsection{Top quark mass dependence}

As can be clearly seen from Fig.~\ref{scale_pdf}, the normalization,
as well as the shape of the $t\bar{t}$ invariant mass distribution
depends on the mass of the top quark. It is then natural to wonder
whether such a rather strong dependence could provide another way to
determine the mass of the top quark. The aim of this subsection is to
provide a quantitative answer, based only on the theoretical
uncertainties.

\begin{figure*}[t]
\begin{center}
    \mbox{
      \subfigure[]{
        \epsfig{file=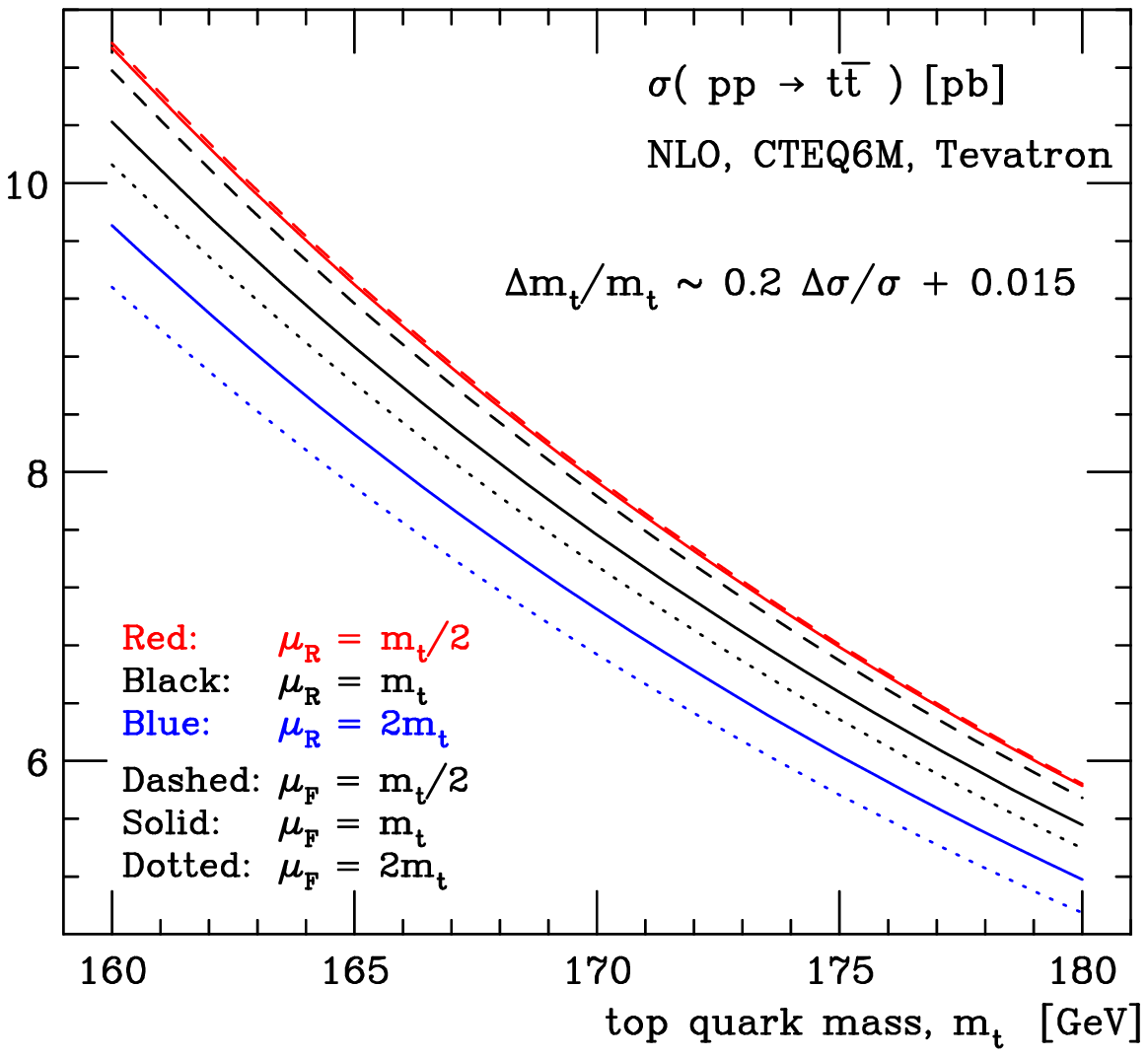, scale=0.6}
      }
      \subfigure[]{
        \epsfig{file=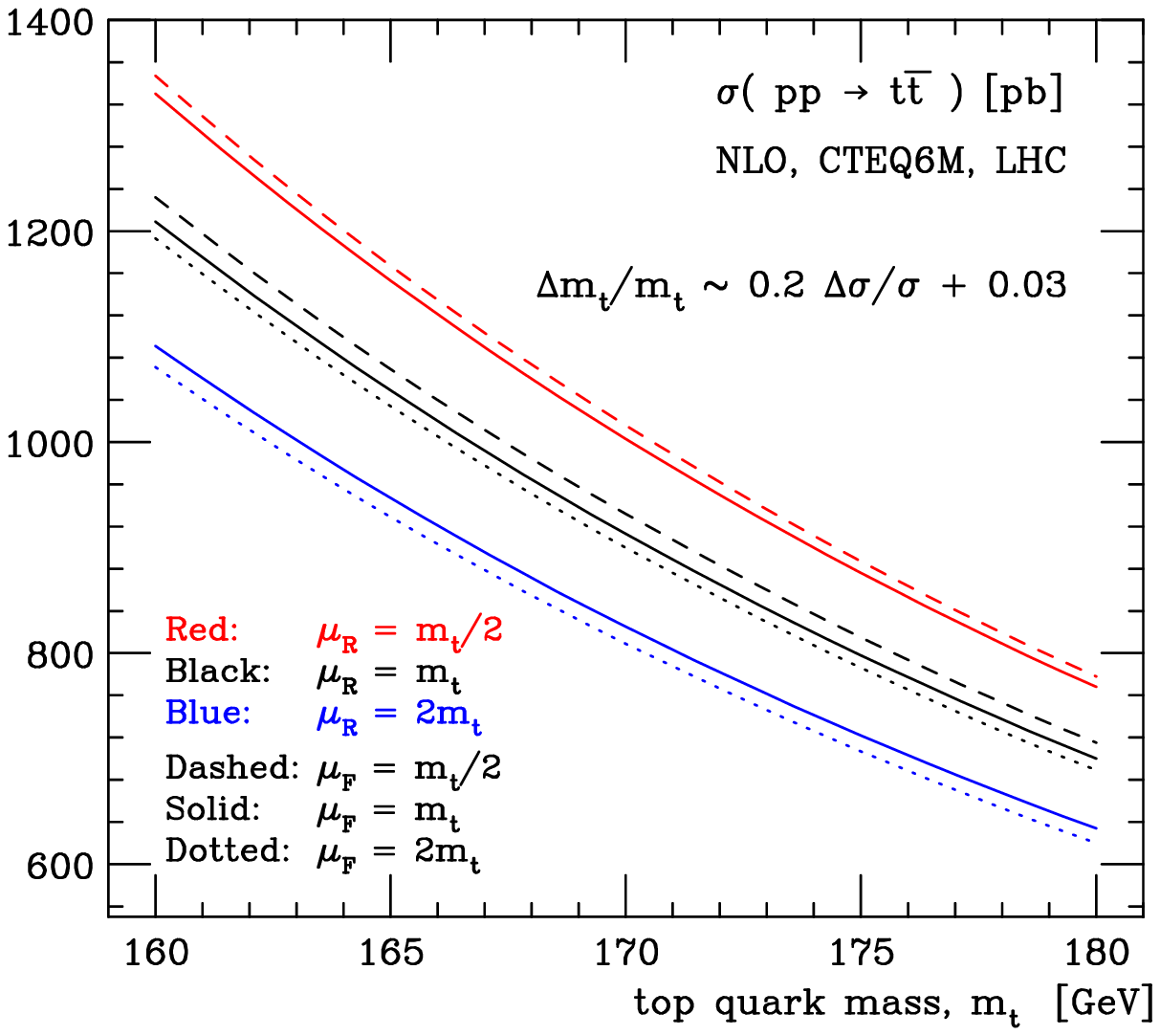, scale=0.6}
      }
    }
\end{center}
  \vspace{-20pt}
      \caption{
The $t\bar{t}$ production
cross section as a function of the top quark mass $m_t$
including scale dependence at the Tevatron (a) and the LHC (b).
}
      \label{cross}
\end{figure*}

In Fig.~\ref{cross} we have plotted the $t\bar{t}$ production cross
section $\sigma$ as a function of the top quark mass at the Tevatron (a)
and the LHC (b). The scale uncertainties, even at the NLO, are rather large.  
Neglecting non-linear terms, a fit to the central curve gives
\begin{equation}
\Delta m_t/m_t\sim 0.2\Delta\sigma/\sigma +0.03 \qquad {\rm (LHC)}.
\label{eq:error}
\end{equation}
This equation relates the relative uncertainty on the measurement of the $t\bar{t}$
cross section to
the relative uncertainty on the top quark mass:
the $\Delta\sigma/\sigma$ term
represents the slope and the constant term the horizontal
spread, {\it i.e.}, the scale uncertainty, of the curves in Fig.~\ref{cross}. This means that a
measurement of the cross section with an uncertainty of 5\% would lead to a
$0.2\times5\%+0.03=4\%$ uncertainty of the top quark mass, the error being mainly
associated with scale variations. 
At the Tevatron the situation is slightly different. The scale dependence 
is milder,
\begin{equation}
\Delta m_t/m_t\sim 0.2\Delta\sigma/\sigma +0.015 \qquad {\rm (Tevatron)}\,,
\end{equation}
and known to be reduced at NLL~\cite{Cacciari:2003fi}, but the 
PDF errors, which are not included in the plot, are not negligible and are found
to be of a similar size~\cite{Cacciari:2003fi}. 

\begin{figure*}[t]
  \begin{center}
    \mbox{
      \subfigure[]{
        \epsfig{file=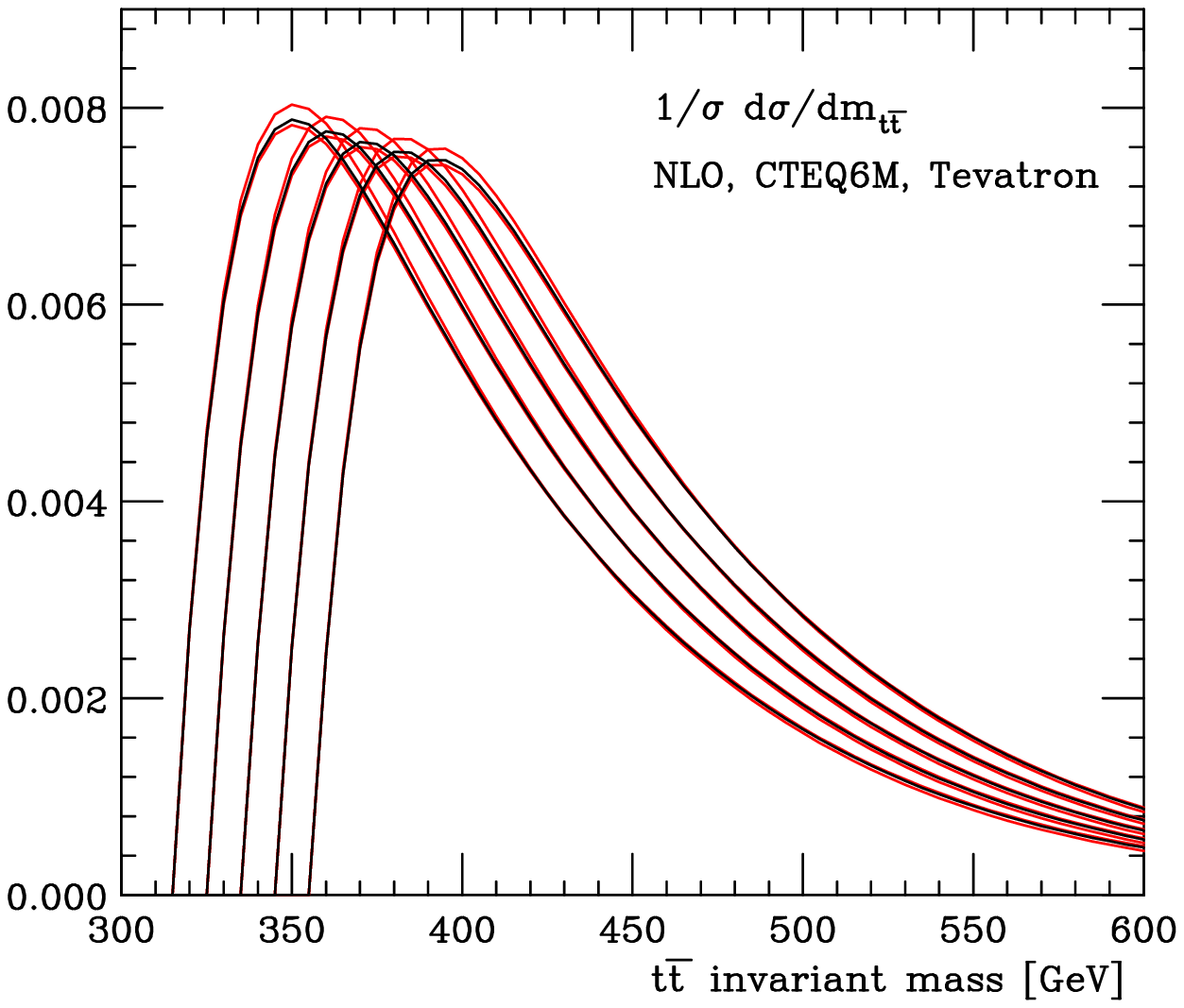, scale=0.6}
      }
      \subfigure[]{
        \epsfig{file=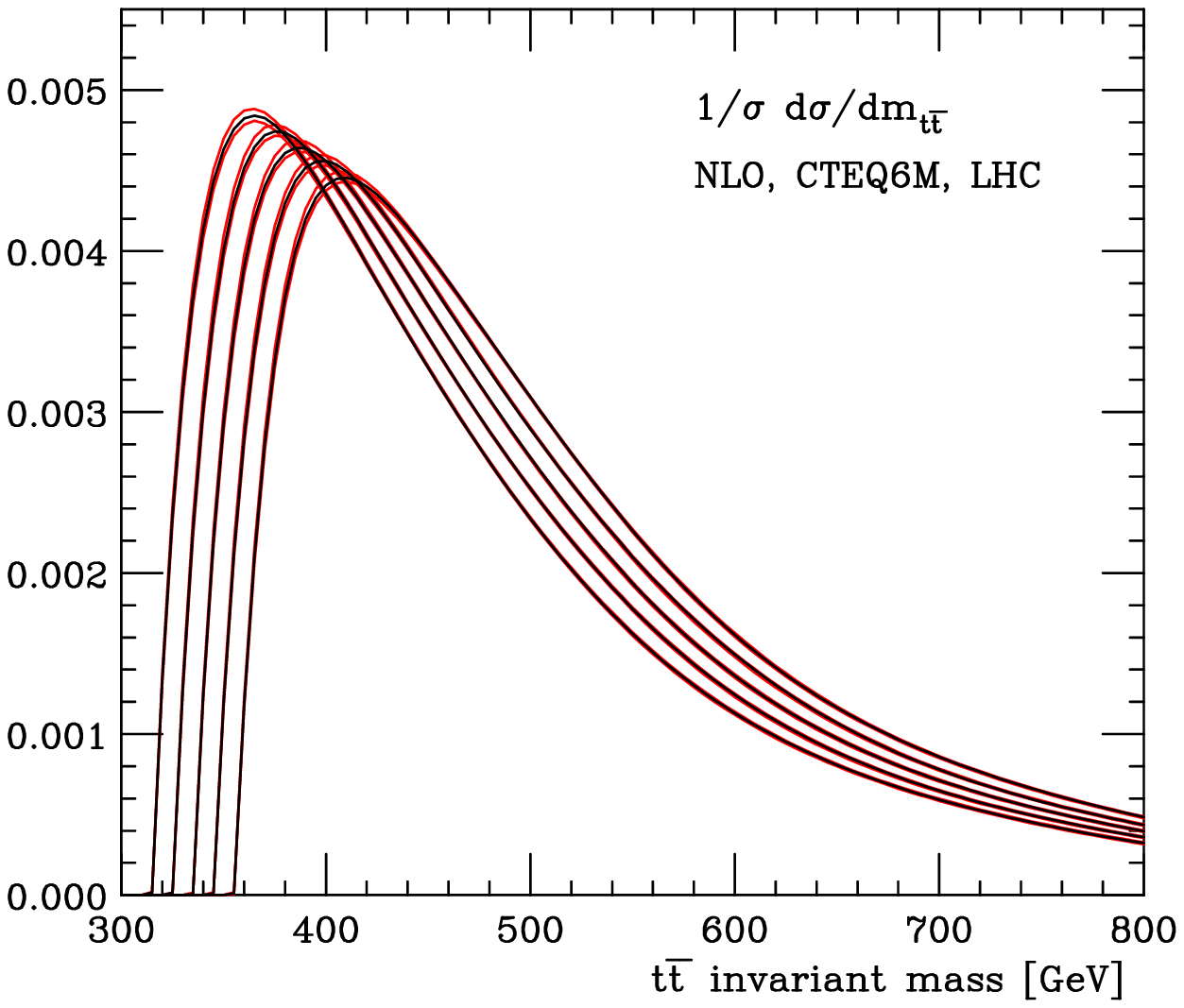, scale=0.6}
      }
    }
  \end{center}
  \vspace{-20pt}
      \caption{The normalized $t\bar{t}$ production cross section as a function of the
$t\bar{t}$ invariant mass, $m_{t\bar{t}}$, for the Tevatron (a) and the LHC (b).
Solid lines from left to right are
for a top quark mass of $m_t=160,\ldots,180$ GeV in steps of 5 GeV, respectively.
The bands spanned by the red lines show the scale uncertainties.
}      \label{normalized}
\end{figure*}
We can therefore conclude that the accuracy of an independent
extraction of the top mass from a measurement of the cross section is
limited by the NLO theoretical uncertainties. Recent work suggests
that inclusion of NNLO corrections could reduce the scale
uncertainties sizably~\cite{Moch:2008qy}.

It is therefore worth investigating whether information on the top
mass can be extracted from some other quantity besides the total cross
section. 
In Fig.~\ref{normalized} the $t\bar{t}$ invariant mass
distributions normalized to unity, $\frac{\partial\sigma}{\partial
m_{t\bar{t}}}\Big|_{\textrm{norm.}}$, are plotted for five different top quark masses,
$m_t=160\ldots 180$ GeV in steps of 5 GeV.
The bands spanned by the red lines show the left-over scale uncertainties which
are sizably reduced compared to Fig.~\ref{scale_pdf}.
We find that the shape of the $m_{t\bar{t}}$ distribution is quite
insensitive to theoretical uncertainties, while retaining a strong
dependence on the top quark mass. It is therefore interesting to consider
whether the invariant mass distribution could provide an independent 
measurement of the top quark mass.

One way to quantify to which extent the shape is sensitive to the 
top mass vs.~the theoretical uncertainties is to perform an analysis 
based on the first few moments of the normalized $t\bar{t}$ invariant 
mass distributions
$\frac{\partial\sigma}{\partial
m_{t\bar{t}}}\Big|_{\textrm{norm.}}$. This approach has the virtue of
being simple and systematic.  Needless to say, alternative quantities,
such as the peak position, or more sophisticated techniques, such as
Kolmogorov tests, could also be employed.

\begin{figure*}[t]
  \begin{center}
    \mbox{
      \subfigure[]{
        \epsfig{file=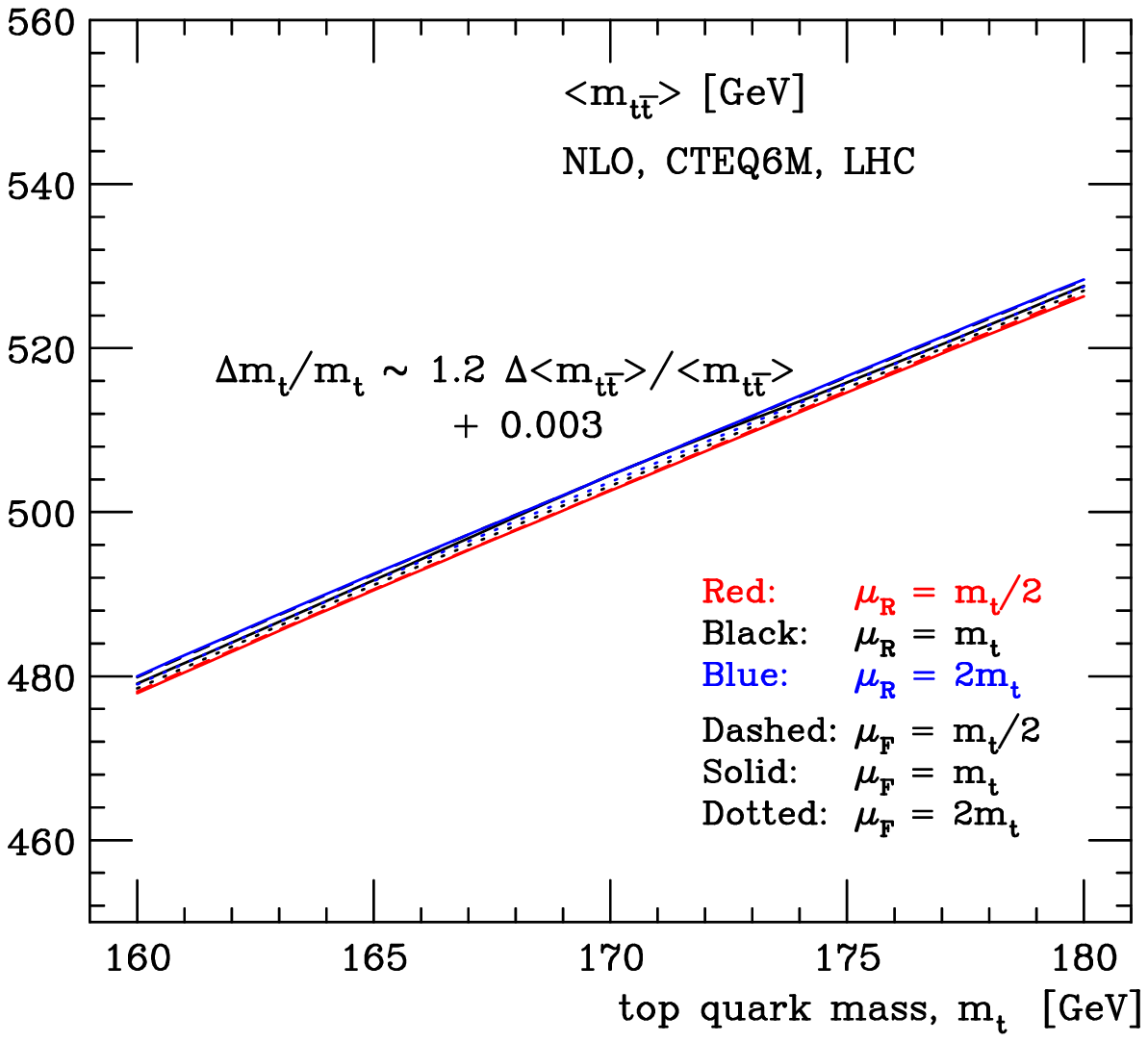, scale=0.6}
      }
      \subfigure[]{
        \epsfig{file=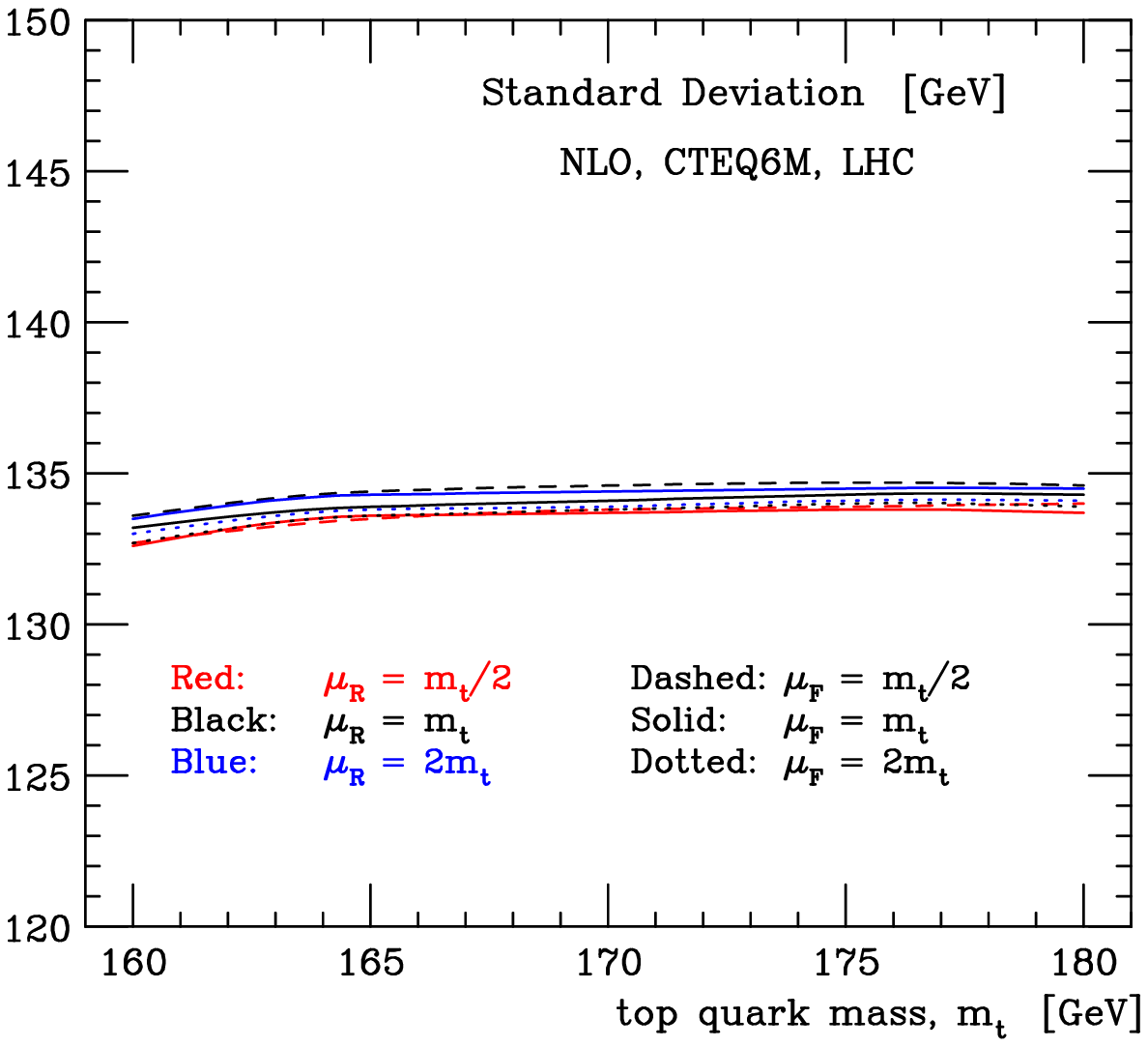, scale=0.6}
      }
    }
    \mbox{
      \subfigure[]{
        \epsfig{file=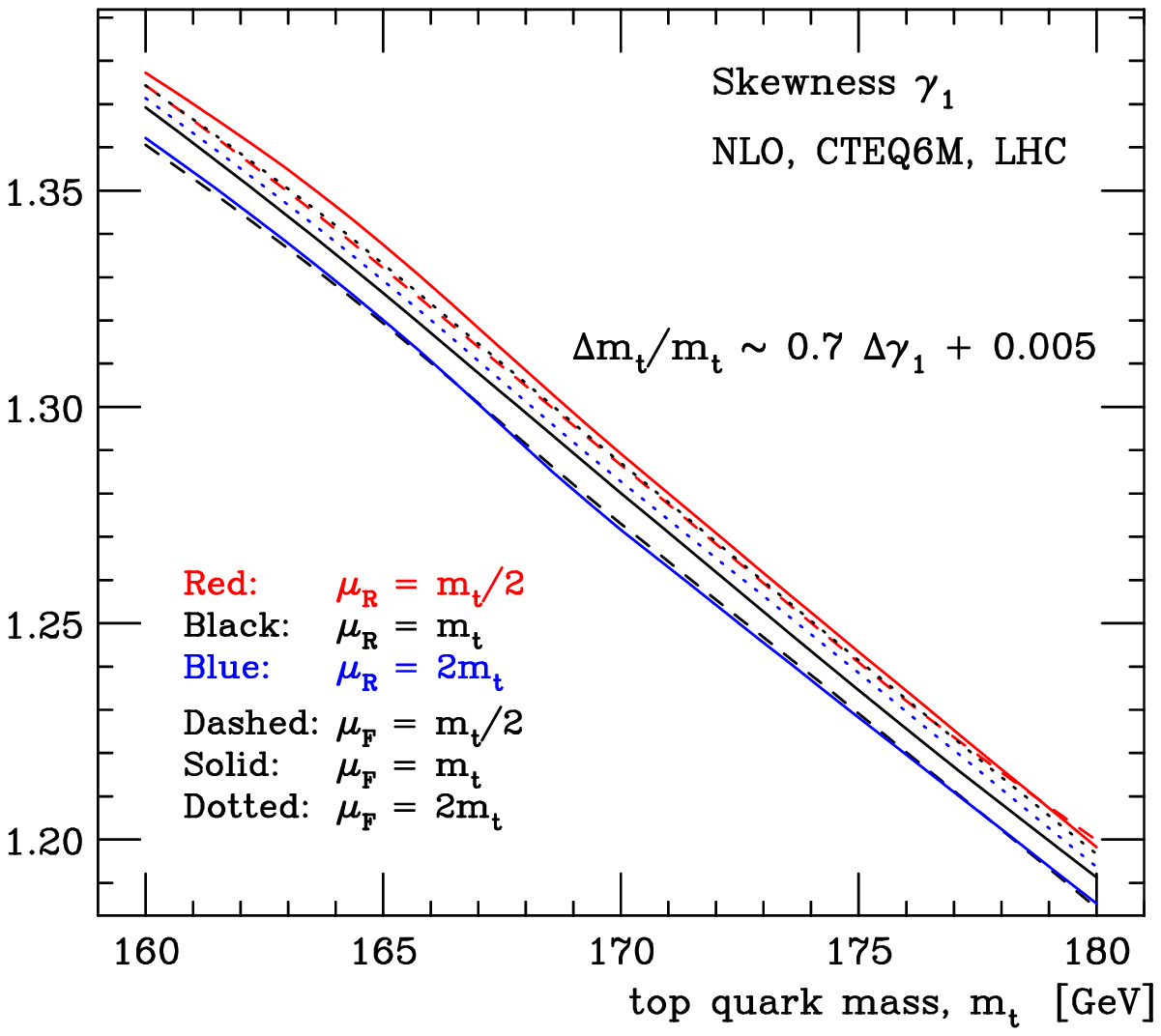, scale=0.6}
      }
      \subfigure[]{
        \epsfig{file=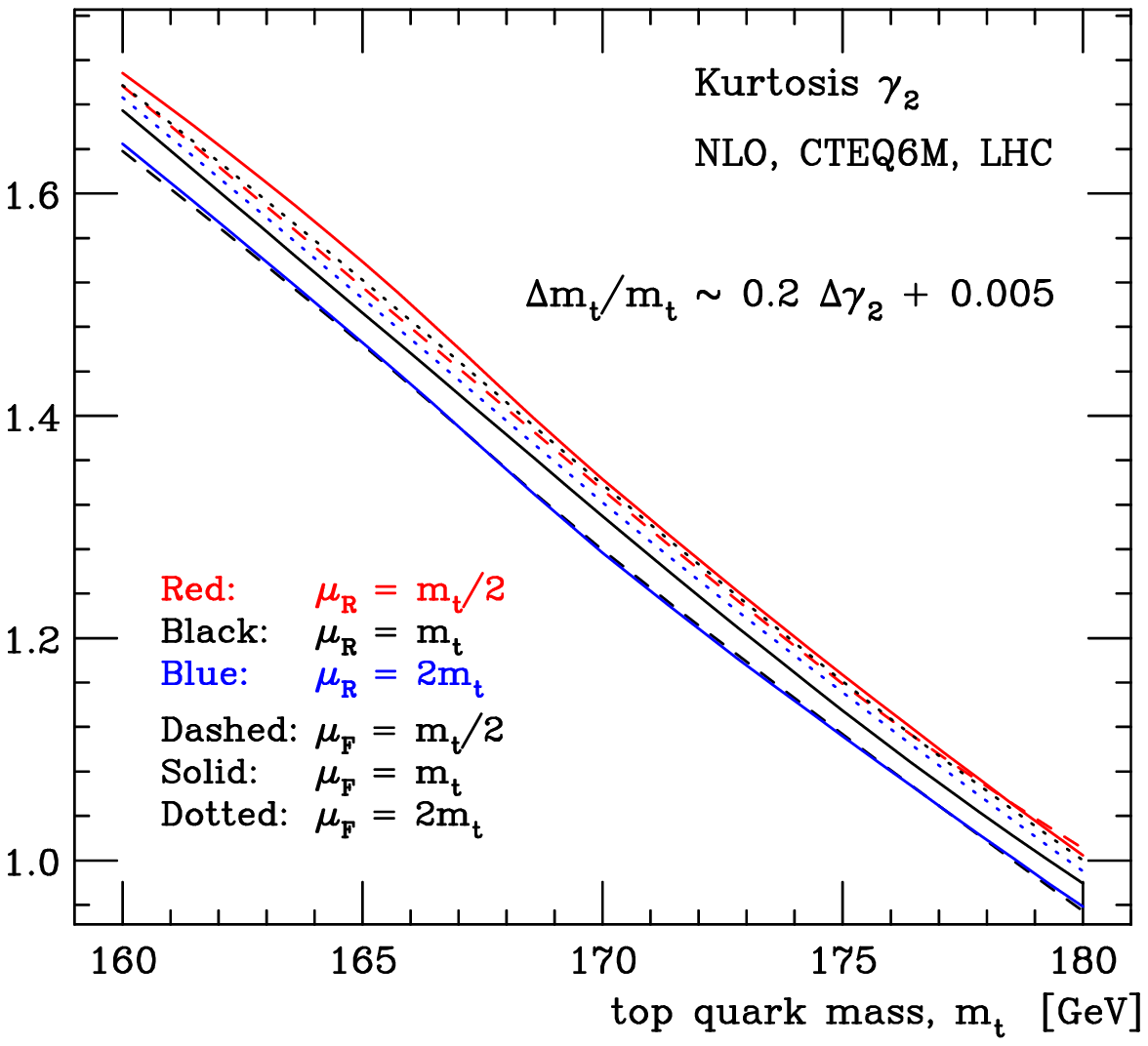, scale=0.6}
      }
    }
  \end{center}
  \vspace{-20pt}
      \caption{The
average value (a),
standard deviation (b),
skewness (c) and
kurtosis (d) of the $t\bar{t}$ invariant mass distribution
as a function of the top quark mass $m_t$
including the scale dependence at the LHC.
}
      \label{moments}
\end{figure*}
In Fig.~\ref{moments} we present the mean value $\langle
m_{t\bar{t}}\rangle$, standard deviation $s$, skewness $\gamma_1$ and
kurtosis $\gamma_2$ of the $t\bar{t}$ invariant mass distributions as
a function of the top quark mass $m_t$ for various scales at the
LHC. We remind that the skewness (kurtosis) is a dimensionless
quantity that gives a measure of asymmetry (peakedness) of a
distribution.  These quantities are defined as
\begin{equation}
\langle m_{t\bar{t}}\rangle=
\int^{m_{\textrm{cutoff}}}\!\textrm{d}m_{t\bar{t}}\,m_{t\bar{t}}
\frac{\partial\sigma}{\partial m_{t\bar{t}}}\Big|_{\textrm{norm.}},\qquad
s=\sqrt{\mu_2},\qquad
\gamma_1=\frac{\mu_3}{\mu_2^{3/2}}\quad\textrm{and}\qquad
\gamma_2=\frac{\mu_4}{\mu_2^{2}}-3,
\end{equation}
respectively. The central moments $\mu_n$ are defined as
\begin{equation}\label{eq:2}
\mu_n=
\int\!\textrm{d}m_{t\bar{t}}\,
\big(m_{t\bar{t}}-\langle m_{t\bar{t}}\rangle\big)^n
\frac{\partial\sigma}{\partial m_{t\bar{t}}}\Big|_{\textrm{norm.}}.
\end{equation}
In our analysis we focus on the low invariant mass region and therefore we 
have limited the $m_{t\bar{t}}$ integrals to $m_{t\bar{t}}< m_{\textrm{cutoff}}=1 \textrm{ TeV}$. The aim of this cut is just to mimick an experimental analysis where the precision on the higher moments would be limited by
the statistics. Since our purpose is only for illustration we do not consider
these effects further. However, we stress that our numerical results 
do retain a significant dependence on this cutoff.  

Due to the small scale uncertainty and the strong linear correlation,
the mean of the $t\bar{t}$ invariant mass distribution, $\langle
m_{t\bar{t}}\rangle$, Fig.~\ref{moments}(a), appears to be an excellent
estimator of the top quark mass.  From the experimental point of view,
one can also hope for smaller uncertainties than those associated to
the measurement of total cross section. In fact, to measure the mean,
many systematics, such as those coming from luminosity or tagging
efficiencies, are much less important. A fit to the mean value shows
that $\Delta m_t/m_t\sim 1.2\Delta\langle m_{t\bar{t}}\rangle/\langle
m_{t\bar{t}}\rangle +0.003$. So, for instance, if the mean value is
measured with a 1\% uncertainty, the uncertainty of the top quark mass
is only 1.3\%, including the scale uncertainties. 

The standard deviation, Fig.~\ref{moments}(b), is almost constant and therefore is not suitable for a top quark mass measurement. In Figs.~\ref{moments}(c)
and \ref{moments}(d) the skewness and the kurtosis for the 
$t\bar{t}$ invariant mass are plotted, respectively. Also here,
the scale uncertainty is reduced, while still slightly larger than
for the mean value, Fig.~\ref{moments}(a). The slopes of the lines
are promising, in particular for the kurtosis, which means that a relatively
large experimental error on the measurement of the kurtosis leads to an
only small error on the top mass measurement.

At this point, we have to stress that the above simple analysis does
not include neither statistical nor systematics effects in the data,
which should also be carefully considered. In particular, the higher monents
such as the skewness and the kurtosis are more sensitive to the tail of
the $m_{t \bar{t}}$ distribution then the lower moments
and therefore more sensitive to statistical and 
systematic effects that affect more strongly this tail.
Eventually, the final
uncertainty on the top quark mass will depend on how well the above
quantities can be measured. It is plausible to expect that a combined
analysis based on the above quantities might lead to an even smaller
uncertainty for the top quark mass.

\begin{figure*}[t]
  \begin{center}
    \mbox{
      \subfigure[]{
        \epsfig{file=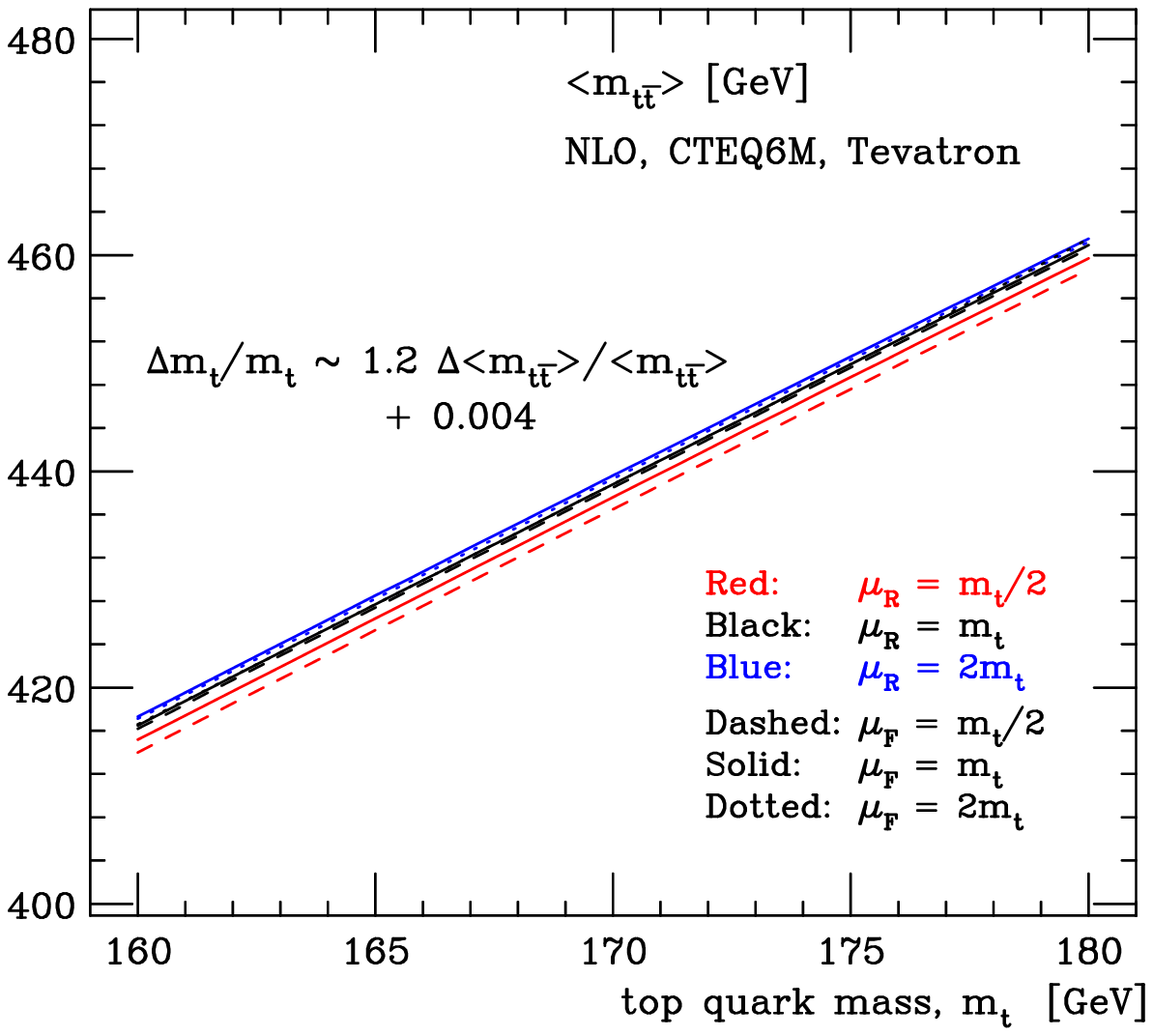, scale=0.6}
      }
      \subfigure[]{
        \epsfig{file=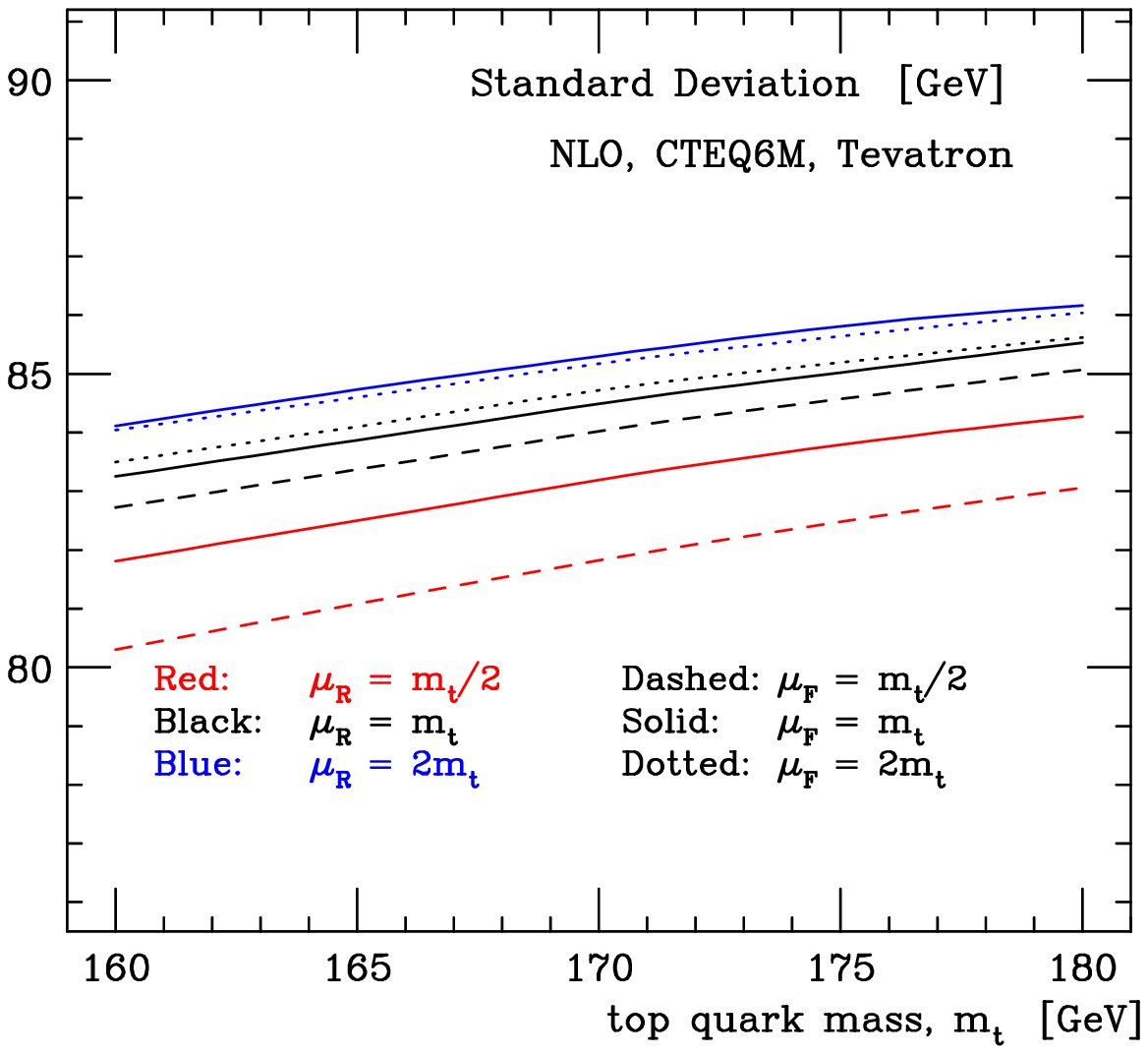, scale=0.6}
      }
    }
    \mbox{
      \subfigure[]{
        \epsfig{file=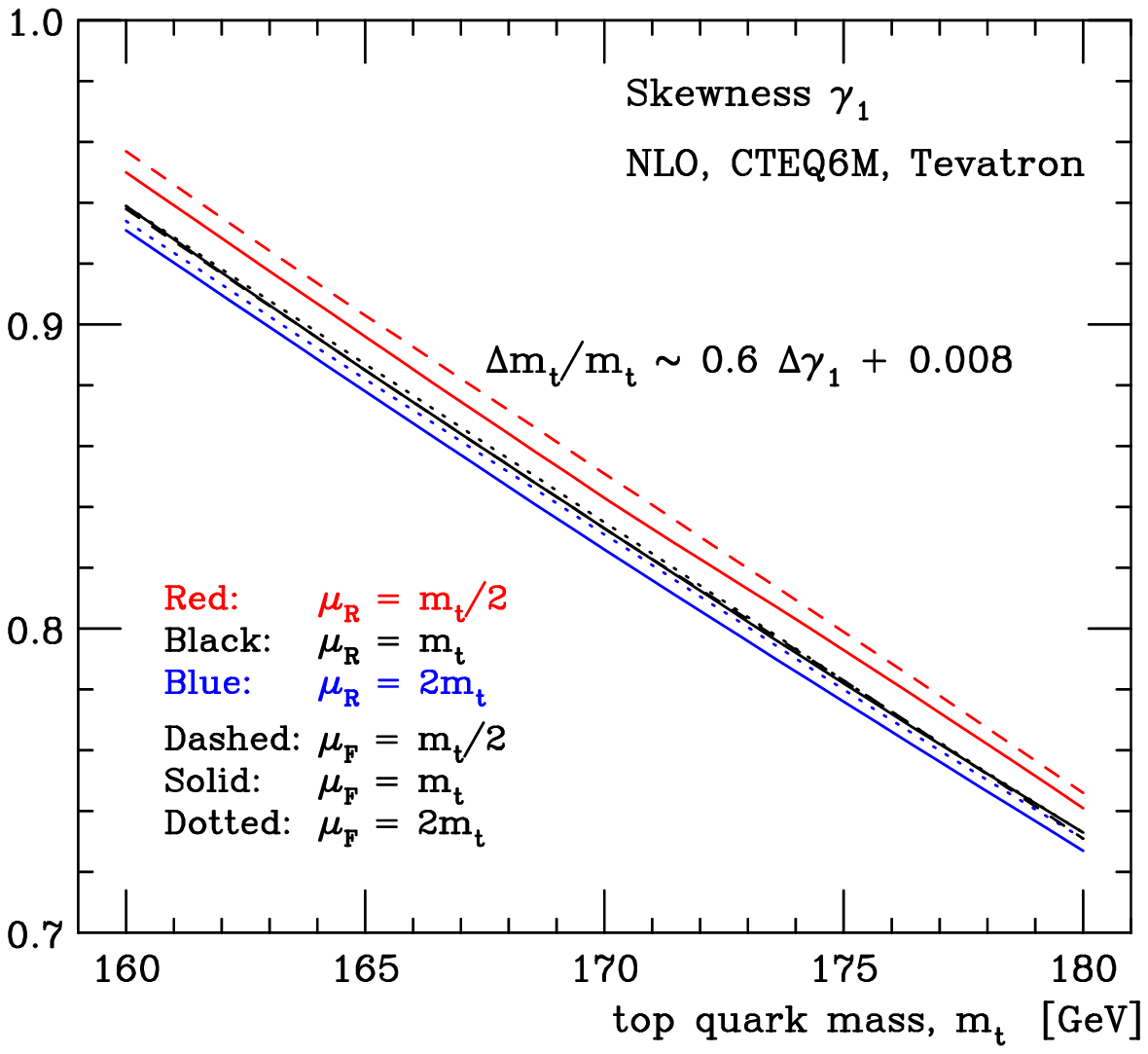, scale=0.6}
      }
      \subfigure[]{
        \epsfig{file=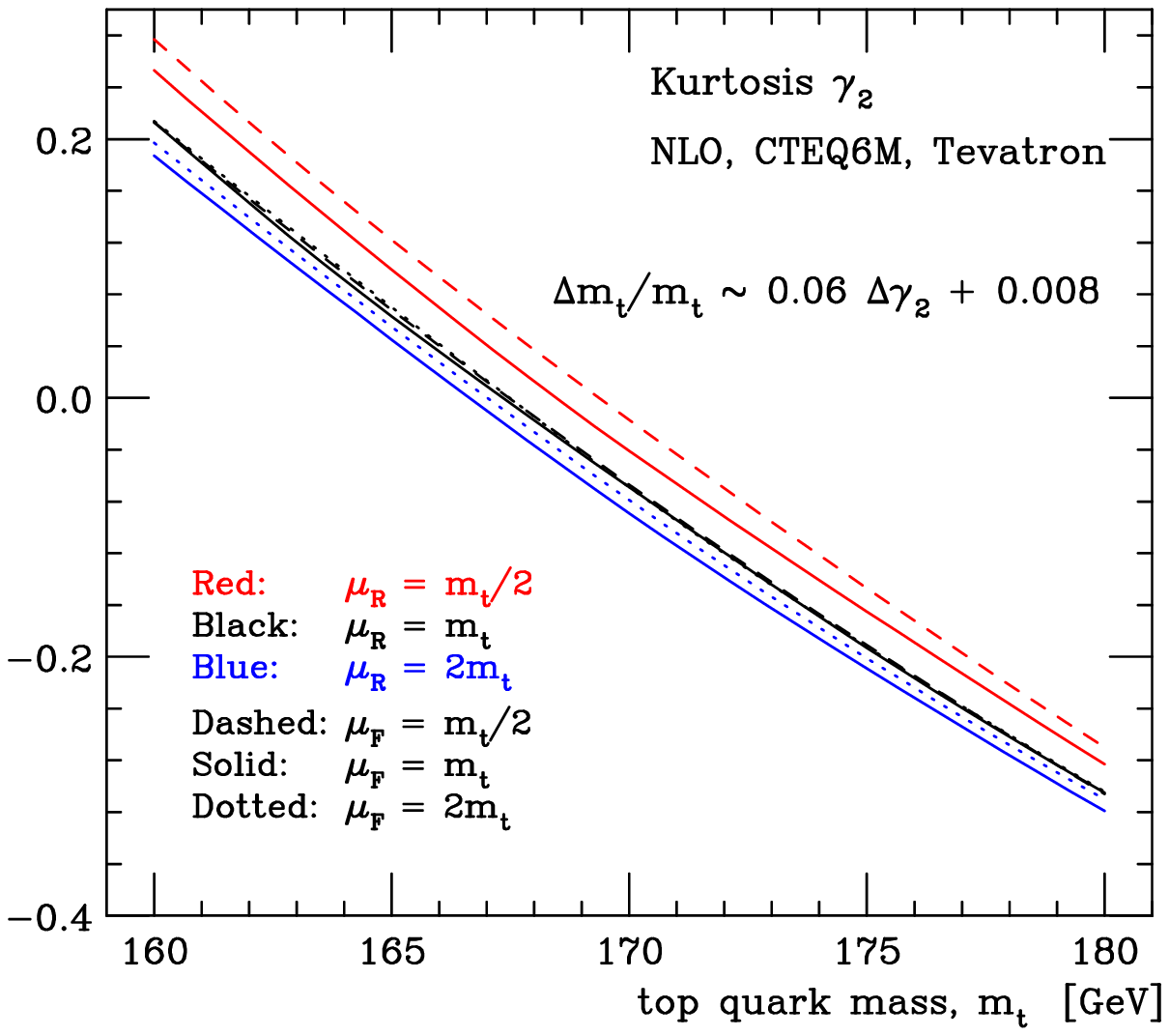, scale=0.6}
      }
    }
  \end{center}
  \vspace{-20pt}
      \caption{The average value (a), standard deviation (b), skewness
(c) and kurtosis (d) of the $t\bar{t}$ invariant mass distribution as
a function of the top quark mass $m_t$ including the scale dependence
at the Tevatron.  For the skewness and kurtosis we restricted the
integration region in Eq.~\ref{eq:2} to $m_{t\bar{t}} < 600$ GeV.
}
      \label{moments_tev}
\end{figure*}

For completeness we show the same analysis performed at the Tevatron
energies, see Fig.~\ref{moments_tev}.
Also in this case, we have used a fixed order NLO
calculation to estimate the scale uncertainties. However, as we have
already mentioned, at the Tevatron the $t \bar{t}$ pairs are produced
almost at threshold, hence a resummed calculation which predicts a
smaller scale uncertainty, is preferred~\cite{Cacciari:2003fi}.

The Tevatron results (Fig.~\ref{moments_tev}) are similar to those
obtained in the LHC study, but the reduction in the scale
uncertainties by analyzing the (higher) moments is smaller compared to
the LHC. The first moment, {\it i.e.}, the mean value, is probably the
best estimator for the top quark mass among all the moments, due to
its small constant value of $0.004$, and the reasonably good
proportionality factor of $1.2$.  The higher moments are more
sensitive to statistical fluctuations and might be less suitable with
a limited sample. The lack of events in the higher invariant mass
regions might give rise to larger errors for the skewness and worse
the kurtosis. In these plots we restrict the $t\bar{t}$ invariant mass
to below $m_{\textrm{cutoff}} = 600$ GeV. We mention that even
though using the fixed order NLO calculation we have overestimated the
scale uncertainties, we have neglected the PDF errors which at the
Tevatron can reach the 6-7\%~\cite{Cacciari:2003fi} and errors coming
from the reconstruction of the $t\bar{t}$ invariant mass from the
(anti-)top quark decay products, see the Appendix.

Finally, we comment about the definition of the top quark mass.
As in the more standard top mass measurements where the top mass is
reconstructed from its decay products, also the top quark invariant
mass is sensitive to extra radiation and to non-perturbative effects
due to confinement, typical of a pole mass~\cite{Smith:1996xz}. In this respect the same
issues and problems in associating a theoretically well-defined
mass to the measurement remain. This is at variance with an extraction
of the top mass from a cross section measurement which can be directly
related to a short distance mass and does not suffer from the same
non-perturbative or extra radiation effects. We stress, however, 
that $m_{t \bar{t}}$ is at least twice the top quark mass,
which would decrease the relative impact of the ambiguities due 
to extra radiation. In addition, typical combinatorial systematics 
associated to the assignment of the jets to the ``right'' tops, 
are absent for $m_{t \bar{t}}$. In this respect, more experimental 
work on the systematics affecting such a measurement would be certainly 
welcome.

\section{Effects from BSM resonances}\label{sec:BSM}

In this section we investigate the effects of (model-independent) new
resonances on the $t\bar{t}$ invariant mass spectrum.  All the
numerical results presented here have been obtained with
MadGraph/MadEvent, through the implementation of a dedicated
``model'', {\tt topBSM}, which is publicly accessible on the MadGraph servers
for on-line event generation and for download.\footnote{Technical
documentation on how to use the model can be found at {\tt
http://cp3wks05.fynu.ucl.ac.be/twiki/bin/view/Software/TopBSM}.}

{\tt topBSM} offers the possibility of  studying a 
wide range of new physics resonances and 
efficiently exploits the flexibility and the possibilities of {\tt MadGraph}:
\begin{itemize}
\item SM effects are consistently included, {\it i.e.}, possibly
non-trivial interference effects between new resonances and the
$t\bar{t}$ background are taken into account. As it will be shown in
the following, in some cases such effects can be important and might
lead to very distinctive signatures (cf. the case of the peak-dip
structure arising in $m_{t\bar t}$ due to the presence of a
(pseudo-)scalar state). In general, they should be always included.
\item The full matrix elements $2\to 6$ including the decays of
the top quarks can be generated, which is crucial for spin correlation
studies.
\item The generated events can be automatically interfaced to parton
showers programs, such as {\tt Pythia}~\cite{Sjostrand:2006za} or 
{\tt Herwig}~\cite{Corcella:2001wc}, to shower and hadronize the events 
after which these events can be processed by a detector simulation for 
full experimental analyses. 
\end{itemize}

We have considered $s$-channel spin-0, spin-1 and spin-2 resonances,
of different color and $CP$ parity, as listed in
Table~\ref{table_BSM}.  The parameters related to each resonance are
simply the mass, the width and the relevant values of the couplings to
standard model particles which enter in the production process (to the
partons and to the top quark).

\begin{table}[h!]
\begin{center}
\begin{tabular}{|c|c|c|l|l|}
\hline
Spin & color & parity $(1,\gamma_5)$& some examples/Ref.\\
\hline
0    & 0     & (1,0)  &SM/MSSM/2HDM, Ref.~\cite{Gaemers:1984sj,Dicus:1994bm,Bernreuther:1997gs}\\
0    & 0     & (0,1)  &MSSM/2HDM, Ref.~\cite{Dicus:1994bm,Bernreuther:1997gs}\\
0    & 8     & (1,0)  &Ref.~\cite{Manohar:2006ga,Gresham:2007ri}\\
0    & 8     & (0,1)  &Ref.~\cite{Manohar:2006ga,Gresham:2007ri}\\
\hline
1    & 0     & (SM,SM)&$Z'$\\
1    & 0     & (1,0)  &vector\\
1    & 0     & (0,1)  &axial vector\\
1    & 0     & (1,1)  &vector-left\\
1    & 0     & (1,-1) &vector-right\\
1    & 8     & (1,0)  &coloron/KK gluon,
                       Ref.~\cite{Simmons:1996fz,Choudhury:2007ux,Dicus:2000hm}\\
1    & 8     & (0,1)  &axigluon, Ref.~\cite{Choudhury:2007ux}\\
\hline
2    & 0     & --    &graviton ``continuum'', Ref.~\cite{ArkaniHamed:1998rs}\\
2    & 0     & --    &graviton resonances, Ref.~\cite{Randall:1999ee}\\
\hline
\end{tabular}
\end{center}
\caption{The BSM particles included in the {\tt topBSM} ``model''. }
\label{table_BSM}
\end{table}

\subsection{Spin-0 resonances}

The first resonances we discuss are spin-0 particles.
We distinguish between color singlet ($\phi$) and color octet ($S^0$), as well
as parity even (scalar) and odd (pseudo-scalar) spin-0 particles.

\subsubsection{Color singlet}

\begin{figure*}[t]
\hfill
\begin{center}
\epsfig{file=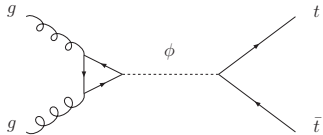, width=0.3\textwidth}
\end{center}
\caption{Feynman diagram for the (pseudo-)scalar contribution to $t\bar{t}$ production.}
\label{scalar}
\end{figure*}

Let us start by considering a color singlet (pseudo-)scalar boson
$\phi$ contributing to the $t\bar{t}$ process $gg\to (\phi \to ) t
\bar{t}$. The Feynman diagram for this loop induced process is
depicted in Fig.~\ref{scalar}. The spin-0 coupling strength to quarks,
\begin{equation}
\label{eq:3}
g_{\phi qq}=a_1i\frac{m_q}{v}+a_2\frac{m_q}{v}\gamma_5,
\end{equation}
is proportional to the quark mass $m_q$.  In analogy with the SM, $v$
is the spin-0 field vacuum expectation value and $a_1$ and $a_2$ are
real proportionality factors for the parity even and odd spin-0
particles, respectively. For the SM Higgs boson $a_1=1$ and $a_2=0$,
while for a pure pseudo-scalar $a_1=0$ and $a_2$ is non-zero.

We do not include scalar production by (anti-)quark annihilation,
$q\bar{q}\to\phi$, because for this cross section to be sizeable
compared to the loop induced gluon fusion process, the branching ratio
for the scalar to $t\bar{t}$ has to be small and can be neglected.

Since we are interested in scalars with strong couplings to the top
quark, we neglect all particles in the loop of Fig.~\ref{scalar}
except for the most heavy quark, {\it i.e.}, the top quark. If the
mass of the spin-0 boson is larger than twice the mass of the top
quark, the loop-induced gluon-gluon-(pseudo-)scalar coupling develops
an imaginary part, which leads to a peak-dip structure for the
interference terms between the QCD background and the signal
\cite{Gaemers:1984sj,Dicus:1994bm,Bernreuther:1997gs}.

The possibility to detect a signal in the $t\bar{t}$ invariant mass
depends on the width of the spin-0 resonance. In general, a scalar
particle couples also to the electroweak bosons. In the SM the decay
rate to $W,Z$ is much larger than the decay rate to $t\bar{t}$, and
therefore the $t\bar t$ channel is suppressed. Moreover, the presence
of a destructive interference between the signal and the QCD
background and the relatively large width of the scalar makes
detection very difficult. An enhanced coupling to top would not help
much because the improvement in the branching ratio would be
compensated by an increase of the total width. In conclusion, there is
little hope to see a SM-like scalar by looking at the $t\bar{t}$
invariant mass spectrum, even if the coupling to top quark were (much)
larger than in the SM.

\begin{figure*}[t]
  \begin{center}
    \mbox{
      \subfigure[]{
        \epsfig{file=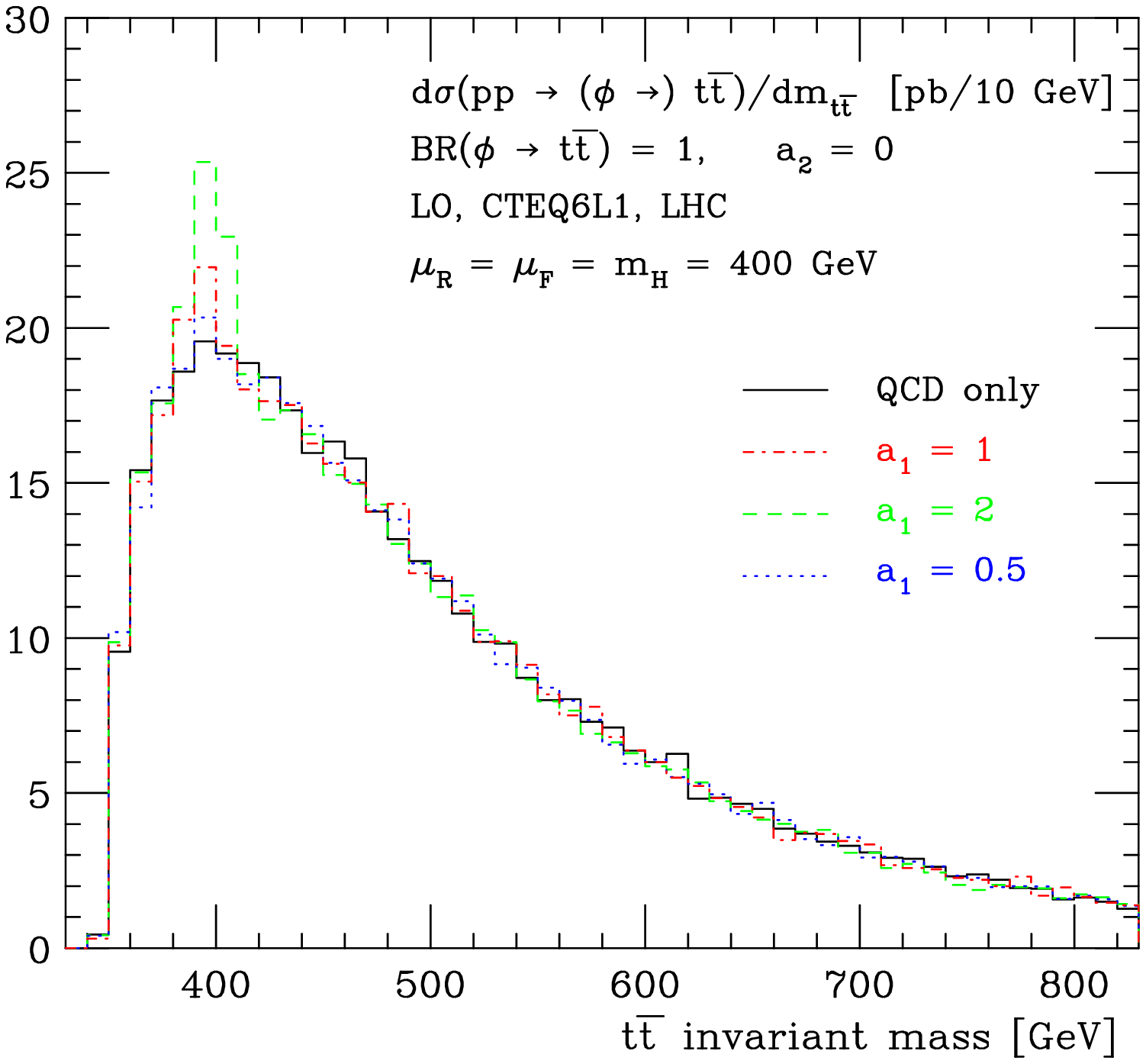,      angle=90, scale=0.45}
      }
      \subfigure[]{
        \epsfig{file=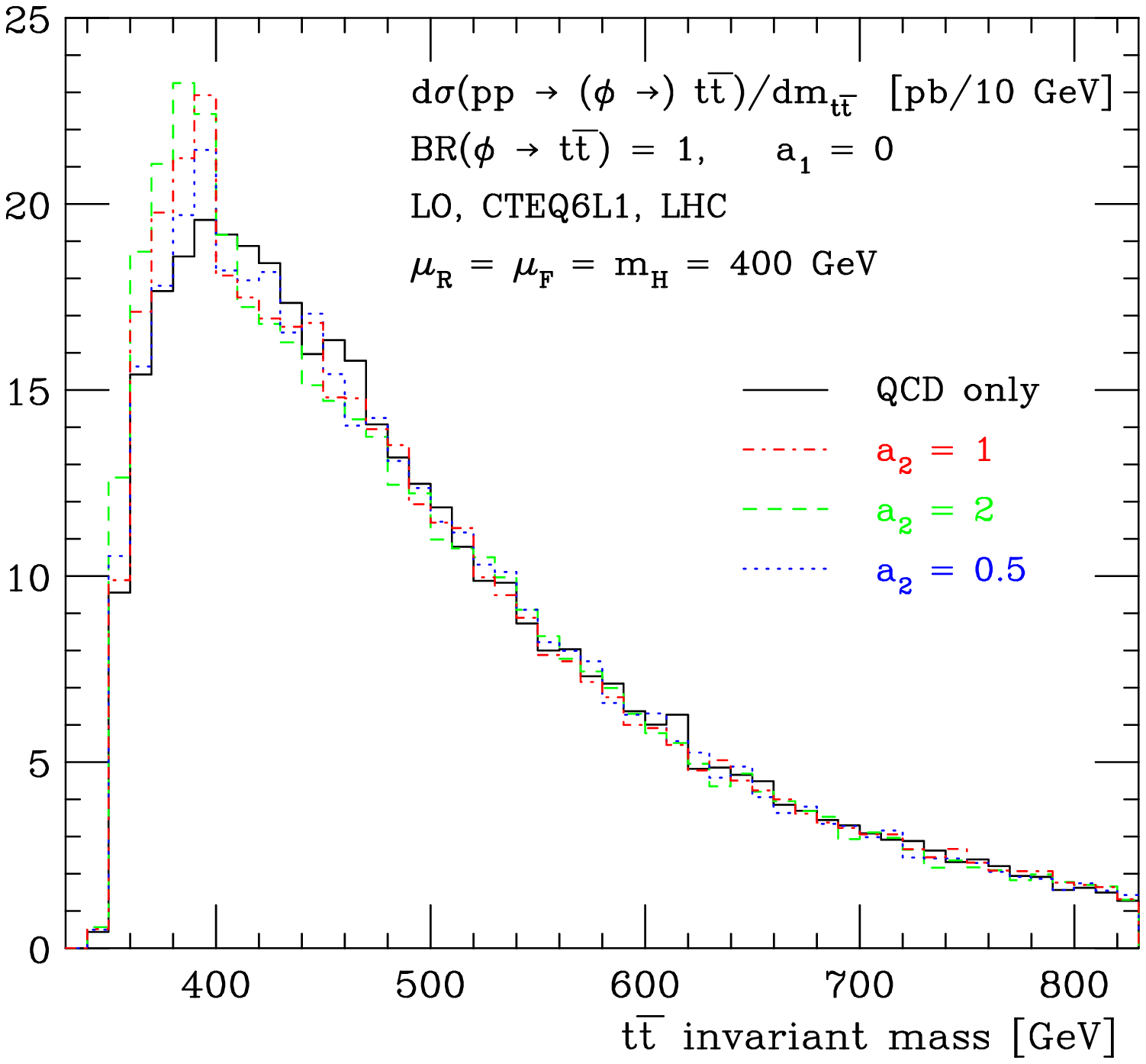, angle=90, scale=0.45}
      }
    }
    \mbox{
      \subfigure[]{
        \epsfig{file=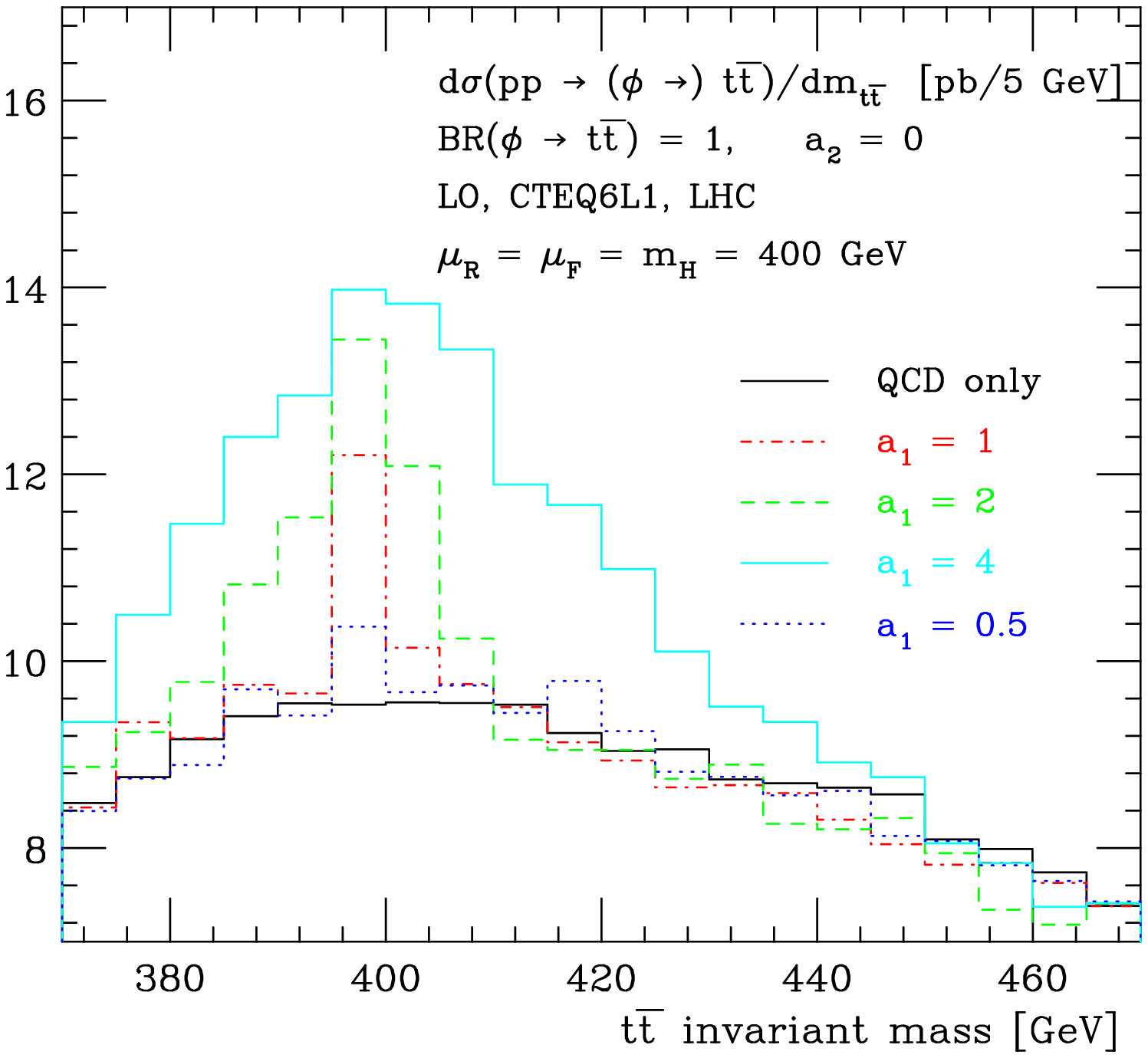, angle=90, scale=0.45}
      }
      \subfigure[]{
        \epsfig{file=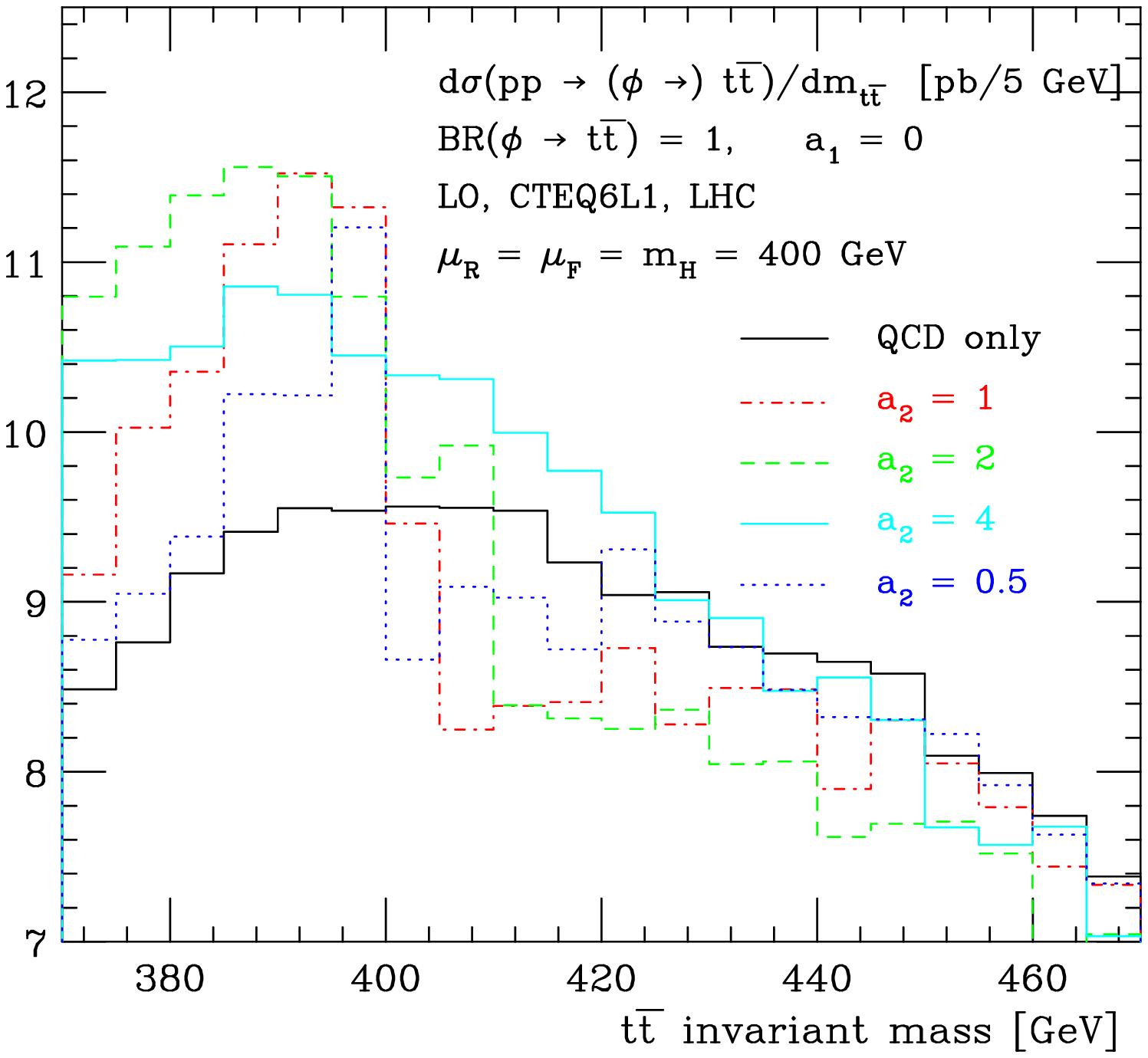, angle=90, scale=0.45}
      }
    }
  \end{center}
  \vspace{-20pt}
      \caption{Invariant $t\bar{t}$ mass spectrum for the 
boson-phobic scalar (\emph{left}) and pseudo-scalar (\emph{right}).
\emph{Bottom}: The interesting region with finer binning.
Different colors represent different coupling strength
of the Higgs to top quarks:
\emph{dot-dashed} for the standard model coupling and
\emph{dotted}, \emph{dashed} and \emph{light solid} for 0.5, 2 and 4 times
the standard model coupling strength, respectively.
\emph{Dark solid} is QCD $t\bar{t}$ production, {\it i.e.}, without the Higgs signal.
All plots were produced using the CTEQ6L1 pdf set with $\mu_R=\mu_F=$ 400 GeV.
No acceptance cuts are applied.}
      \label{heft_PD}
\end{figure*}

On the other hand, the case of a pseudo-scalar or a `boson-phobic'
scalar resonance that does not couple to the heavy vector bosons is
more promising. For such a state, the branching ratio to $t\bar{t}$
can be taken unity, BR($\phi\to t\bar{t}$) $=1$, {\it i.e.}, the total
width of the scalar spin-0 resonance is equal to the SM partial width
to $t\bar{t}$. SUSY models with this feature can be constructed
\cite{Djouadi:2005gj}.  The smaller widths of the pseudo-scalar and
the boson-phobic scalar give a narrow resonance peak in the $t\bar{t}$
invariant mass spectrum.  The interference between the signal and the
QCD $t\bar{t}$ production leads to a dip in $t\bar{t}$ production at
an invariant mass just above the mass of the spin-0 particle.  In this
case the signal together with the interference terms sum to the
characteristic peak-dip structure, Fig.~\ref{heft_PD}.

The \emph{dot-dashed} line in Fig.~\ref{heft_PD} shows the effect of a
400 GeV color singlet spin-0 particle on the $t\bar{t}$ invariant mass
spectrum with couplings $a_1=1$, $a_2=0$ and $a_1=0$, $a_2=1$ for the
\emph{left} and the \emph{right} plots, respectively.  Comparing with
the QCD $t\bar{t}$ production, the \emph{dark solid} line, a peak-dip
structure is visible when the spin-0 particle is a pseudo-scalar,
$a_1=0$ and $a_2=1$.  In the case where it is a scalar, $a_1=1$ and
$a_2=0$, there is only a peak and a very small dip.

If the coupling to the top quark is enhanced, the peak as well as the
dip becomes broader due to the larger decay width. The peak increases
in the case the spin-0 is a scalar, but remains the same for the
pseudo-scalar.  The \emph{dashed} line shows the effect of enhancing
the $ttH$ coupling by a factor of two. If the coupling to the top
quarks is taken even larger, the increasing width of the
(pseudo-)scalar starts to dominate the effects on the invariant
mass. This results in the disappearance of the dip, as shown by the
\emph{light solid} line in Fig.~\ref{heft_PD}.

In the case where the coupling to the top is smaller than in the SM,
the peak of the scalar gets smaller and the dip completely disappears.
The effect of varying the coupling for the pseudo-scalar are much
smaller.  Even if the coupling to top quarks is reduced by a factor of
two, $a_1=0$ and $a_2=0.5$, a very clear peak-dip structure is still
visible, as shown by the \emph{dotted} line in Fig.~\ref{heft_PD}.

\subsubsection{Color octet}

\begin{figure*}[t]
\hfill
\begin{center}
  \mbox{
    \subfigure[]{
      \epsfig{file=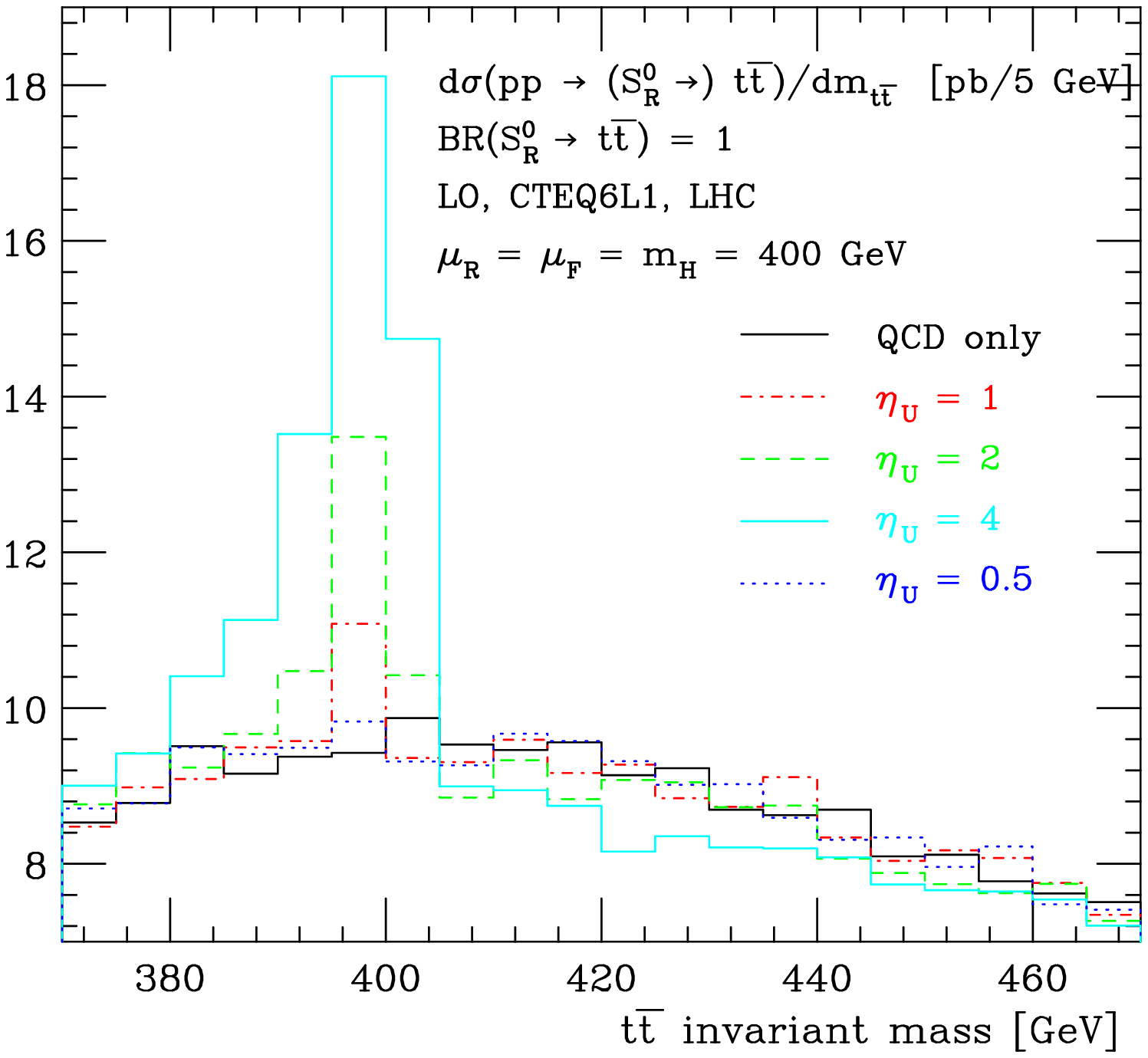, angle=90, scale=0.45}
    }\quad
    \subfigure[]{
      \epsfig{file=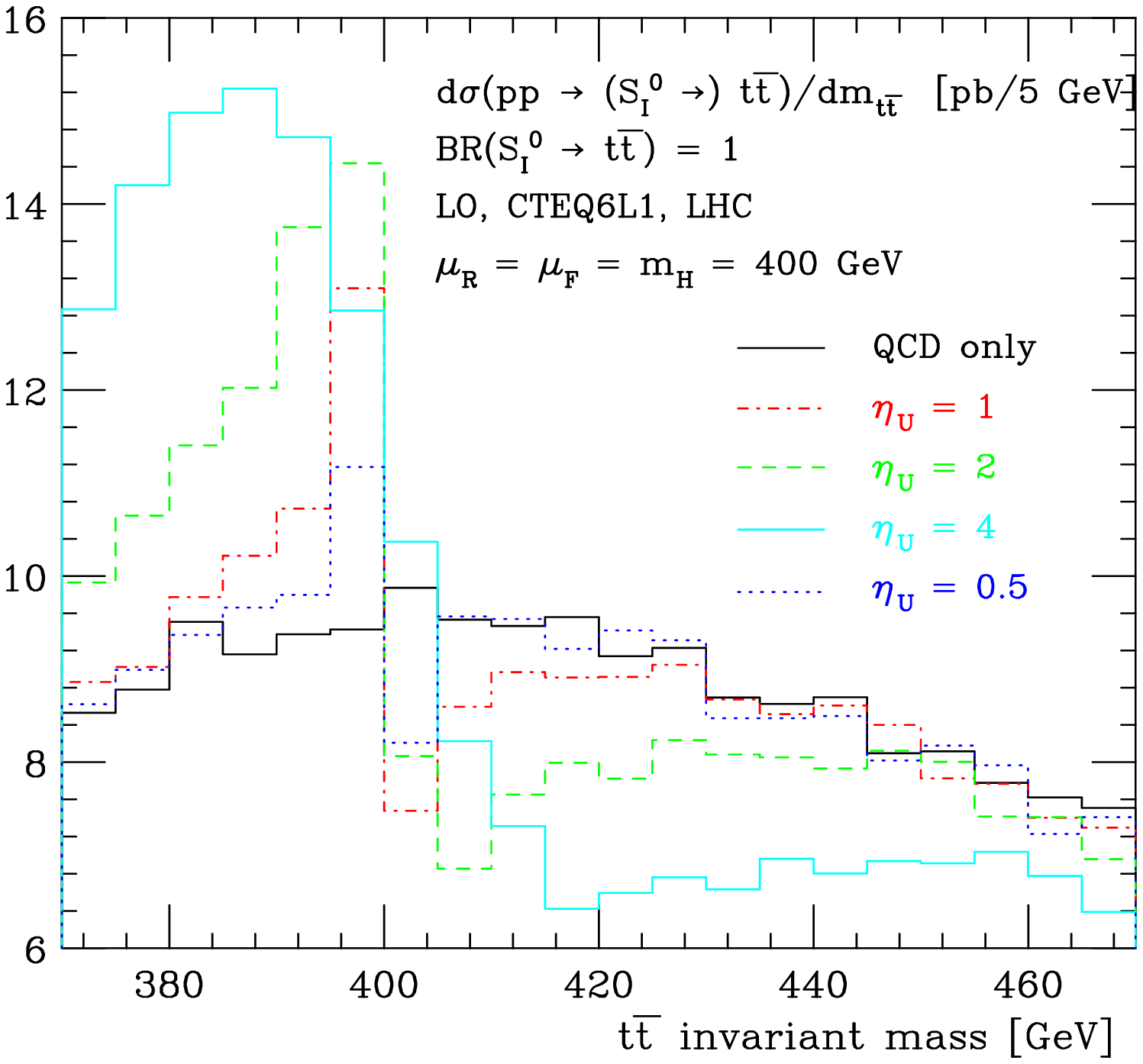, angle=90, scale=0.45}
    }
  }
\vspace{-20pt}
\end{center}
\caption{Invariant $t\bar{t}$ spectrum for $pp\to t\bar{t}$ including a
$s$-channel $S_R^0$ scalar color octet (a) and a pseudo-scalar scalar $S_I^0$
color octet (b) with masses $m_{S_R^0}=m_{S_I^0}=$ 400 GeV.
\emph{Dark solid} line is QCD $t\bar{t}$ production,
\emph{dash-dotted} line is with standard coupling between the scalar and $t\bar{t}$,
\emph{dashed}, \emph{light solid} and \emph{dotted}
the coupling is 2, 4 and 0.5 times as large, respectively.
All plots were produced using the CTEQ6L1 pdf set with $\mu_R=\mu_F=$ 400 GeV.
No acceptance cuts are applied.}
\label{scalar_octet}
\end{figure*}

The case of a color octet resonance is very similar.  
Here we shall study scalar $S_R^0$ and a pseudo-scalar $S_I^0$
color octets, similar to those introduced in
Refs.~\cite{Manohar:2006ga,Gresham:2007ri}. In these models the
(pseudo-)scalar color octet couples only to quarks, with the same 
SM coupling but for the color
\begin{equation}
g_{S_R^0qq}=\eta_U i\frac{m_q}{v}T^a_{ij},
\qquad\qquad\textrm{and}\qquad g_{S_I^0qq}=\eta_U \gamma_5\frac{m_q}{v}T^a_{ij},
\end{equation}
where $\eta_U$ is a coupling proportionality factor and of order 1.
The production and decay mechanism for the (pseudo-)scalar color octet are similar
to the `peak-dip' color singlets, {\it i.e.}, the resonance is produced through a top quark
loop by gluon-gluon fusion, and the decay is mainly to top quarks. We find that 
compared to the `peak-dip' color singlet the `signal' cross section is 5/72 times smaller,
{\it i.e.}, $\sigma(gg\to S^0_{R,I}\to t\bar{t})=\frac{5}{72}\sigma(gg\to H\to t\bar{t})$,
the interference between signal and background is 5/12 times smaller and the
width of the (pseudo-)scalar color octet is 6 times smaller than 
the width of the `peak-dip' \mbox{(pseudo-)scalar} color singlet. 
In Fig.~\ref{scalar_octet}(a) the $t\bar{t}$ invariant mass
is plotted in a model with a color octet scalar of a mass of 400 GeV,
and in Fig.~\ref{scalar_octet}(b) for the pseudo-scalar.
As expected, the results are very similar to the color singlet,
see the lower plots of Fig.~\ref{heft_PD}. The same `peak-dip'
structure is also present for the color octet but it is more pronounced. 
This is mainly due to the smaller width of the (pseudo-)scalar color octet. 

We conclude this section by mentioning that pseudo-scalar singlet and octet resonances
could also arise from bound states of meta-stable gluinos in split SUSY
scenarios~\cite{Cheung:2004ad}. Also in this case $gg\to\phi$ would be the dominant
production channel and it could be described within the same framework. 

\subsection{Spin-1 resonances}

In this section we discuss a spin-1 resonance produced by $q\bar{q}$
annihilation. This resonance can either be a color singlet or a color
octet. For the color octet case we distinguish between a vector and an
axial-vector. Although both the vector and the axial-vector interfere
with the QCD $t\bar{t}$ production, only the vector shows interference
effects in the $t\bar{t}$ invariant mass spectrum.

\begin{figure*}[t]
\hfill
\begin{center}
  \epsfig{file=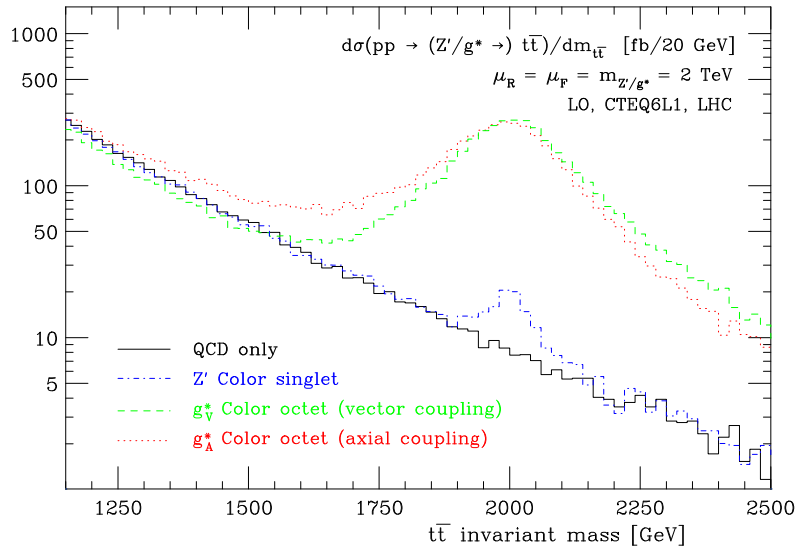, angle=90, width=0.65\textwidth}
\vspace{-20pt}
\end{center}
\caption{Invariant $t\bar{t}$ spectrum for $pp\to t\bar{t}$ including a
$s$-channel $Z'$ color singlet vector boson
and color octet (axial) vector bosons with masses $m_X=$ 2000 GeV that couples with
standard model strength to quarks.
\emph{Solid} QCD $t\bar{t}$ production,
\emph{dotdashed} with a color singlet ($Z'$),
\emph{dotted} with a color octet axial vector (axigluon $g^*_A$),
\emph{dashed} with a color octet vector boson (KK gluon/coloron $g^*_V$).
All plots were produced using the CTEQ6L1 pdf set with $\mu_R=\mu_F=$ 2000 GeV.
No cuts were applied in making any of the plots.}
\label{zprime_X}
\end{figure*}

Including an $s$-channel color singlet vector boson (a ``model-independent'' $Z'$) in the
$t\bar{t}$ production process gives a simple peak in the invariant
mass spectrum as can be seen from the {\it dot-dashed} line in Fig.~\ref{zprime_X}. The precise
width and height of the peak depends on the model parameters in the
model for the $Z'$. As a benchmark we show a $Z'$ vector boson with
mass $m_{Z'}=$ 2 TeV that couples with the same strength to fermions
as a standard model $Z$ boson. The interference effects with the SM
$Z$ boson can be neglected in the $t\bar{t}$ channel, so the peak is
independent of the parity of the coupling.

In general, for the color octet spin-1 particles the interference with
the SM $t\bar{t}$ production cannot be neglected.  Two cases are to be
considered: a color octet vector particle ({\it e.g.}, a KK gluon \cite{Dicus:2000hm} or
coloron \cite{Choudhury:2007ux}), and an axial-vector particle ({\it e.g.},
an axigluon \cite{Frampton:1987ut,Frampton:1987dn,Choudhury:2007ux}).
It is natural to assume a coupling strength equal to the strong (QCD) coupling $g_s$
for their coupling to quarks.

In Fig.~\ref{zprime_X} the effects of a color octet spin-1 particle
on the $t\bar{t}$ invariant mass spectrum are presented. The
interference effects of the axial vector ({\it dotted} line) with the QCD
$t\bar{t}$ production does not change the shape of the $t\bar{t}$
invariant mass spectrum. Hence the effects of the color octet axial vector and the
color singlet are very similar, apart from the size due to the different
coupling constant.

The interference of the color octet vector particle  with the QCD $t\bar{t}$
production does effect the $t\bar{t}$ invariant mass
distribution. There is negative interference in the invariant mass
region below the resonance mass and positive interference for heavier
invariant masses. This slightly changes the shape of the peak as can
be seen from Fig.~\ref{zprime_X}. Other quantities, such as 
the charge asymmetry between the top and the anti-top quarks 
could be more sensitive to axial vectors~\cite{Antunano:2007da} and could
help their discovery at higher invariant masses.

\subsection{Spin-2 resonances}

The interactions between spin-2 particles, or \emph{gravitons}, and
ordinary matter is in general Planck suppressed, which makes it
impossible to see effects of the gravitons at TeV energies.
There are,
however, models with extra dimensions where the contributions from
the gravitons might be large enough to make a discovery at the LHC.  
In this case a model-independent approach is not really appropriate.
Instead we consider two scenarios that have distinct signals in the $t\bar{t}$
invariant mass. 
First the ADD model~\cite{ArkaniHamed:1998rs,ArkaniHamed:1998nn},
where the effect of a the large number of graviton KK states
contributing to a cross sections could be important and, secondly, 
the RS model \cite{Randall:1999ee,Randall:1999vf} where
only a limited number of KK modes contribute, but the coupling
constant itself is enhanced by a large ``warp'' factor.

In the so-called ADD models~\cite{ArkaniHamed:1998rs,ArkaniHamed:1998nn} 
all SM fields are confined to a four-dimensional brane, letting only gravitons
propagate through the bulk. The extra $n$ bulk dimensions are compactified on a
$n$-torus with a radius $R$. If the radius $R$ is large enough,
(of the order of 0.1 mm for 2 extra dimensions) the $(4+n)$ dimensional Planck
scale can be as small as the TeV scale.

Due to the fact that the radius of the extra dimensions is large, 
the graviton KK states can be almost degenerate in mass. So, although all
graviton couplings are Planck suppressed, the sum
of all the KK states can contribute significantly to the $t\bar{t}$ 
invariant mass spectrum. All states are summed up to the cutoff scale
$M_S$, defined by $\lambda^2R^n=8\pi(4\pi)^{n/2}\Gamma(n/2)M_S^{-n+2}$,
where $\lambda$ is related to the four dimensional
Newton's constant $\lambda=\sqrt{16\pi G_N}$.

\begin{figure*}[t]
\hfill
\begin{center}
  \epsfig{file=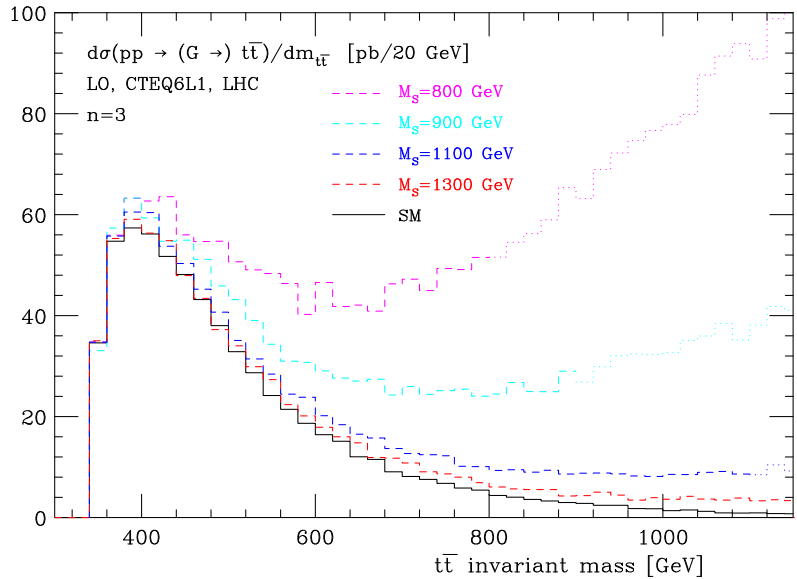, angle=90, width=0.65\textwidth}
\end{center}
\vspace{-20pt}
\caption{Invariant $t\bar{t}$ spectrum for $pp\to t\bar{t}$ including
$s$-channel gravitons. The distributions show the effect 
of the almost degenerate tower of KK gravitons in the ADD model with
$n=3$ extra dimensions and, from top to bottom, with a cutoff scale
$M_S=$ 800, 900, 1100 and 1300 GeV. The bottom line are contributions from SM only.
We used CTEQ6L1 and set the scales to $\mu_R=\mu_F=m_t$.}
\label{spin-2_ADD}
\end{figure*}

The effect of this tower of graviton states
on the $t\bar{t}$ invariant mass distribution is 
plotted in Fig.~\ref{spin-2_ADD} in the case of 3 extra dimensions and for
4 different cutoff scales. Due to the sum over all nearly degenerate 
resonances, there is no single resonance peak in the invariant mass distribution.
It is also clear that the distribution is only valid well below the cutoff scale $M_S$,
otherwise unitarity violating effects become sizable.

In the so-called RS model \cite{Randall:1999ee,Randall:1999vf} there is one
extra dimension postulated that is compactified to a $\mathbf{S}^1/\mathbf{Z}_2$
orbifold.  There are two branes on specific points of the orbifold: a
``Planck'' brane at $\phi=0$ and a ``TeV'' brane at $\phi=\pi$ where
the physical SM fields are confined. The bulk space is warped in such
a way that the (reduced) Planck mass is warped down on the ``TeV'' brane
to $\Lambda=\overline{M}_{\textrm{pl}}e^{-\pi\kappa R}$. The gauge
hierarchy problem ($\Lambda=\mathcal{O}(1 \textrm{ TeV})$) is now
solved with only a minor fine-tuning of $\kappa R\simeq 12$.
After KK compactification of the massless graviton field, the coupling
constant of KK gravitons with matter is given by the inverse of $\Lambda$.

A prediction in the RS model is that the masses of
the KK modes $m_n$ are given by $m_n^2=x_n\kappa e^{-\pi\kappa R}$,
where $x_n$ are the positive zero's of the Bessel function $J_1(x)$.
If one of the masses is given, all the others are fixed, which could
give rise to a series of resonances in the $t\bar{t}$ invariant mass spectrum.

\begin{figure*}[t]
\hfill
\begin{center}
  \epsfig{file=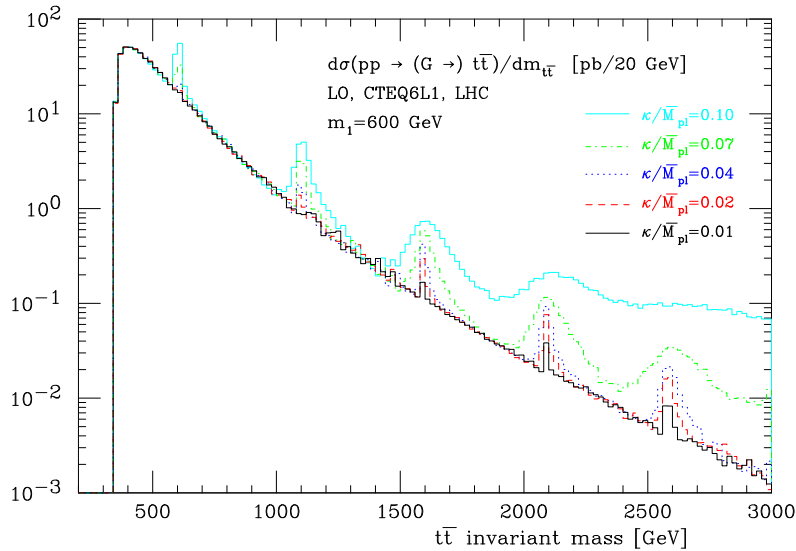, angle=90, width=0.65\textwidth}
\end{center}
\vspace{-20pt}
\caption{Invariant $t\bar{t}$ spectrum for $pp\to t\bar{t}$ including
$s$-channel gravitons. The distribution shows the effect of a couple
of KK resonances in the RS extra dimensions model. The mass of the
first KK mode is $m_1=600$ GeV and the colored lines represent various
choices for the ratio $\kappa/\overline{M}_{\textrm{pl}}$.  We used
CTEQ6L1 and set the scales to $\mu_R=\mu_F=m_t$.}
\label{spin-2_RS}
\end{figure*}

In Fig.~\ref{spin-2_RS} the effect of a series of KK graviton modes on
the $t\bar{t}$ invariant mass spectrum is shown with $m_1=600$ GeV and
for various ratios $\kappa/\overline{M}_{\textrm{pl}}$. The resonances
are clearly visible over the QCD background.  Higher KK states are
characterized by larger widths.

\section{Spin information from (anti-)top quark directions}
\label{sec:spin_info}

A useful, yet simple, quantity sensitive to the spin of the
intermediate heavy state into a $t\bar{t}$ pair, is the 
Collins-Soper angle $\theta$ \cite{Collins:1977iv}.
This angle is similar to the angle between the top quark and the beam
direction, but minimizes the dependence on initial state radiation.
$\theta$ is defined as follows. Let $p_A$ and $p_B$ be the momenta of the
incoming hadrons in the rest frame of the top-antitop pair. If the
transverse momentum of the top-antitop pair is non-zero, then $p_A$
and $p_B$ are not collinear. The angle $\theta$ is defined to be the
angle between the axis that bisects the angle between $p_A$ and $p_B$
and the top quark momentum in the $t\bar{t}$ rest frame.

\subsection{Standard Model}

\begin{figure*}[t]
\hfill
\begin{center}
  \epsfig{file=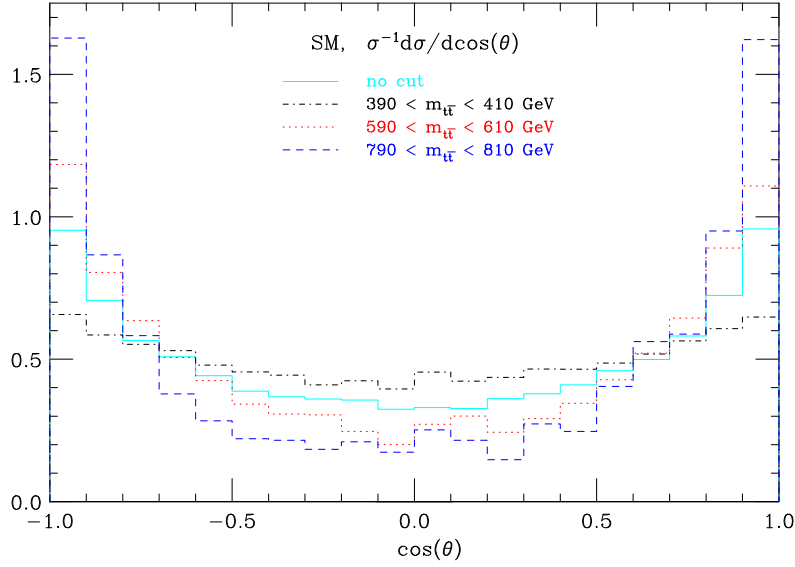, angle=90, width=0.65\textwidth}
\end{center}
\vspace{-10pt}
\caption{Normalized distribution for $\cos\theta$, where $\theta$ is
the Collins-Soper angle, for SM production at the LHC.  Different
lines represent different cuts on $t\bar{t}$ invariant mass.}
\label{backgr_spin_info}
\end{figure*}

The distribution of $\theta$ in the SM is plotted in
Fig.~\ref{backgr_spin_info}. Also plotted in the same figure are the
distributions with cuts on the $t\bar{t}$ invariant mass spectrum as
backgrounds to narrow resonances.

A simple analytic calculation confirms this behavior. The matrix element squared for the 
initial state $q\bar{q}$ to the SM $t\bar{t}$ contribution in terms of the Collins-Soper
angle $\cos\theta$ is proportional to
\begin{equation}
|\mathcal{M}(q\bar{q}\to t\bar{t})|^2\sim
s (1+\cos^2\theta)+4m_t^2(1-\cos^2\theta),
\end{equation}
where $s$ is the center of mass energy squared, $s=(p_q+p_{\bar{q}})^2$.
For the $gg$ initial state we have
\begin{multline}
|\mathcal{M}(gg\to t\bar{t})|^2\sim\\
\frac{s(7+9\cos^2\theta)-36m_t^2\cos^2\theta}{\big(sc_-+4m_t^2\cos^2\theta\big)^2}
\bigg[s^2c_+c_-+2sm_t^2\Big(3c_-^2+c_+^2\Big)
-4m_t^4\Big(3c_-^2+c_+^2+c_-\Big)\bigg],
\end{multline}
where $c_+=1+\cos^2\theta$ and $c_-=1-\cos^2\theta$.

\subsection{Spin-0 resonances}

\begin{figure*}[t]
\hfill
\begin{center}
  \mbox{
    \subfigure[]{
      \epsfig{file=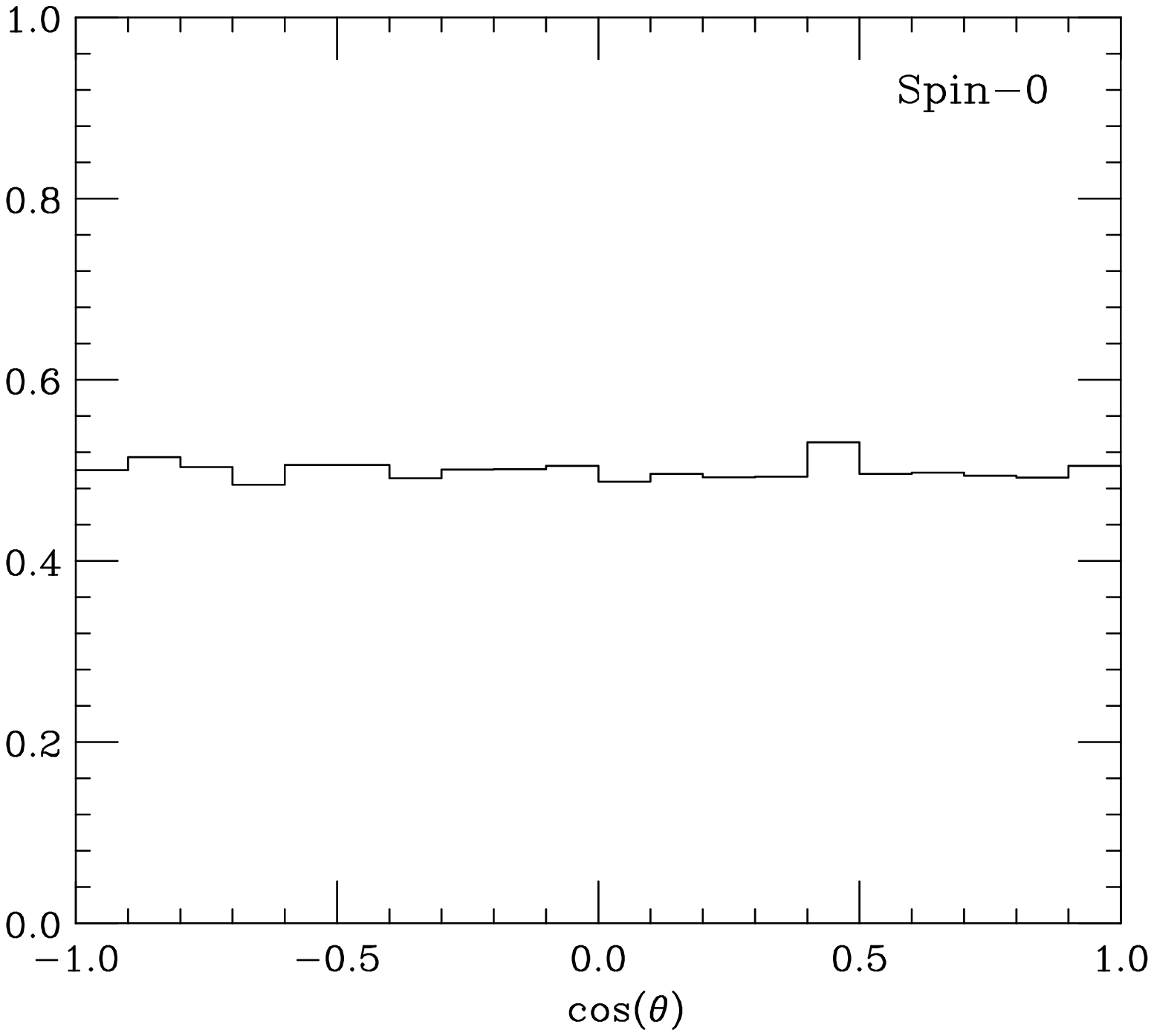, angle=90, width=0.45\textwidth}
    }\quad
    \subfigure[]{
      \epsfig{file=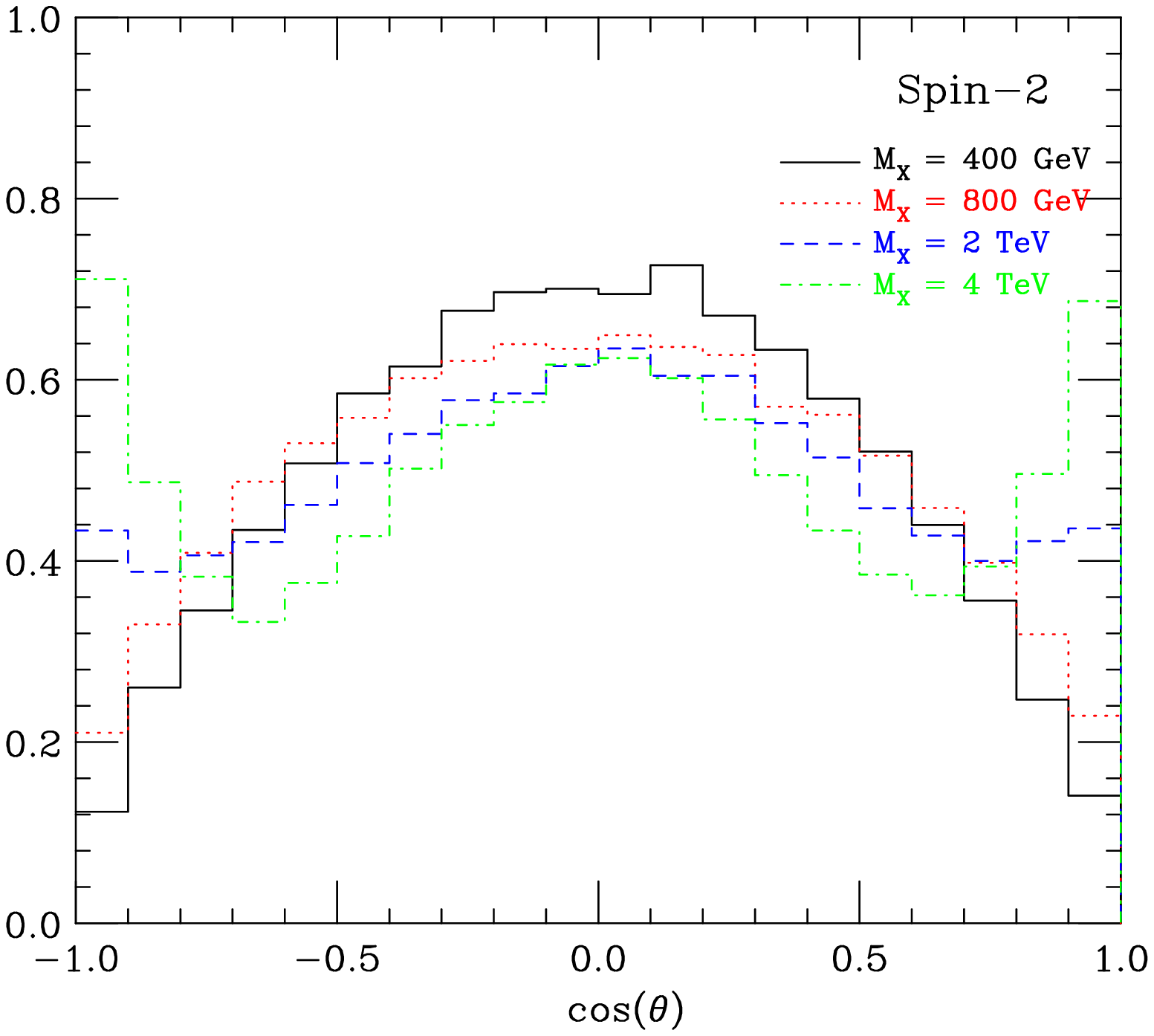, angle=90, width=0.45\textwidth}
    }
  }
  \mbox{
    \subfigure[]{
      \epsfig{file=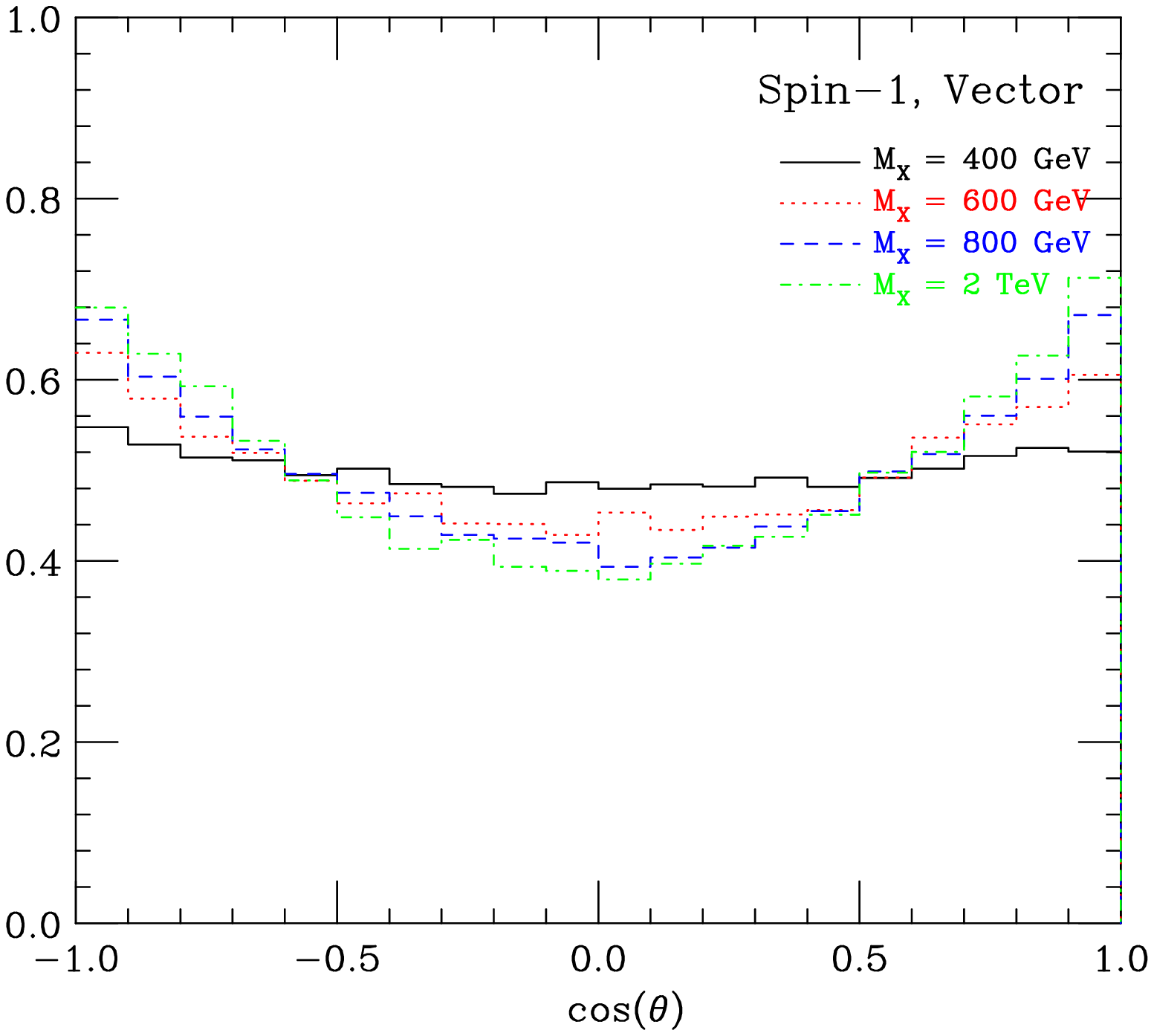, angle=90, width=0.45\textwidth}
    }\quad
    \subfigure[]{
      \epsfig{file=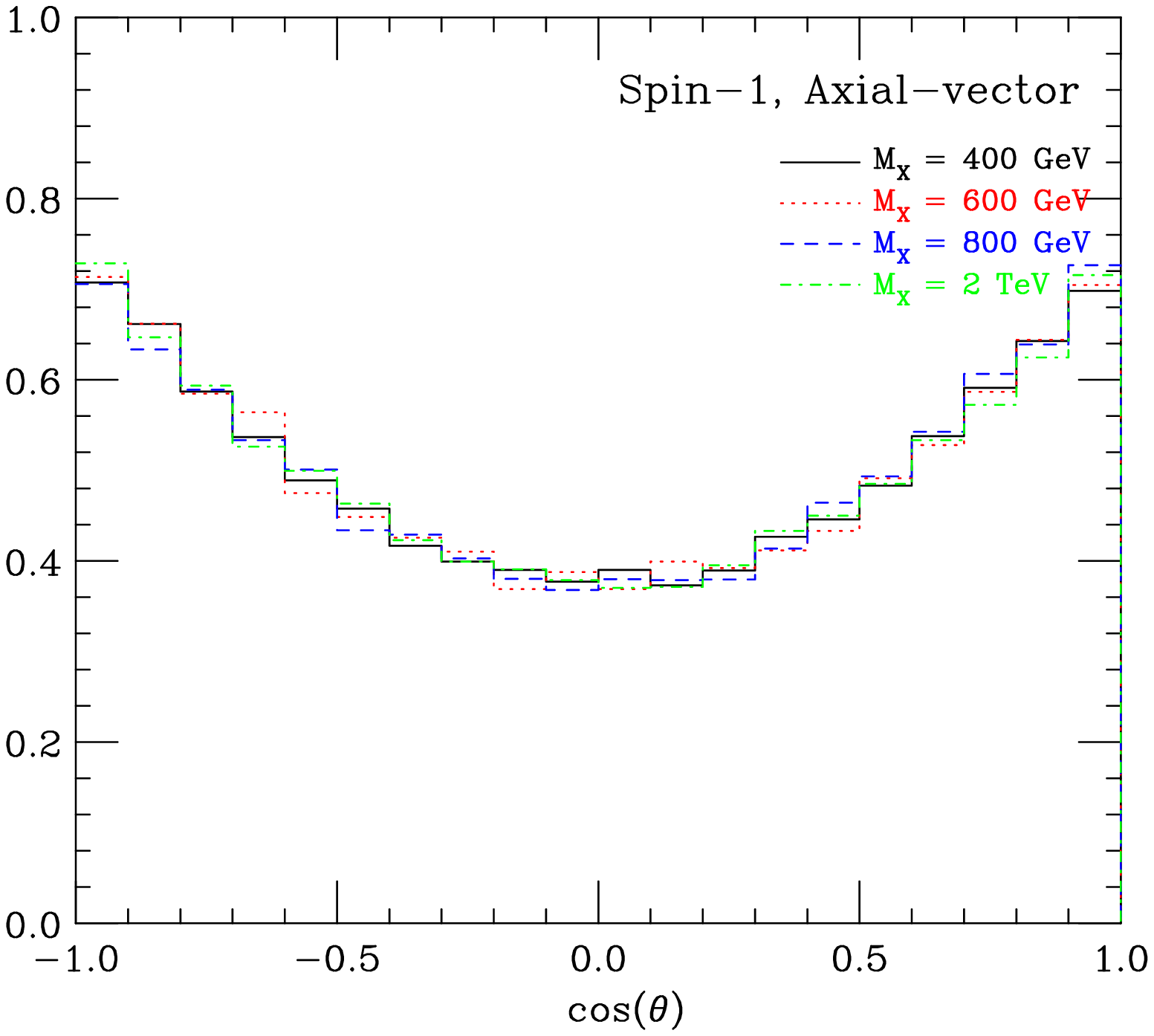, angle=90, width=0.45\textwidth}
    }
  }
\end{center}
\vspace{-20pt}
\caption{Normalized distributions for $\cos\theta$, where $\theta$ is the Collins-Soper angle,
for spin-0 (a), spin-2 (b), vector (c) and axial-vector (d) resonances
of mass $M_X$.
All plots were produced using the CTEQ6L1 pdf set with $\mu_R=\mu_F=M_X$.}
\label{spin_info}
\end{figure*}

In Fig.~\ref{spin_info}(a) the normalized cross section as a function of  $\cos\theta$ 
is plotted for a spin-0 resonance. The distribution is independent of the
mass and parity of the resonance. The matrix element squared for the spin-0
resonance $H$ is at leading order proportional to
\begin{equation}
|\mathcal{M}(gg/q\bar{q}\to H \to t\bar{t})|^2\sim
(|a_1|^2+|a_2|^2)p_t\cdot p_{\bar{t}}-(|a_1|^2-|a_2|^2)m_t^2,
\end{equation}
where $p_t$ and $p_{\bar{t}}$ are the momenta of the top and anti-top quarks, respectively,
and $a_1$ and $a_2$ are the coupling constants, see Eq.~\ref{eq:3},
for the scalar and pseudo-scalar, respectively. The matrix element squared is
clearly independent of the angle $\cos\theta$, which explains the flat
distribution.

\subsection{Spin-1 resonances}

For a generic spin-1 resonance $Z'$ the matrix element squared is proportional to
\begin{multline}
|\mathcal{M}(q\bar{q}\to Z' \to t\bar{t})|^2\sim
2(|a_L|^4+|a_R|^4)p_q\cdot p_{\bar{t}}\,\, p_{\bar{q}}\cdot p_{t}+
4|a_L|^2|a_R|^2p_q\cdot p_{t}\,\, p_{\bar{q}}\cdot p_{\bar{t}}+\\
m_t^2(|a_L|^2+|a_R|^2)(a_La_R^*+a_Ra_L^*)p_q\cdot p_{\bar{q}},
\end{multline}
where $a_L$ and $a_R$ are the left and right handed part of the couplings of the $Z'$
resonance to quarks,
{\it i.e.}, $g_{Z'q\bar{q}}\sim a_L\frac{1-\gamma_5}{2}+a_R\frac{1+\gamma_5}{2}$ and
where $p_q$ and $p_{\bar{q}}$ are the momenta of the incoming quark and anti-quark, respectively.
In terms of $\cos\theta$ the matrix element squared is proportional to
\begin{equation}\label{eq:15}
|\mathcal{M}(q\bar{q}\to Z' \to t\bar{t})|^2\sim
(|a_L|^2+|a_R|^2)(s-4m_t^2)(1+\cos^2\theta)+
4m_t^2|a_L+a_R|^2.
\end{equation}
The normalized $\cos\theta$ distribution
is independent of the mass of the resonance for a axial vector, $a_R=-a_L$
(see Fig.~\ref{spin_info}(d)), while
for a pure vector resonance the dependence is maximal Fig.~\ref{spin_info}(c).
However, for heavy resonances,
$M_X\gtrsim 800$ GeV the difference between the curves for the vector and the axial-vector
is less then 8\% which makes it challenging to get any information
about parity of the coupling from this distribution. In
Ref.~\cite{Barger:2006hm} a similar polar angle has been studied. That polar
angle is also sensitive to the chirality of the coupling. However, the
Collins--Soper angle used here has the advantage that it minimizes the
effects from initial state radiation.

\subsection{Spin-2 resonances}

In the case of the spin-2 resonance $G_{\mu\nu}$, both the $q\bar{q}$ and $gg$ initial states
contribute. The matrix element squared for the $q\bar{q}$ initial state is
proportional to
\begin{multline}
|\mathcal{M}(q\bar{q}\to G_{\mu\nu} \to t\bar{t})|^2\sim
s(1-3\cos^2\theta+4\cos^4\theta)+
16m_t^2\cos^2\theta(1-\cos^2\theta),
\end{multline}
and for the $gg$ initial state
\begin{equation}
|\mathcal{M}(gg\to G_{\mu\nu} \to t\bar{t})|^2\sim
\Big[s(1+\cos^2\theta)+4m_t^2(1-\cos^2\theta)\Big](1-\cos^2\theta).
\end{equation}
The large differences in the distributions for the spin-2 resonances
between light compared to heavy spin-2 particles,
see Fig.~\ref{spin_info}(b),
is due to the fact that the relatively light \mbox{spin-2} particles are mainly produced
by gluon fusion, while the very heavy spin-2 particles by quark-antiquark annihilation.

\section{Spin correlations in (anti-)top-quark decays}\label{sec:spin_corr}

In the standard model, the semi-weak top-quark decay width is rather
large $\Gamma \approx 1.5 \textrm{ GeV} > \Lambda_{QCD}$ and top
quarks do not form bound hadronic states. At present, we do not have
any direct measurement of the top width and the formation of top
hadrons is not excluded.  This could happen for example, if $V_{tb}$
were much smaller than what is predicted in the standard model, as
discussed in Ref.~\cite{Alwall:2006bx}.  Note, however, that even if
this were to happen, the information on the spin of the top quark
would be anyway fully inherited by its decay
products~\cite{Willenbrock:2002ta}, as spin-flip would occur at time
scales of the order $m_t/\Lambda_{QCD}^2$, {\it i.e.}, much later than the
lifetime of the top quark.  In this respect, spin correlation effects
are a very robust probe of new physics entering in the production
cross section.

For standard model leptonic top decays, the directions of the leptons
are 100\% correlated with the polarization of the top quarks.  The
spin analyzing power of the direction of the $b$ quark ($W^+$ boson) is
not as good, around $-0.4$ ($0.4$).  In hadronic top decays the
anti-down (or anti-strange) quarks coming from the $W^+$ boson decay
have the same full spin analyzing power as the lepton. On the other
hand, the up (or charm) quarks have a spin analyzing power of only
$-0.3$, {\it i.e.}, the same as the neutrino in leptonic decays.  For
the decay of anti-top quarks or spin-down top quarks, all spin
analyzing powers change sign. The angular distributions of the two
down-type fermions (leptons in leptonic top decays or jets coming from
down-type quarks in hadronic $W$ decays) give maximal information about
the spin of the (anti-)top quarks in $t\bar{t}$
events~\cite{Jezabek:1994zv,Czarnecki:1990pe}.

In studies on the spin correlations in $t\bar{t}$ production,
two distributions are usually considered~\cite{Beneke:2000hk,Bernreuther:2004jv}.
First the distribution
\begin{equation}\label{eq2}
\frac{1}{\sigma}\frac{d^2\sigma}{d\cos\theta_+d\cos\theta_-}
=\frac{1}{4}\Big(1-A\cos\theta_+\cos\theta_-+b_+\cos\theta_++b_-\cos\theta_- \Big),
\end{equation}
where $\theta_+$ ($\theta_-$) is the angle between the $t$ ($\bar{t}$)
direction in the $t\bar{t}$ center of momentum frame and the $f_d^+$
($f_d^-$) direction in the $t$ ($\bar{t}$) rest frame, where $f_d^+$
($f_d^-$) is the down-type fermion coming from the $W^+$ ($W^-$)
decay.  For a parity conserving $t\bar{t}$ production mechanism, such
as QCD, the parameters $b_+$ and $b_-$ vanish. In practice, the way to
construct these angles is first to construct the $t$ and $\bar{t}$
four-momenta in the laboratory frame. Then perform a rotation-free
boost from the laboratory frame to the $t\bar{t}$ center of momentum
frame to define the $t$ ($\bar{t}$) direction in the $t\bar{t}$ center
of momentum frame. Thirdly, boost the down-type fermion momenta, {\it
i.e.}, the lepton in leptonic top decays and the down-type quark in
hadronic $W$ decays, from the $t\bar{t}$ center of momentum
rotation-free to the $t$ and $\bar{t}$ rest frames.  If the $t$ and
$\bar{t}$ rest frames are constructed directly by boosting from the
laboratory frame a Wigner rotation has to be taken into
account~\cite{Bernreuther:2004jv}.

Defining the angles $\theta_+$ and $\theta_-$ as described above,
corresponds to studying spin correlations of the $t\bar{t}$ pair 
in the helicity basis. 

It is important to stress that in spin correlation studies it is
mandatory to reconstruct the top and the anti-top quark momenta.  In
the case of a double leptonic decay, two neutrino's are emitted and
the full reconstruction of the event becomes non trivial.  Imposing
kinematic constraints, such as the known top and W masses, a
constrained system of equations for the neutrino momenta can be set
up.  In general multiple solutions arise and the best solution can be
only obtained on a statistical basis~\cite{Dalitz:1992np,Beneke:2000hk,Sonnenschein:2006ud,Arai:2007ts,PRL.95.022001,CMSdi,ATLASdi}. In the appendix
some of the issues for the reconstruction are discussed in more detail.
Alternatively, the single leptonic $t\bar{t}$ decay could be used by
letting one of the jets of the hadronically decayed (anti-)top quark
play the role of the lepton. Ideally, one would like to use the jet
coming from the down-type quark, because it has the same (maximum)
spin analyzing power as the lepton. In experiments one cannot easily
distinguish between up- and down-type quarks jets on event-by-event
basis, and the analyzing power gets averaged $(1-0.3)/2\approx0.35$.
However, improvements can be achieved by exploiting the fact that
down-type quark jets have in general a smaller transverse momentum
than the up-type quark jets.  Using the least energetic (non-$b$) jet
from the hadronic top decay increases the spin-analyzing power from
$0.35$ to approximately $0.5$
\cite{Bernreuther:2004jv,Mahlon:1995zn}. For illustrational purposes
in the following we assume that the top quark momenta are correctly
reconstructed and the spin analyzing power is maximal.

\begin{figure*}[t]
  \begin{center}
    \mbox{
      \subfigure[]{
        \epsfig{file=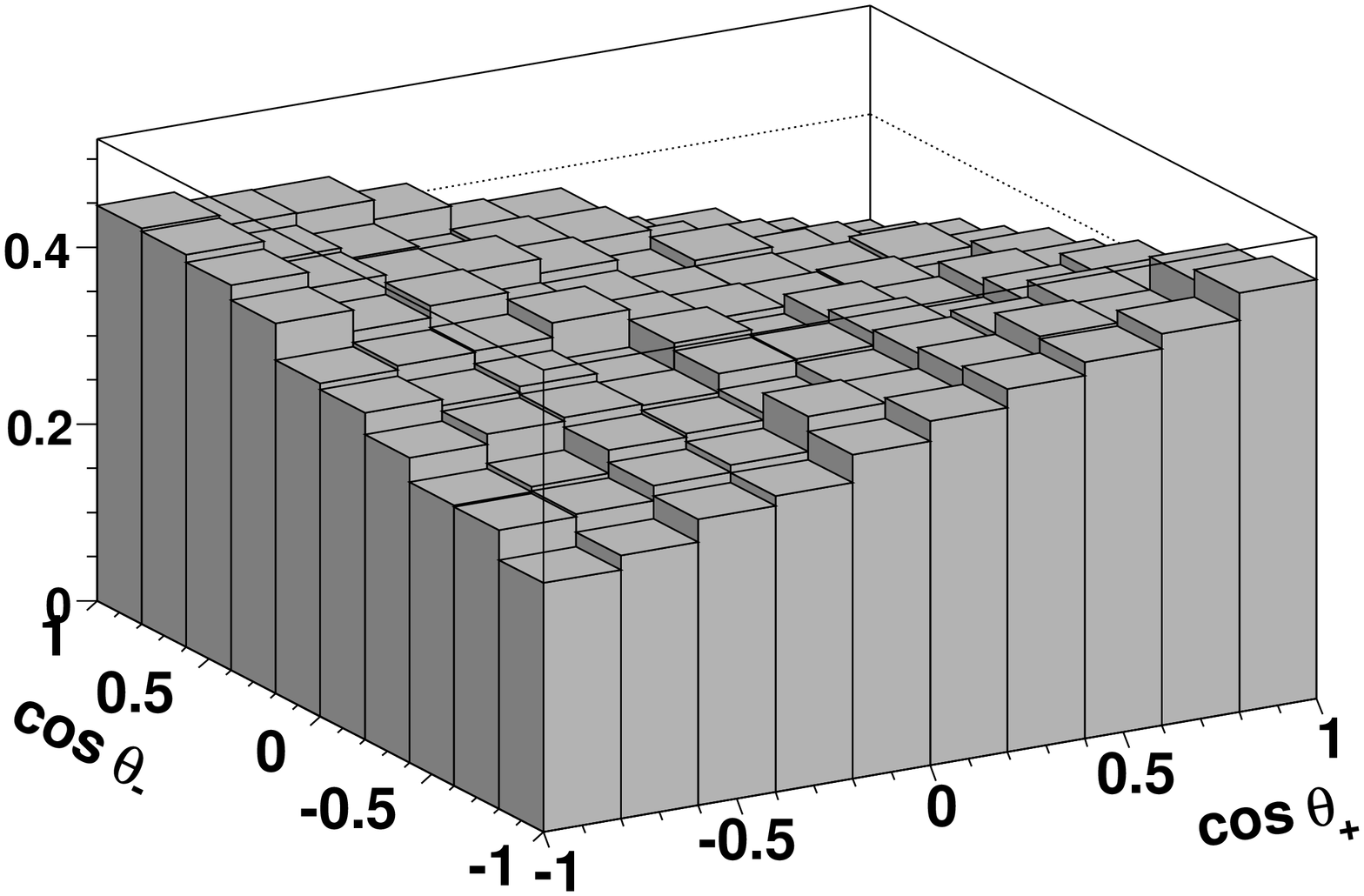, angle=-90, width=0.4\textwidth, clip=}
      }
    }
    \mbox{
      \subfigure[]{
        \epsfig{file=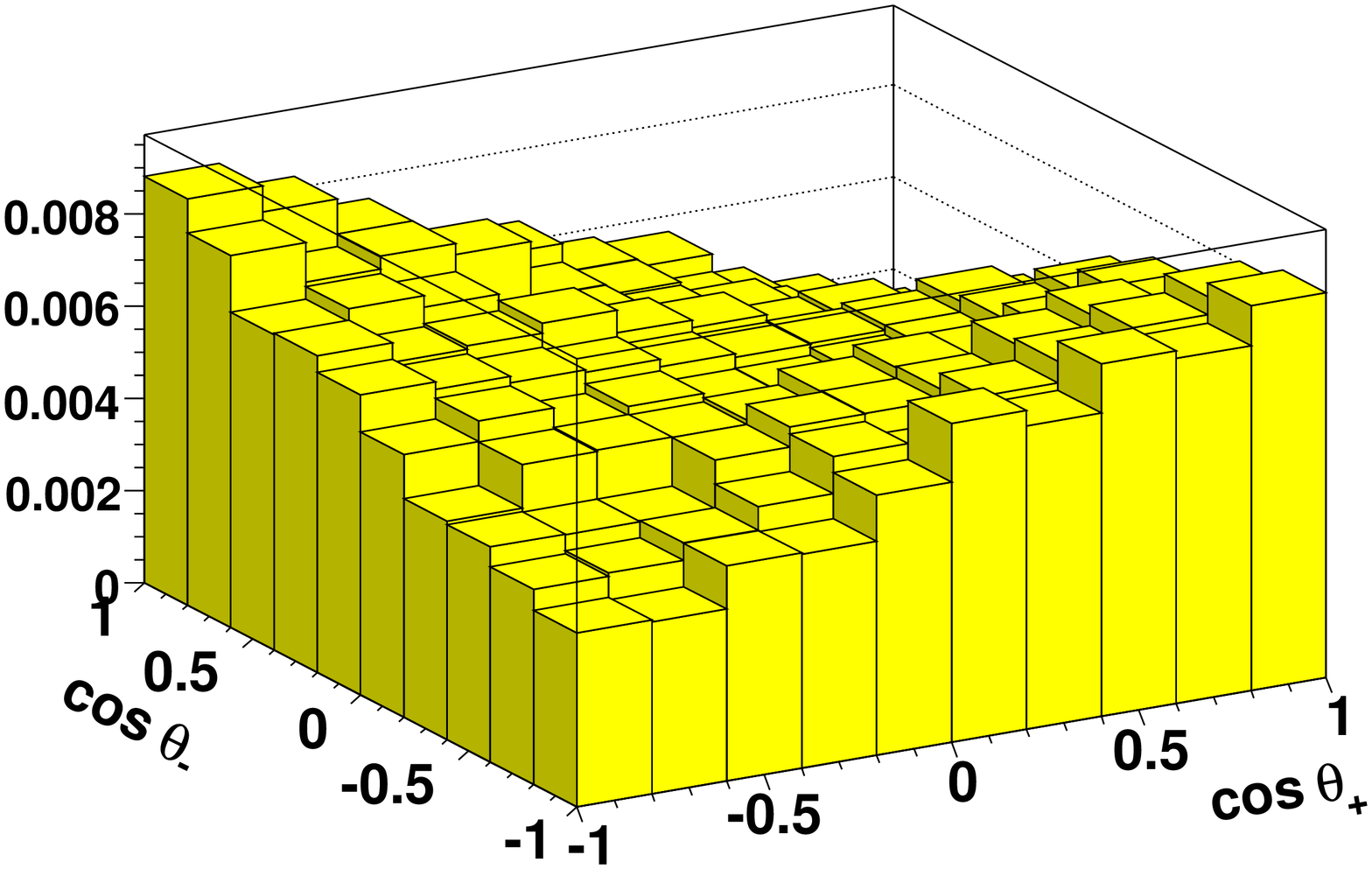, angle=-90, width=0.35\textwidth, clip=}
      }\qquad
      \subfigure[]{
        \epsfig{file=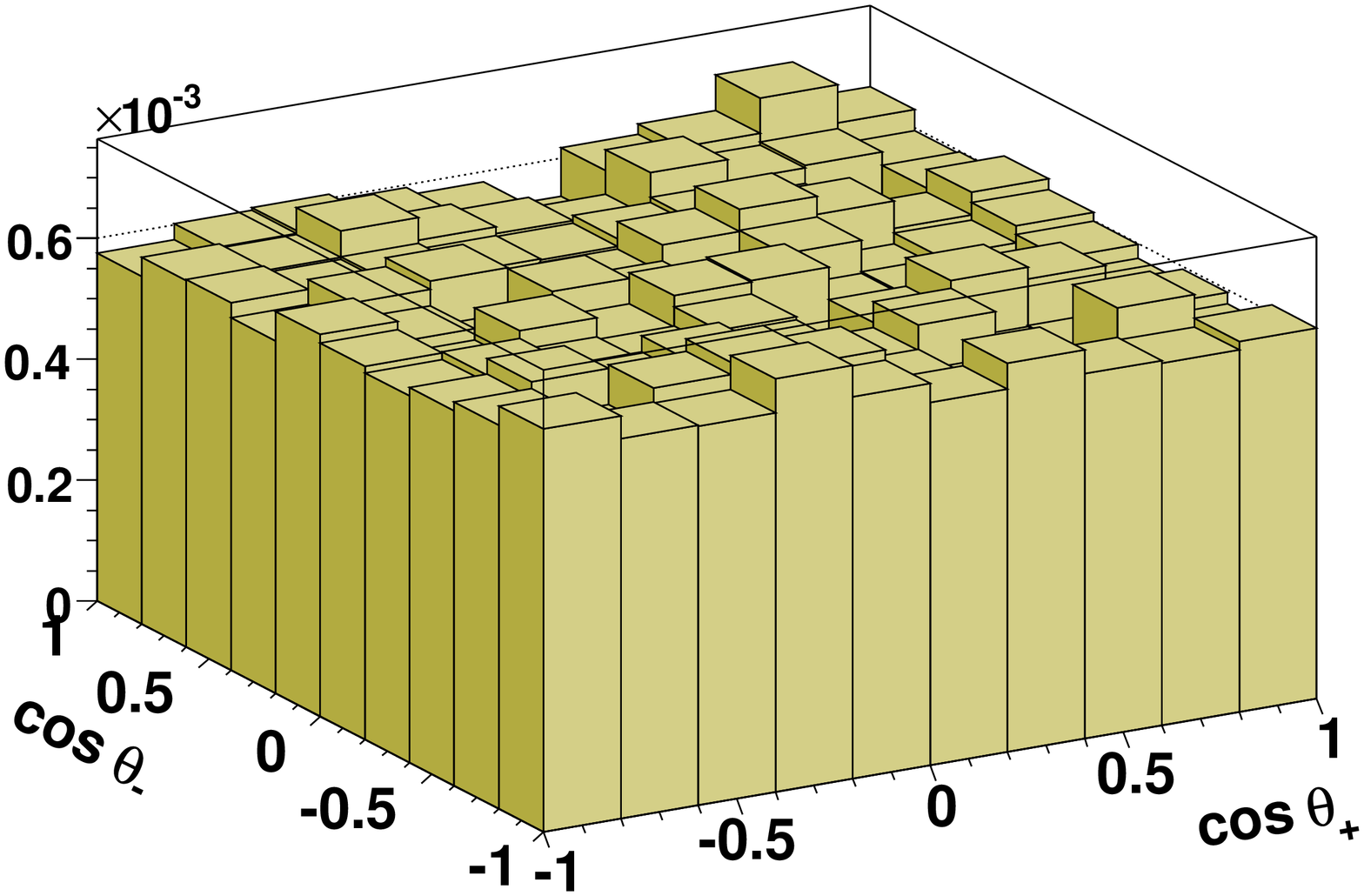, angle=-90, width=0.35\textwidth, clip=}
      }
    }
  \end{center}
\vspace{-20pt}
  \caption{The distribution $\frac{1}{\sigma}\frac{d^2\sigma}{d\cos\theta_+d\cos\theta_-}$
for SM $t\bar{t}$ production at the LHC, using pdf set CTEQ6L1, without applying cuts (a),
and for the regions $390<m_{t\bar{t}}<410$ GeV (b) and $790<m_{t\bar{t}}<810$ GeV (c).}
\label{sm_spin_3D}
\end{figure*}

In Fig.~\ref{sm_spin_3D} this distribution is plotted for QCD $t\bar{t}$
production.

\begin{figure*}[t]
  \begin{center}
    \mbox{
      \subfigure[]{
        \epsfig{file=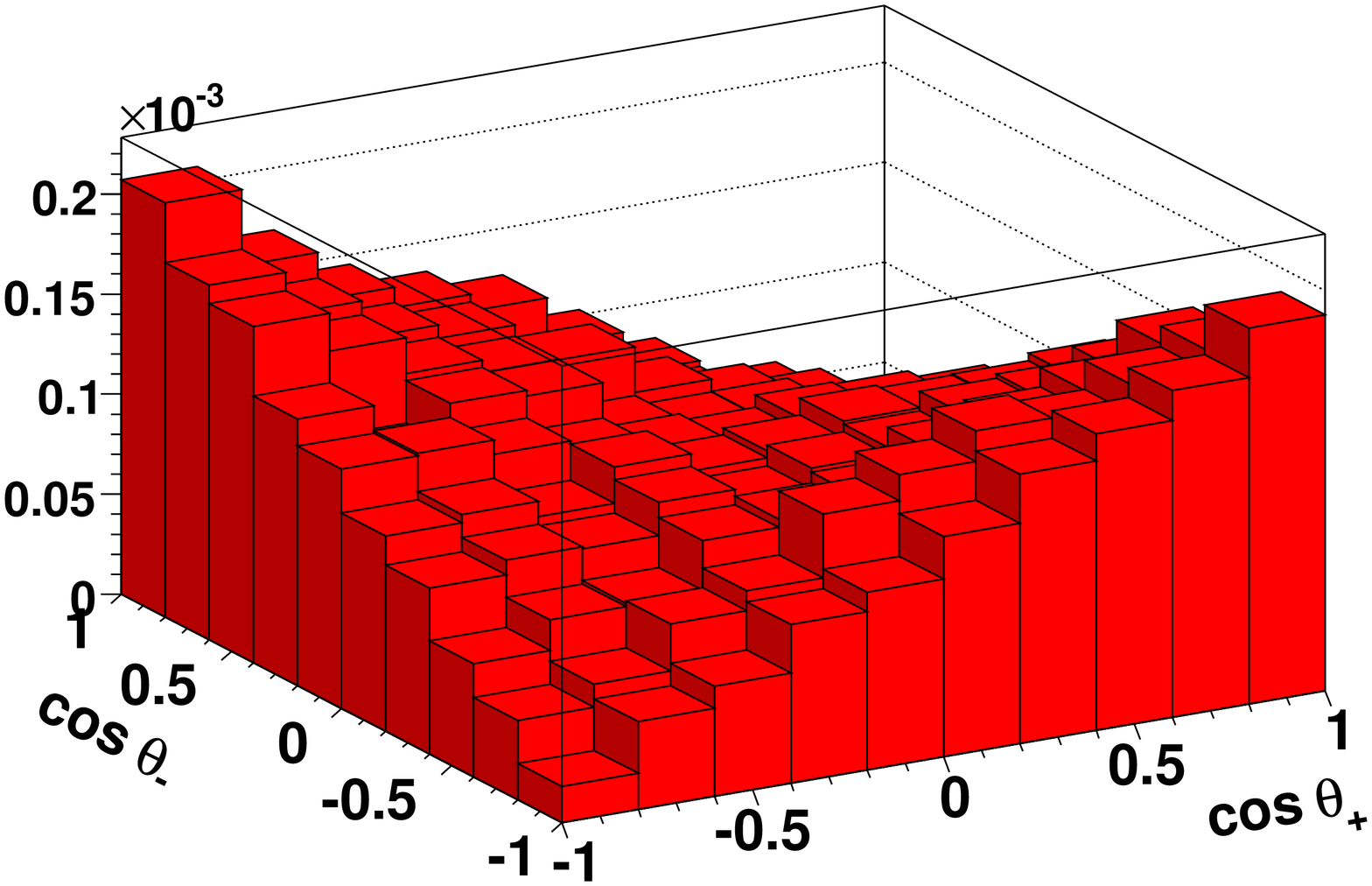, angle=-90, width=0.3\textwidth, clip=}
      }\qquad
      \subfigure[]{
        \epsfig{file=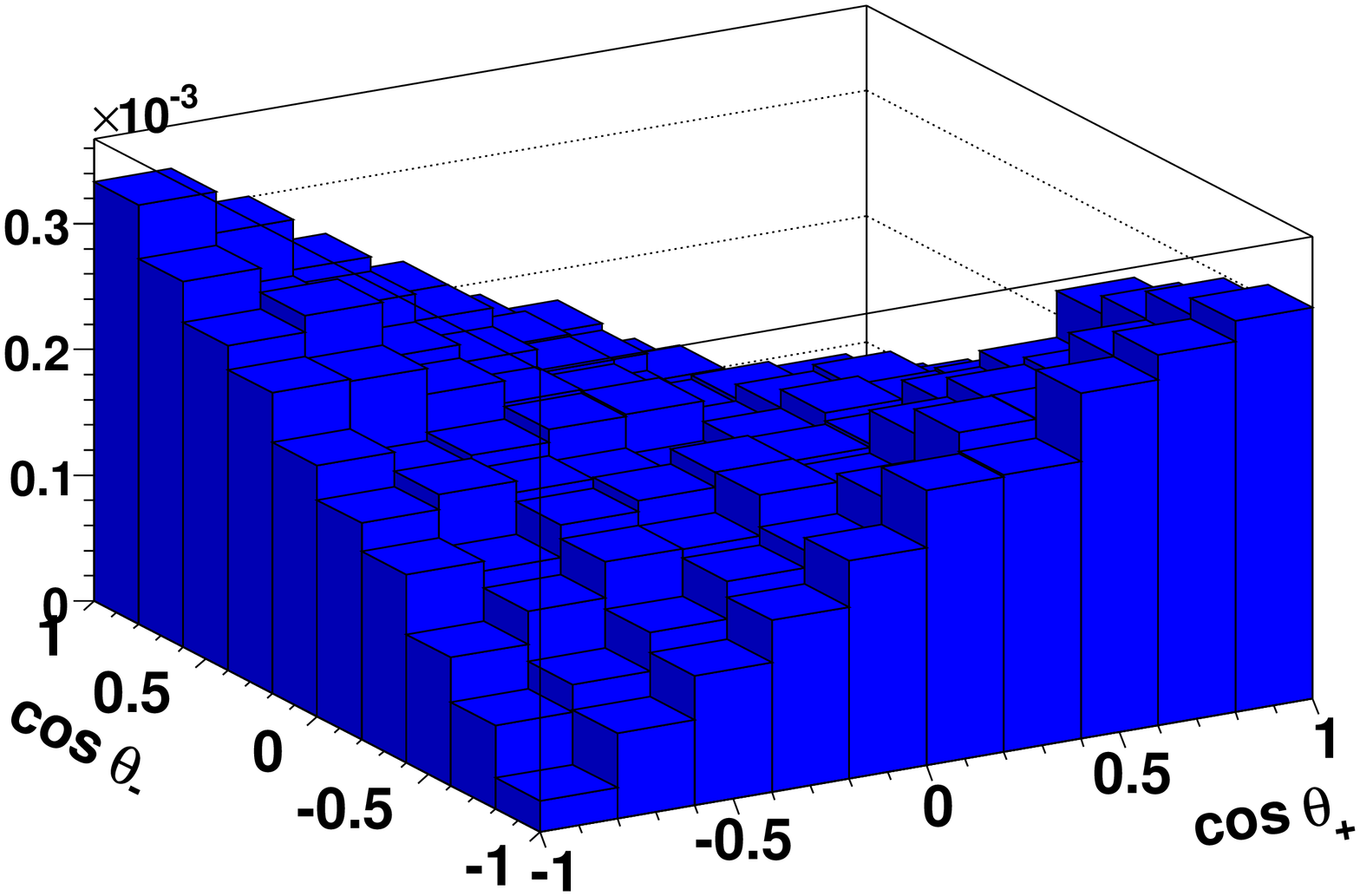, angle=-90, width=0.3\textwidth, clip=}
      }
    }
    \mbox{
      \subfigure[]{
        \epsfig{file=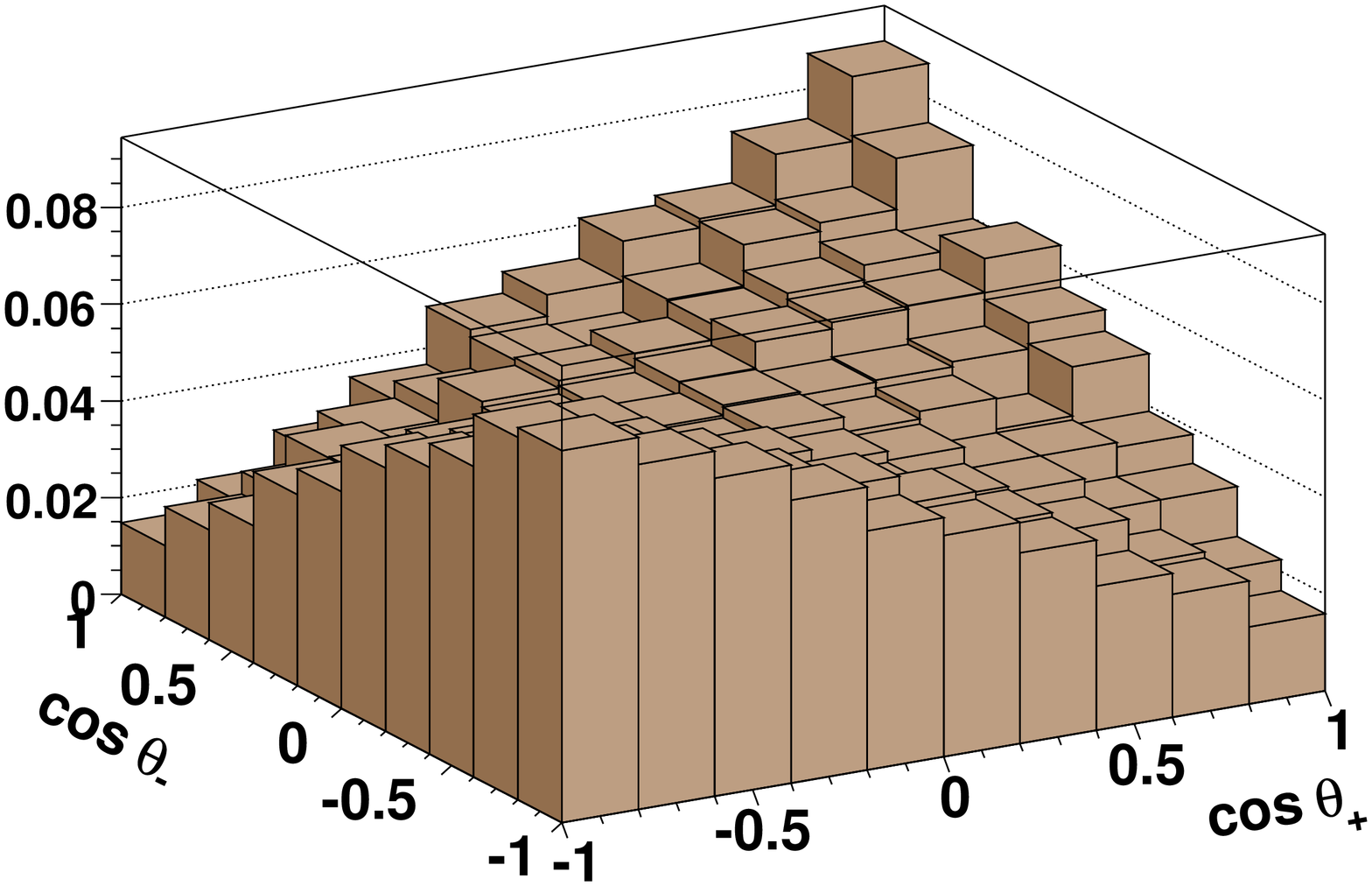, angle=-90, width=0.3\textwidth, clip=}
      }\qquad
      \subfigure[]{
        \epsfig{file=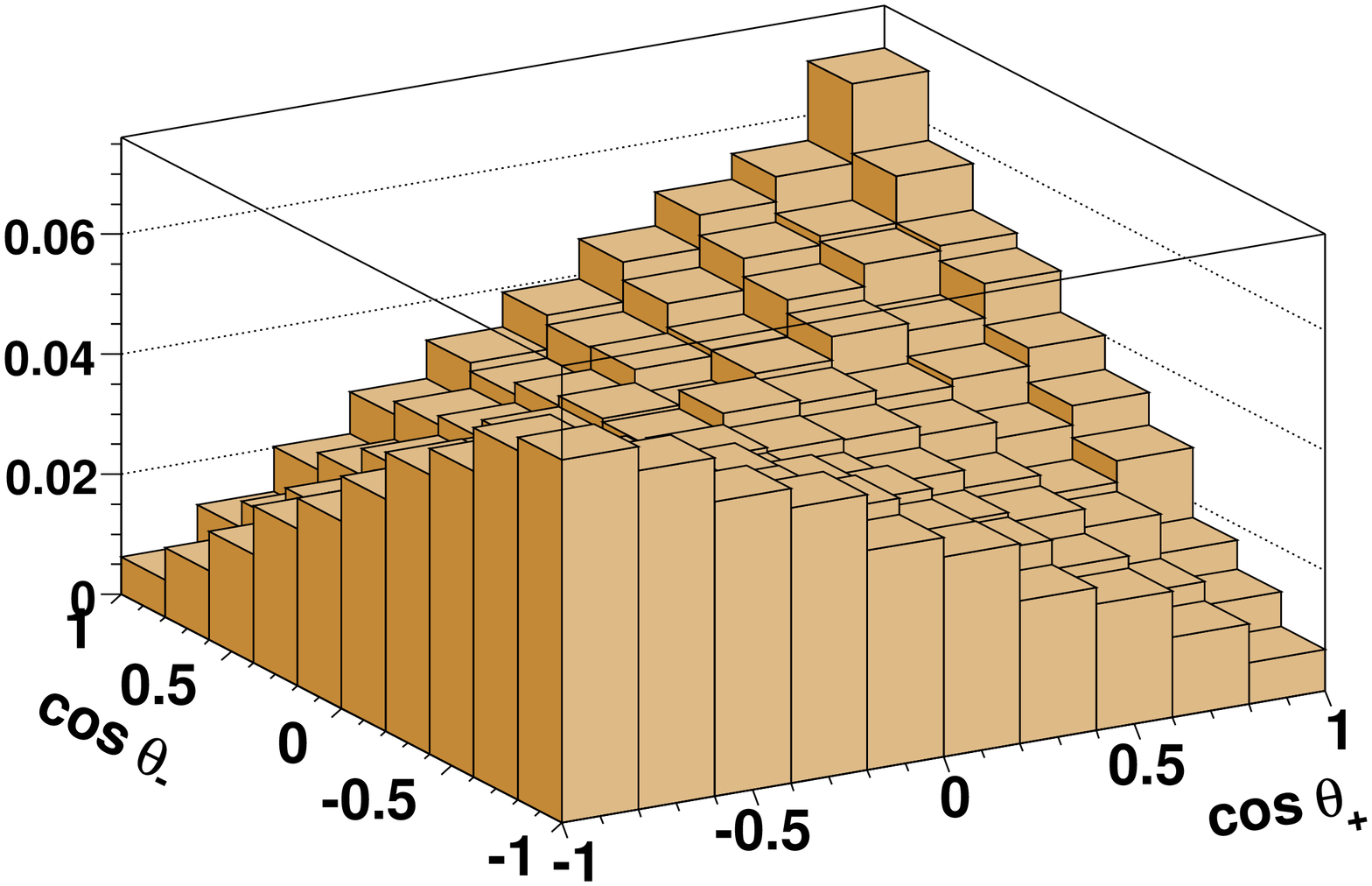, angle=-90, width=0.3\textwidth, clip=}
      }
    }
    \mbox{
      \subfigure[]{
        \epsfig{file=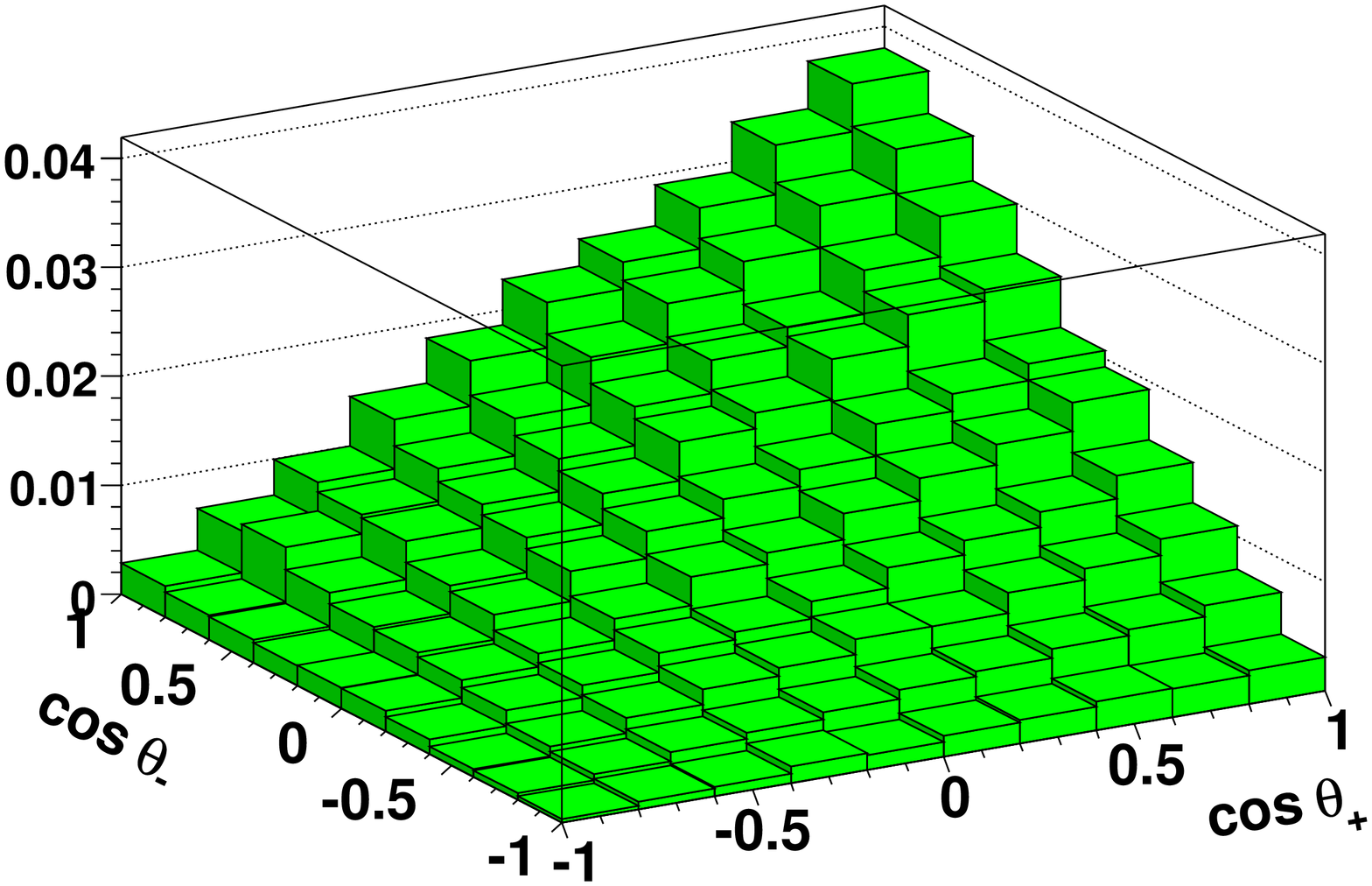, angle=-90, width=0.3\textwidth, clip=}
      }\qquad
      \subfigure[]{
        \epsfig{file=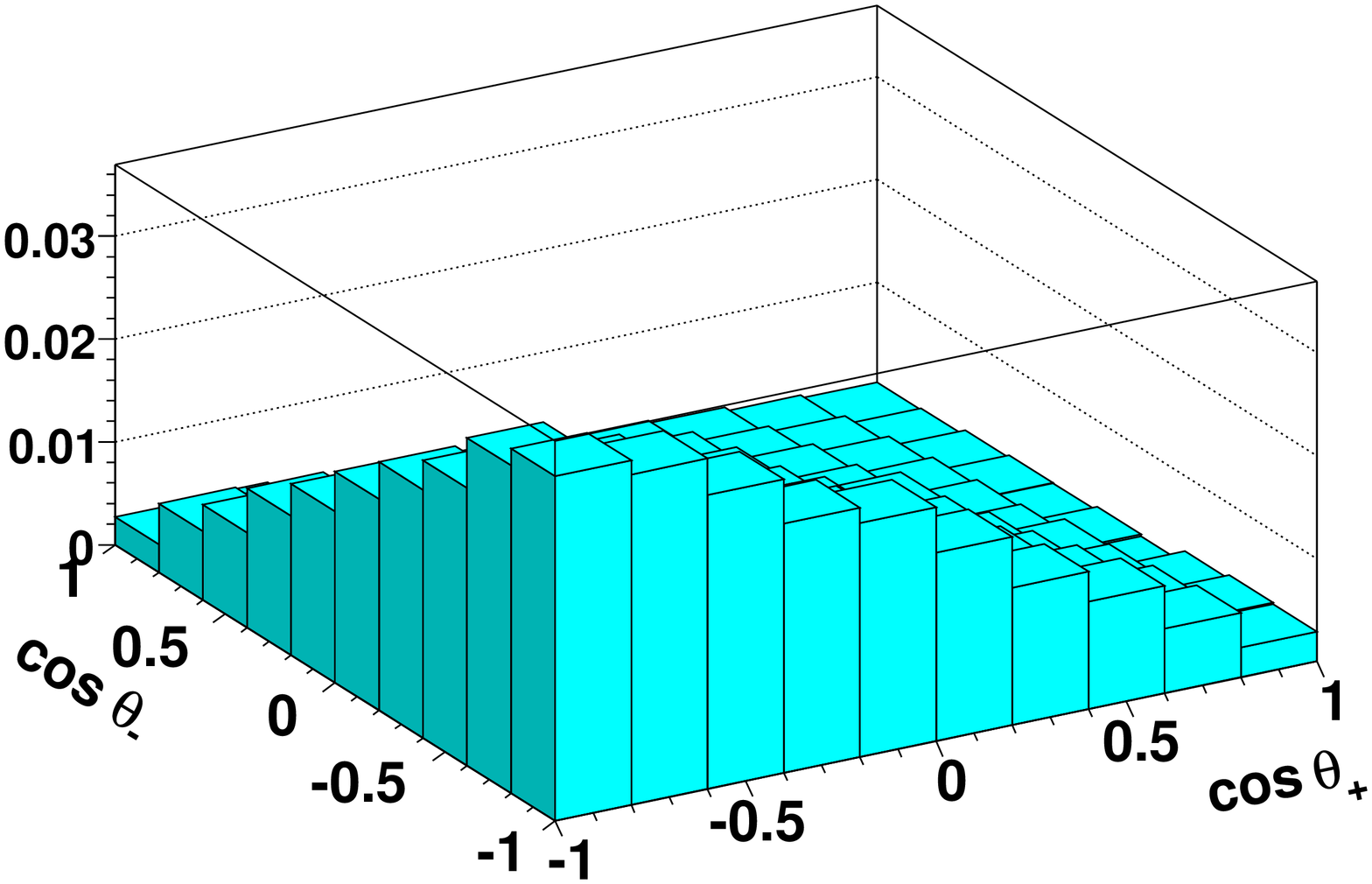, angle=-90, width=0.3\textwidth, clip=}
      }
    }
    \mbox{
      \subfigure[]{
        \epsfig{file=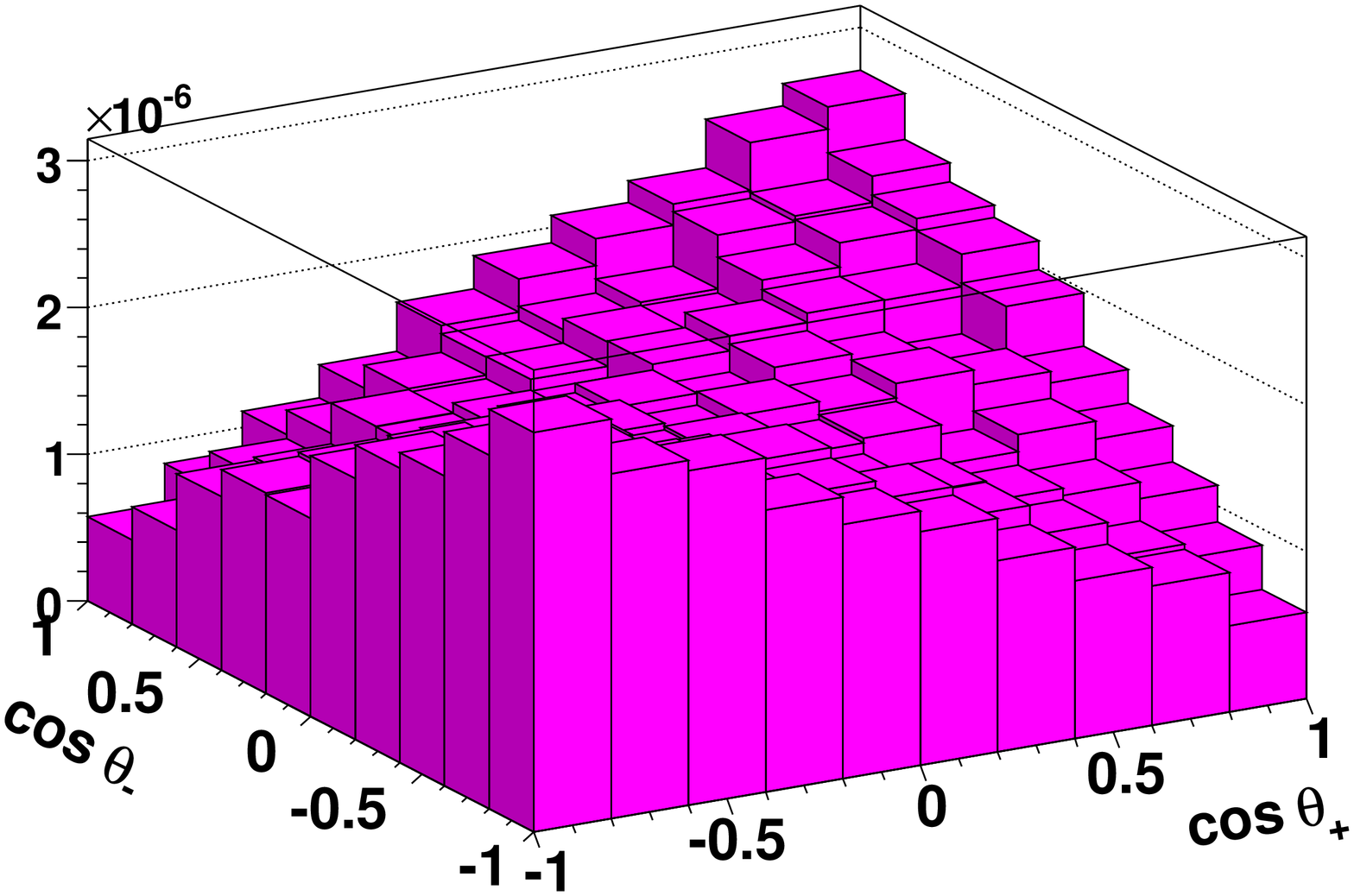, angle=-90, width=0.3\textwidth, clip=}
      }
    }
  \end{center}
\vspace{-20pt}
      \caption{
The distribution $\frac{1}{\sigma}\frac{d^2\sigma}{d\cos\theta_+d\cos\theta_-}$ for
(a) scalar,
(b) pseudo-scalar,
(c) vector,
(d) axial-vector,
(e) vector-left,
(f) vector-right,
(g) spin-2.
$M_X=$ 800 GeV at the LHC, using the pdf set CTEQ6L1. No cuts were applied.
}
      \label{spin_3D_800}
\end{figure*}

The differences among the various  $t\bar{t}$ production mechanisms are
manifest. In Fig.~\ref{spin_3D_800} the distributions
are plotted for resonance masses of 800 GeV for 
the following states:
\begin{itemize}
\item Scalar boson (a), 
\item Pseudo-scalar boson (b),
\item Vector boson (c),
\item Axial-vector boson (d),
\item Vector-left boson (e),
\item Vector-right boson (f),
\item Spin-2 boson (g).
\end{itemize}

With the vector-left and vector-right we understand a spin-1 vector
boson that couples \emph{only} to left- or right-handed fermions,
respectively. We choose very narrow resonances by taking the width of
resonances to be 1\% of the mass, {\it i.e.},~8 GeV for a mass of
800 GeV. We do not include the SM QCD $t\bar{t}$
production background in these plots.

In Table~\ref{table_shape} the distributions are fitted to Eq.~(\ref{eq2})
and compared with analytic computations. For the sake of simplicity, 
in the analytic computations the off-diagonal elements of the spin 
correlations matrix in the helicity basis are neglected. 
This means that the interference between different 
top quark spins are not included. In fact, the interference effects are negligible 
and the fitted values agree very well with the analytic computations. 
For completeness we also included the
the numbers for a smaller resonance mass, $M_X=$ 400 GeV, where the effects from the
mass of the top play a larger role.

\begin{table*}[t]
\begin{center}
\begin{tabular}{|l|c|c c|c c|c c|}
\hline
resonance    &mass (GeV)&$A$ calc.&$A$ fit.&$b_+$ calc.&$b_+$ fit.&$b_-$ calc.&$b_-$ fit.\\
\hline
sm           &  --      &    0.319&   0.304&        0&   0.008&        0&  -0.003\\
sm           & $390<m_{t\bar{t}}<410$ &  0.501&   0.532&        0&   0.004&        0&   0.005\\
sm           & $790<m_{t\bar{t}}<810$ & -0.061&  -0.051&        0&  -0.014&        0&  -0.011\\
\hline
scalar       & 400      &        1&   0.972&        0&   0.005&        0&   0.007\\
pseudo-scalar& 400      &        1&   0.966&        0&   0.007&        0&   0.002\\
vector       & 400      &   -0.449&  -0.432&        0&   0.008&        0&  -0.004\\
axial-vector & 400      &       -1&  -0.990&        0&  -0.004&        0&   0.002\\
vector-left  & 400      &   -0.531&  -0.536&    0.605&   0.607&    0.605&   0.600\\
vector-right & 400      &   -0.531&  -0.558&   -0.605&  -0.604&   -0.605&  -0.610\\
spin-2       & 400      &       --&  -0.348&        0&   0.001&        0&   0.006\\
\hline
scalar       & 800      &        1&   0.985&        0&  -0.015&        0&   0.004\\
pseudo-scalar& 800      &        1&   0.978&        0&  -0.004&        0&  -0.004\\
vector       & 800      &   -0.826&  -0.819&        0&   0.008&        0&   0.005\\
axial-vector & 800      &       -1&  -1.001&        0&   0.008&        0&   0.008\\
vector-left  & 800      &   -0.900&  -0.912&    0.945&   0.955&    0.945&   0.946\\
vector-right & 800      &   -0.900&  -0.884&   -0.945&  -0.938&   -0.945&  -0.943\\
spin-2       & 800      &       --&  -0.743&        0&   0.022&        0&   0.013\\
\hline
\end{tabular}
\end{center}
\caption{$\frac{1}{\sigma}\frac{d^2\sigma}{d\cos\theta_+d\cos\theta_-}=\frac{1}{4}
(1-A\cos\theta_+\cos\theta_-+b_+\cos\theta_++b_-\cos\theta_-)$.
For a top mass of 175 GeV. In the analytic calculation of the parameters,
the interference between the various top spins is neglected.}
\label{table_shape}
\end{table*}

\begin{figure*}[t]
\begin{center}
      \epsfig{file=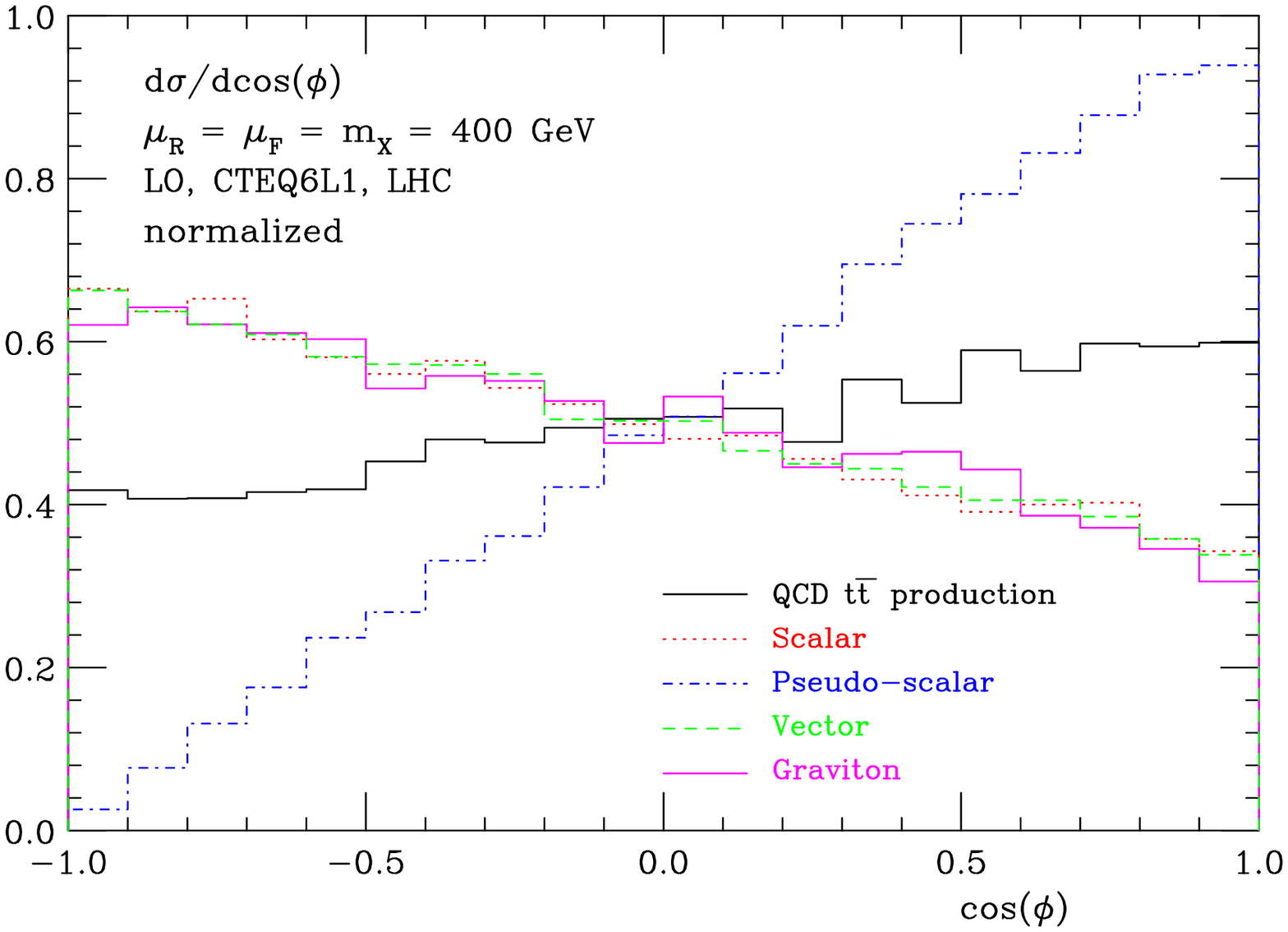,angle=90, width=0.65\textwidth}
\end{center}
\vspace{-20pt}
\caption{
Distribution for the angle $\phi$, defined in the text,
for different $t\bar{t}$ production mechanisms at the LHC.
\emph{Dark Solid} line is the SM $t\bar{t}$ production,
\emph{dotted line} is $t\bar{t}$ production through a scalar,
\emph{dot-dashed} line is $t\bar{t}$ production through a pseudo-scalar,
\emph{dashed} line is $t\bar{t}$ production through a vector (this is independent of the coupling).
The \emph{light solid} line is $t\bar{t}$ production through a graviton.
The plots are normalized. The pdf set CTEQ6l1 is used with $M_X=$ 800 GeV and $\mu_R=\mu_F=$ 800 GeV.
}
\label{phi}
\end{figure*}

The second angle, which is commonly considered when studying
 spin correlations in $t\bar{t}$ production is $\phi$,
{\it i.e.}, the angle between the directions of the $f_d^+$ and $f_d^-$ 
in the $t$ and $\bar{t}$ rest frames, respectively. The distribution
\begin{equation}\label{eq3}
\frac{1}{\sigma}\frac{d\sigma}{d\cos\phi}=\tfrac{1}{2}(1-D\cos\phi),
\end{equation}
for this angle is plotted in Fig.~\ref{phi} for a resonance mass
of 800 GeV.

The distributions for the angle $\phi$ are the same for production 
through a scalar and a  vector boson. 
The distribution for the pseudo-scalar, on the other hand, 
is completely different from the one for the scalar
and the vector boson~\cite{Bernreuther:1997gs}.
Also, the angular distribution for SM $t\bar{t}$ 
production is different from the other production mechanisms.
In the case of a spin-1 state, the $\phi$ distribution is independent 
of the type of coupling to the top: it makes no difference whether
it is pure vector,  an axial-vector, a left-handed or right-handed couplings.

\section{Conclusions}\label{sec:con}

Top physics is entering the precision phase at the Tevatron and will
be one of the leading priorities at the LHC. The importance of the top quark
in the quest for the mechanism of EWSB and new physics is due to the
convergence of two factors. First the LHC will be a top factory with
tens of millions of top quarks produced in the first years of running at the
nominal low luminosity, $2\times 10^{33}$ cm${}^{-2}$ s${}^{-1}$, allowing studies at
an unprecedented level of accuracy.  Second, due to its large mass, top
is the optimal probe for new physics at TeV scales. Many different
models, including the SM, predict the existence of heavy states that
preferably couple to the top quark, and that could affect its SM
couplings via loop corrections or be directly produced at the LHC.

Given the large number of the models proposed and their complexity, a
``top-down'' approach, {\it e.g.}, model parameter scanning, will not
be practical, in particular if comparison of many different channels
and observables at once will be necessary.  As an alternative, a
simpler and more pragmatic ``bottom-up'' approach could be employed,
whereby one identifies specific observables which can be developed as
tools for discriminating generic features of new physics resonances,
thus keeping the analysis as model independent as possible.

In this paper we have presented an example on how such a study could
be performed for the invariant mass distribution of the $t\bar t$
pair.

As a first step we have assessed the accuracy of the best theoretical
predictions available for $t \bar t$ production at hadron colliders.
We have found that the shape of the distribution is under good
 theoretical control, especially at low invariant mass values,
suggesting also the possibility of a precise top mass extraction.

We have then identified the features of new physics scenarios, namely
the existence of heavy bosonic resonances of various spin, color and
parity, that could show up in the $m_{t\bar t}$ distribution, and
implemented them in the MadGraph/MadEvent package.
The full matrix elements, $pp \to X \to t\bar t \to 6f$, $X$ being a
spin-0, spin-1 or spin-2 particle, particles with arbitrary masses,
width, color and couplings, have been automatically generated by
MadGraph. The effects due to the interference with the $pp \to t\bar
t$ SM process are included when relevant.

The strategy to gain information on new physics is then
straightforward and consists of three successive steps:
\begin{enumerate}
\item The discovery of the resonance (and the determination
of its mass and width) which could appear as
a sharp or broad peak or as a more distinctive peak-dip structure in
very specific cases.  In this measurement the key aspect will be the
experimental resolution in the $m_{t\bar t}$ reconstruction.
\item The identification of the spin of the resonance, which can be inferred
from the angular distribution of the top and the anti-top.
\item Information on the couplings of the resonance to the the top anti-top
pair, which can be obtained by measuring the spin correlations of the
top anti-top pair (for this last step the full matrix matrix element
$2 \to 6$ is required).
\end{enumerate}

In conclusion, we have outlined a simple strategy and provided the
necessary Monte Carlo tools to search for new resonances in $t \bar t$
events. We look forward to more detailed experimental analyses.

\section*{Acknowledgments}
We are thankful to Johan Alwall for his help and numerous
contributions at all stages of this project and to Tony Liss for
stimulating conversations, comments and suggestions. We are in debt to
Dave Rainwater for support in the development of the spin-2 HELAS
subroutines. We would also like to thank all the members of CP3 for
the great atmosphere and environment that foster our efforts and, in
particular, Keith Hamilton and Andrea Giammanco for making comments on
the manuscript. This work is partially supported by the Belgian
Federal Science Policy (IAP 6/11).

\section*{Appendix: reconstruction issues in $t\bar{t}$ events}
\label{sec:appendix}

In this appendix we address some of the issues arising in the
reconstruction of the $t\bar{t}$ events and in particular their
different impact in the three-step analyses proposed above.

For a generic $2\to 6$ process, where the final state particles masses
are known, 16 independent variables are needed to determine the
kinematics of the event: the six final state particle three-momenta, $
\{{\bf p}_i\}$, the two energy fractions carried by the initial state
partons, $x_{a,b}$, minus the overall momentum conservation which
reduces the number of independent variables by 4, leads to $6\times
3+2-4=16$.

From the measured angles and energies of the final state particles, together
with constraints from $W$ boson and top quark masses, a system of
equations can be set up to solve for the 16 unknown variables on an
event-by-event basis.

The three decay modes of the $t\bar{t}$ pairs face each their own
challenges for detection and reconstruction. Around 44\% of the
$t\bar{t}$ pairs decay hadronically, 30\% of the $t\bar{t}$ pairs
decay single-leptonically and 5\% double leptonic (not including
tau's)~\cite{Beneke:2000hk}\footnote{The remaining 21\% are events including decays to
tau's, which are not be considered here.}. These
three channels offer very different challenges related to the detection of
final state particles and the reconstruction the (anti-)top quark momenta,
which we will now address channel-by-channel.

The \emph{fully-hadronic} decays have the advantage that in principle
the momenta of all the final state particles can be determined,
leading to $6\times 3=18$ measurements. Together with the four mass
constraints the system of equations for the 16 independent variables
is over-constrained. Such constraints can be used in two ways. First
they can be used to extract information, typically the jet energies,
that have bad detector resolution. For example, measuring only the
angles $(\theta, \phi$) of the six jets and including all the
constraints from the top quark and $W$ boson masses would already
provide the required 16 independent quantities (althought with
combinatorics). Alternatevely, the constraints from the masses can
also be used to solve the combinatorics in reconstructing the $W$
boson and (anti-)top quark momenta.  Combinatorics, affect each of the
three steps in the analysis proposed in this paper in a different
way. In the first step, {\it i.e.}, the measurement of the $t\bar{t}$
invariant mass, Sec.~\ref{sec:BSM}, combinatorics play no role: the
invariant mass can be calculated by summing all the final state
momenta irrespective of assigning jets to top or anti-top quarks,
$W^+$ or $W^-$ bosons. In the second step, {\it i.e.} the measurement
of the spin of an intermediate resonance, Sec.~\ref{sec:spin_info},
there is in principle a 12-fold ambiguity (assuming $b$-tagging) in
assigning the correct ($b$-)jets to the top or anti-top quark. These
ambiguities could be solved (in case of a very good jet energy
resolution) or anyway allieviated by using the constraints from the
top quark and $W$ masses. In the third step, {\it i.e.}, the
measurement of the spin correlations of the top anti-top pair,
Sec.~\ref{sec:spin_corr}, not all ambiguities can be solved: in any
case is not possibly to uniquely identify on event-by-event basis
which of the two jets come from the down-type quarks in the $W$ boson
decays.  Experimentally, the fully-hadronic decay is difficult to
trigger and extract from multi-jet backgrounds, which makes this
channel challgenging for BSM physics studies.~\cite{CMSTDR}

The \emph{single-lepton} decay channel is much more promising.  The
single lepton in the final state greatly improves the possibility for
triggering on these events and extracting it from backgrounds compared
to the fully-hadronic decay mode.  The presence of a missing neutrino
in the final state entails that only $5\times 3=15$ independent
measurements can be obtained, one short of 16 necessary. The missing
information can be recovered by including a constraint coming from,
{\it e.g.}, the $W$ boson mass (up to a two-fold
quadratic ambiguity, which can be solved in various ways, {\it e.g.},
see Ref.~\cite{Barger:2006hm}). Using also the constraint from the top
mass removes the ambiguity for assigning the correct $b$-jet to the
top quark needed for second and third step in our analysis.  For the
third step, however, it is still non-trivial to solve the ambiguity
coming from assigning the correct jet to the down-type quark in the
non-leptonic $W$ decay. Several methods have been proposed, including,
for example choosing the least energetic
non-$b$-jet~\cite{Bernreuther:2004jv,Mahlon:1995zn}. Given its rate
and the various reconstruction studies and possibilities, the
single-lepton channel is the most straightforward search channel for
BSM physics in $t\bar{t}$ events.

In the \emph{double-lepton} decay mode there are two missing
neutrino's in the final state. This makes the reconstruction of the
full event kinematics challenging but certainly not
impossible~\cite{Dalitz:1992np,Beneke:2000hk,Sonnenschein:2006ud,Arai:2007ts,PRL.95.022001,CMSdi,ATLASdi}.
There are four visible particles in the final state, two $b$-jets and
two opposite sign leptons, leading to $4\times 3=12$ independent
measurements.  The additional four constraints from the $W$ boson and
top quark masses are just enough to set up a system of non-linear
equations to solve for the necessary 16 variables.  It can be shown
that this system has up to eight solutions~\cite{Sonnenschein:2006ud},
of which, in general, only a few are physical and can be discarded or
included based on their likelihood.  It has to be noted that each
solution has no further ambiguities and the event is completely
reconstructed. For this reason, despite the small branching ratio, this
channel competes in reach with the single-lepton in the studies of the
spin correlation studies in
$t\bar{t}$~\cite{Beneke:2000hk,Borjanovic:2004ce}.


\begin{thebibliography}{99}
\bibitem{Hill:1991at}
  C.~T.~Hill,
  Phys.\ Lett.\  B {\bf 266}, 419 (1991).

\bibitem{Hill:1994hp}
  C.~T.~Hill,
  Phys.\ Lett.\  B {\bf 345}, 483 (1995)
  [arXiv:hep-ph/9411426].

\bibitem{Hill:1993hs}
  C.~T.~Hill and S.~J.~Parke,
  Phys.\ Rev.\  D {\bf 49}, 4454 (1994)
  [arXiv:hep-ph/9312324].

\bibitem{Harris:1999ya}
  R.~M.~Harris, C.~T.~Hill and S.~J.~Parke,
  arXiv:hep-ph/9911288.

\bibitem{Dobrescu:1997nm}
  B.~A.~Dobrescu and C.~T.~Hill,
  Phys.\ Rev.\ Lett.\  {\bf 81}, 2634 (1998)
  [arXiv:hep-ph/9712319].

\bibitem{Chivukula:1998wd}
  R.~S.~Chivukula, B.~A.~Dobrescu, H.~Georgi and C.~T.~Hill,
  Phys.\ Rev.\  D {\bf 59}, 075003 (1999)
  [arXiv:hep-ph/9809470].

\bibitem{He:2001fz}
  H.~J.~He, C.~T.~Hill and T.~M.~P.~Tait,
  Phys.\ Rev.\  D {\bf 65}, 055006 (2002)
  [arXiv:hep-ph/0108041].

\bibitem{Hill:2002ap}
  C.~T.~Hill and E.~H.~Simmons,
  Phys.\ Rept.\  {\bf 381}, 235 (2003)
  [Erratum-ibid.\  {\bf 390}, 553 (2004)]
  [arXiv:hep-ph/0203079].

\bibitem{ArkaniHamed:2001nc}
  N.~Arkani-Hamed, A.~G.~Cohen and H.~Georgi,
  Phys.\ Lett.\  B {\bf 513}, 232 (2001)
  [arXiv:hep-ph/0105239].

\bibitem{ArkaniHamed:2002pa}
  N.~Arkani-Hamed, A.~G.~Cohen, T.~Gregoire and J.~G.~Wacker,
  JHEP {\bf 0208}, 020 (2002)
  [arXiv:hep-ph/0202089].

\bibitem{ArkaniHamed:2002qx}
  N.~Arkani-Hamed, A.~G.~Cohen, E.~Katz, A.~E.~Nelson, T.~Gregoire and J.~G.~Wacker,
  JHEP {\bf 0208}, 021 (2002)
  [arXiv:hep-ph/0206020].

\bibitem{ArkaniHamed:2002qy}
  N.~Arkani-Hamed, A.~G.~Cohen, E.~Katz and A.~E.~Nelson,
  JHEP {\bf 0207}, 034 (2002)
  [arXiv:hep-ph/0206021].

\bibitem{Low:2002ws}
  I.~Low, W.~Skiba and D.~Smith,
  Phys.\ Rev.\  D {\bf 66}, 072001 (2002)
  [arXiv:hep-ph/0207243].

\bibitem{Han:2003wu}
  T.~Han, H.~E.~Logan, B.~McElrath and L.~T.~Wang,
  Phys.\ Rev.\  D {\bf 67}, 095004 (2003)
  [arXiv:hep-ph/0301040].

\bibitem{Azuelos:2004dm}
  G.~Azuelos {\it et al.},
  Eur.\ Phys.\ J.\  C {\bf 39S2}, 13 (2005)
  [arXiv:hep-ph/0402037].

\bibitem{Schmaltz:2005ky}
  M.~Schmaltz and D.~Tucker-Smith,
  Ann.\ Rev.\ Nucl.\ Part.\ Sci.\  {\bf 55}, 229 (2005)
  [arXiv:hep-ph/0502182].

\bibitem{ArkaniHamed:1998rs}
  N.~Arkani-Hamed, S.~Dimopoulos and G.~R.~Dvali,
  Phys.\ Lett.\  B {\bf 429}, 263 (1998)
  [arXiv:hep-ph/9803315].

\bibitem{Randall:1999ee}
  L.~Randall and R.~Sundrum,
  Phys.\ Rev.\ Lett.\  {\bf 83}, 3370 (1999)
  [arXiv:hep-ph/9905221].

\bibitem{Fitzpatrick:2007qr}
  A.~L.~Fitzpatrick, J.~Kaplan, L.~Randall and L.~T.~Wang,
  JHEP {\bf 0709}, 013 (2007)
  [arXiv:hep-ph/0701150].

\bibitem{Arai:2004yd}
  M.~Arai, N.~Okada, K.~Smolek and V.~Simak,
  Phys.\ Rev.\  D {\bf 70}, 115015 (2004)
  [arXiv:hep-ph/0409273].

\bibitem{Arai:2007ts}
  M.~Arai, N.~Okada, K.~Smolek and V.~Simak,
  Phys.\ Rev.\  D {\bf 75}, 095008 (2007)
  [arXiv:hep-ph/0701155].

\bibitem{McMullen:2001zj}
  C.~D.~McMullen and S.~Nandi,
  arXiv:hep-ph/0110275.

\bibitem{Agashe:2003zs}
  K.~Agashe, A.~Delgado, M.~J.~May and R.~Sundrum,
  JHEP {\bf 0308}, 050 (2003)
  [arXiv:hep-ph/0308036].

\bibitem{Agashe:2006hk}
  K.~Agashe, A.~Belyaev, T.~Krupovnickas, G.~Perez and J.~Virzi,
  Phys.\ Rev.\  D {\bf 77}, 015003 (2008)
  [arXiv:hep-ph/0612015].

\bibitem{Lillie:2007yh}
  B.~Lillie, L.~Randall and L.~T.~Wang,
  JHEP {\bf 0709}, 074 (2007)
  [arXiv:hep-ph/0701166].

\bibitem{Djouadi:2007eg}
  A.~Djouadi, G.~Moreau and R.~K.~Singh,
  Nucl.\ Phys.\  B {\bf 797}, 1 (2008)
  [arXiv:0706.4191 [hep-ph]].

\bibitem{Ghavri:2006kc}
  R.~Ghavri, C.~D.~McMullen and S.~Nandi,
  Phys.\ Rev.\  D {\bf 74}, 015012 (2006)
  [arXiv:hep-ph/0602014].

\bibitem{Burdman:2006gy}
  G.~Burdman, B.~A.~Dobrescu and E.~Ponton,
  Phys.\ Rev.\  D {\bf 74}, 075008 (2006)
  [arXiv:hep-ph/0601186].

\bibitem{Lillie:2007ve}
  B.~Lillie, J.~Shu and T.~M.~P.~Tait,
  Phys.\ Rev.\  D {\bf 76}, 115016 (2007)
  [arXiv:0706.3960 [hep-ph]].

\bibitem{Agashe:2007ki}
  K.~Agashe {\it et al.},
  Phys.\ Rev.\  D {\bf 76}, 115015 (2007)
  [arXiv:0709.0007 [hep-ph]].

\bibitem{Stelzer:1994ta}
  T.~Stelzer and W.~F.~Long,
  Comput.\ Phys.\ Commun.\  {\bf 81}, 357 (1994)
  [arXiv:hep-ph/9401258].

\bibitem{Maltoni:2002qb}
  F.~Maltoni and T.~Stelzer,
  JHEP {\bf 0302}, 027 (2003)
  [arXiv:hep-ph/0208156].

\bibitem{Alwall:2007st}
  J.~Alwall {\it et al.},
  JHEP {\bf 0709}, 028 (2007)
  [arXiv:0706.2334 [hep-ph]].

\bibitem{Barger:2006hm}
  V.~Barger, T.~Han and D.~G.~E.~Walker,
  Phys.\ Rev.\ Lett.\  {\bf 100}, 031801 (2008)
  [arXiv:hep-ph/0612016].

\bibitem{Baur:2007ck}
  U.~Baur and L.~H.~Orr,
  Phys.\ Rev.\  D {\bf 76}, 094012 (2007)
  [arXiv:0707.2066 [hep-ph]].

\bibitem{Cacciari:2003fi}
  M.~Cacciari, S.~Frixione, M.~L.~Mangano, P.~Nason and G.~Ridolfi,
  JHEP {\bf 0404}, 068 (2004)
  [arXiv:hep-ph/0303085].

\bibitem{Campbell:1999ah}
  J.~M.~Campbell and R.~K.~Ellis,
  Phys.\ Rev.\  D {\bf 60}, 113006 (1999)
  [arXiv:hep-ph/9905386].

\bibitem{Pumplin:2002vw}
  J.~Pumplin, D.~R.~Stump, J.~Huston, H.~L.~Lai, P.~Nadolsky and W.~K.~Tung,
  JHEP {\bf 0207}, 012 (2002)
  [arXiv:hep-ph/0201195].

\bibitem{Bonciani:1998vc}
  R.~Bonciani, S.~Catani, M.~L.~Mangano and P.~Nason,
  Nucl.\ Phys.\  B {\bf 529}, 424 (1998)
  [arXiv:hep-ph/9801375].

\bibitem{Beneke:2000hk}
  M.~Beneke {\it et al.},
  arXiv:hep-ph/0003033.

\bibitem{matteo}
  M.~Cacciari, private communication.

\bibitem{Frixione:2003ei}
  S.~Frixione, P.~Nason and B.~R.~Webber,
  JHEP {\bf 0308}, 007 (2003)
  [arXiv:hep-ph/0305252].

\bibitem{Kuhn:2006vh}
  J.~H.~Kuhn, A.~Scharf and P.~Uwer,
  Eur.\ Phys.\ J.\  C {\bf 51}, 37 (2007)
  [arXiv:hep-ph/0610335].

\bibitem{Bernreuther:2006vg}
  W.~Bernreuther, M.~Fuecker and Z.~G.~Si,
  Phys.\ Rev.\  D {\bf 74}, 113005 (2006)
  [arXiv:hep-ph/0610334].

\bibitem{Ferrari:2007qf}
  P.~Ferrari,
  arXiv:0705.3021 [hep-ex].

\bibitem{Baur:2008uv}
  U.~Baur and L.~H.~Orr,
  arXiv:0803.1160 [hep-ph].

\bibitem{Moch:2008qy}
  S.~Moch and P.~Uwer,
  Phys.\ Rev.\  D {\bf 78}, 034003 (2008)
  [arXiv:0804.1476 [hep-ph]].

\bibitem{Smith:1996xz}
  M.~C.~Smith and S.~S.~Willenbrock,
  Phys.\ Rev.\ Lett.\  {\bf 79}, 3825 (1997)
  [arXiv:hep-ph/9612329].

\bibitem{Sjostrand:2006za}
  T.~Sjostrand, S.~Mrenna and P.~Skands,
  JHEP {\bf 0605}, 026 (2006)
  [arXiv:hep-ph/0603175].

\bibitem{Corcella:2001wc}
  G.~Corcella {\it et al.},
  arXiv:hep-ph/0201201.

\bibitem{Gaemers:1984sj}
  K.~J.~F.~Gaemers and F.~Hoogeveen,
  Phys.\ Lett.\  B {\bf 146}, 347 (1984).

\bibitem{Dicus:1994bm}
  D.~Dicus, A.~Stange and S.~Willenbrock,
  Phys.\ Lett.\  B {\bf 333}, 126 (1994)
  [arXiv:hep-ph/9404359].

\bibitem{Bernreuther:1997gs}
  W.~Bernreuther, M.~Flesch and P.~Haberl,
  Phys.\ Rev.\  D {\bf 58}, 114031 (1998)
  [arXiv:hep-ph/9709284].

\bibitem{Manohar:2006ga}
  A.~V.~Manohar and M.~B.~Wise,
  Phys.\ Rev.\  D {\bf 74}, 035009 (2006)
  [arXiv:hep-ph/0606172].

\bibitem{Gresham:2007ri}
  M.~I.~Gresham and M.~B.~Wise,
  Phys.\ Rev.\  D {\bf 76}, 075003 (2007)
  [arXiv:0706.0909 [hep-ph]].

\bibitem{Simmons:1996fz}
  E.~H.~Simmons,
  Phys.\ Rev.\  D {\bf 55}, 1678 (1997)
  [arXiv:hep-ph/9608269].

\bibitem{Choudhury:2007ux}
  D.~Choudhury, R.~M.~Godbole, R.~K.~Singh and K.~Wagh,
  Phys.\ Lett.\  B {\bf 657}, 69 (2007)
  [arXiv:0705.1499 [hep-ph]].

\bibitem{Dicus:2000hm}
  D.~A.~Dicus, C.~D.~McMullen and S.~Nandi,
  Phys.\ Rev.\  D {\bf 65}, 076007 (2002)
  [arXiv:hep-ph/0012259].

\bibitem{Djouadi:2005gj}
  A.~Djouadi,
  Phys.\ Rept.\  {\bf 459}, 1 (2008)
  [arXiv:hep-ph/0503173].

\bibitem{Cheung:2004ad}
  K.~Cheung and W.~Y.~Keung,
  Phys.\ Rev.\  D {\bf 71}, 015015 (2005)
  [arXiv:hep-ph/0408335].

\bibitem{Frampton:1987ut}
  P.~H.~Frampton and S.~L.~Glashow,
  Phys.\ Rev.\ Lett.\  {\bf 58}, 2168 (1987).

\bibitem{Frampton:1987dn}
  P.~H.~Frampton and S.~L.~Glashow,
  Phys.\ Lett.\  B {\bf 190}, 157 (1987).

\bibitem{Antunano:2007da}
  O.~Antunano, J.~H.~Kuhn and G.~Rodrigo,
  Phys.\ Rev.\  D {\bf 77}, 014003 (2008)
  [arXiv:0709.1652 [hep-ph]].

\bibitem{ArkaniHamed:1998nn}
  N.~Arkani-Hamed, S.~Dimopoulos and G.~R.~Dvali,
  Phys.\ Rev.\  D {\bf 59}, 086004 (1999)
  [arXiv:hep-ph/9807344].

\bibitem{Randall:1999vf}
  L.~Randall and R.~Sundrum,
  Phys.\ Rev.\ Lett.\  {\bf 83}, 4690 (1999)
  [arXiv:hep-th/9906064].

\bibitem{Collins:1977iv}
  J.~C.~Collins and D.~E.~Soper,
  Phys.\ Rev.\  D {\bf 16}, 2219 (1977).

\bibitem{Alwall:2006bx}
  J.~Alwall {\it et al.},
  Eur.\ Phys.\ J.\  C {\bf 49}, 791 (2007)
  [arXiv:hep-ph/0607115].

\bibitem{Willenbrock:2002ta}
  S.~Willenbrock,
  arXiv:hep-ph/0211067.

\bibitem{Jezabek:1994zv}
  M.~Jezabek and J.~H.~Kuhn,
  Phys.\ Lett.\  B {\bf 329}, 317 (1994)
  [arXiv:hep-ph/9403366].

\bibitem{Czarnecki:1990pe}
  A.~Czarnecki, M.~Jezabek and J.~H.~Kuhn,
  Nucl.\ Phys.\  B {\bf 351}, 70 (1991).

\bibitem{Bernreuther:2004jv}
  W.~Bernreuther, A.~Brandenburg, Z.~G.~Si and P.~Uwer,
  Nucl.\ Phys.\  B {\bf 690}, 81 (2004)
  [arXiv:hep-ph/0403035].

\bibitem{Mahlon:1995zn}
  G.~Mahlon and S.~J.~Parke,
  Phys.\ Rev.\  D {\bf 53}, 4886 (1996)
  [arXiv:hep-ph/9512264].

\bibitem{Dalitz:1992np}
  R.~H.~Dalitz and G.~R.~Goldstein,
  Phys.\ Lett.\  B {\bf 287}, 225 (1992).

\bibitem{Sonnenschein:2006ud}
  L.~Sonnenschein,
  Phys.\ Rev.\  D {\bf 73}, 054015 (2006)
  [arXiv:hep-ph/0603011].

\bibitem{PRL.95.022001}
  D.~E.~Acosta {\it et al.}  [CDF Collaboration],
  Phys.\ Rev.\ Lett.\  {\bf 95}, 022001 (2005)
  [arXiv:hep-ex/0412042].

\bibitem{CMSdi}
  M. Baarmand {\it et al.}, CMS NOTE 2006/111 (2006).

\bibitem{ATLASdi}
  K.~Smolek and V.~Simak, ATL-PHYS-2003-012 (2003).

\bibitem{CMSTDR}
  G.~L.~Bayatian {\it et al.}  [CMS Collaboration],
  J.\ Phys.\ {\bf G34}, 995 (2007).

\bibitem{Borjanovic:2004ce}
  I.~Borjanovic {\it et al.},
  Eur.\ Phys.\ J.\  C {\bf 39S2}, 63 (2005)
  [arXiv:hep-ex/0403021].

\end{thebibliography}
\end{document}